\documentclass[%
reprint,
onecolumn,
superscriptaddress,
nofootinbib,
amsmath,amssymb,
aps,
pre,
]{revtex4-1}

\usepackage{enumerate}
\usepackage{graphicx}
\usepackage{dcolumn}
\usepackage{bm}
\usepackage[caption=false]{subfig}
\usepackage{amssymb,amsfonts,amsmath}
\usepackage{mathtools,color} 
\usepackage{float}

\usepackage[normalem]{ulem}

\def\be{\begin{equation}}
\def\ee{\end{equation}}
\def\nn{\nonumber}

\def\bea{\begin{eqnarray}}
\def\eea{\end{eqnarray}}
\def\beas{\begin{eqnarray*}}
\def\eeas{\end{eqnarray*}}
\def\bsplit{\begin{split}}
\def\esplit{\end{split}}

\def\nn{\nonumber}
\def\f{\frac}

\def\l{\left(}
\def\r{\right)}
\def\L{\left[}
\def\R{\right]}
\def\bs{\boldsymbol}
\def\mbf{\mathbf}

\def\a{\alpha}

\def\abs#1{|#1|}
\def\norm#1{\|#1\|}
\def\Eja{E^{(j)}_{\alpha}}
\def\cjk{c^{(j)}_k}

\newcommand{\rri}{\affiliation{Raman Research Institute, Bangalore 560080, India}}
\newcommand{\curie}{\affiliation{Laboratoire Physico Chimie Curie, Institut Curie, CNRS UMR168, 75005 Paris, France}}
\newcommand{\columbia}{\affiliation{Industrial Engineering and Operations Research, Columbia University, New York 10027, USA}}
\newcommand{\simons}{\affiliation{Simons Centre for the Study of Living Machines, National Centre for Biological Sciences (TIFR), Bangalore 560065, India}}

\begin{document}


\title{
Glycan processing in the Golgi --  optimal information coding and constraints on cisternal number and enzyme specificity
}


\author{Alkesh Yadav}
\rri

\author{Quentin Vagne}
\curie

\author{Pierre Sens}
\curie

\author{Garud Iyengar}
\columbia

\author{Madan Rao}
\simons
\email{madan@ncbs.res.in}

\date{\today}





\begin{abstract} 
Many proteins that undergo sequential enzymatic
modification in the Golgi cisternae are displayed at the plasma membrane
as cell
identity markers. The modified proteins, called glycans, 
represent a molecular code.
The fidelity of this {\it glycan code} is measured by how accurately the glycan synthesis machinery realises
 the desired target glycan distribution for a particular cell type and niche.
In this paper, we quantitatively analyse
the tradeoffs between the number of cisternae and the number and
specificity of enzymes, 
in order to synthesize a prescribed target glycan distribution of a
certain complexity.
We find that to synthesize 
complex 
distributions, such as those observed in real
cells, one needs to have  multiple cisternae and precise enzyme partitioning in the 
Golgi. Additionally, 
for fixed number of enzymes and cisternae, there is an optimal level of
specificity of enzymes that achieves 
the  
target distribution with high fidelity.
Our results show how  
the complexity of the target glycan distribution places functional 
constraints on the 
Golgi cisternal number and enzyme specificity.
\end{abstract}

\maketitle 

\section{Introduction}
\label{sec:intro}
A majority of the proteins synthesized in the endoplasmic
reticulum (ER) are transferred to the Golgi 
cisternae 
for further chemical modification by glycosylation~\cite{mboc}, a process that 
sequentially and covalently attaches sugar moieties to proteins, 
catalyzed by a set of enzymatic reactions within the ER and the Golgi cisternae. These enzymes, called
glycosyltransferases, 
are localized in the ER and cis-medial and trans Golgi cisternae in a specific manner~\cite{varki,Cummings2014}.
Glycans, the final products of this glycosylation assembly line are
delivered to the plasma membrane (PM) conjugated with proteins, whereupon they engage in
multiple cellular functions, 
including immune recognition, 
cell identity markers, cell-cell adhesion and cell
signalling~\cite{varki,Cummings2014,varki2017,Drickamer1998,Gagneux&Varki1999}.   
This \emph{glycan code}~\cite{gabius2018,dwek}, representing
information~\cite{winterburn1972} about the cell, is 
generated dynamically, 
following the biochemistry of sequential enzymatic reactions and the biophysics of secretory
transport~\cite{varki2017,varki1998,parashuraman2019}.

In this paper, we will focus on the role of glycans as markers of cell identity. 
For the glycans to play this role, they must inevitably represent a
molecular code~\cite{gabius2018,varki2017,parashuraman2019}.  
While the functional consequences of glycan alterations have been well
studied, 
the glycan code has 
remained an enigma~\cite{gabius2018,parashuraman2019,Bard&Chia2016,Russo2013}. In this paper,
we study one aspect of molecular coding, namely  
the {\it fidelity} of this molecular code generation. While it has been recognised that
fidelity of the glycan code is necessary for reliable cellular
recognition~\cite{demetriou}, a quantitative measure of fidelity of the
code and its 
implications for cellular structure and organization are lacking.

There are  two aspects of the cell-type specific glycan code that have an
important bearing on quantifying fidelity. The first is that extant 
glycan distributions have high {\it complexity}, owing to evolutionary pressures arising from (a) reliable cell type identification amongst a large set of different cell types in a complex organism, the
  preservation and diversification of ``self-recognition''~\cite{Drickamer1998}, (b)
  pathogen-mediated selection
  pressures~\cite{varki,varki2017,Gagneux&Varki1999}, and (c) 
  {\it herd immunity} within a heterogenous population of cells of a
  community~\cite{WillsGreen1995} or within a single
  organism~\cite{Drickamer1998}.
Here, we will interpret this to mean that the
{\it target distribution} of glycans of a given cell type is complex; in
Sect.\,\ref{sec:hypotest} we  define a quantitative measure for complexity and 
demonstrate its implications
in the context of {\it human} T-cells. 
The second is that the cellular machinery for the synthesis of glycans,
which involves sequential chemical processing via cisternal  resident
enzymes and cisternal transport, 
is subject to variation and noise~\cite{varki2017,varki1998,parashuraman2019}; the {\it synthesized glycan distribution} is, therefore, a function of cellular parameters
such as the number and specificity of enzymes,
inter-cisternal transfer rates, and number of cisternae. 
We will discuss an explicit model of the cellular synthesis machinery in Sect.\,\ref{sec:seqchem}.

In this paper, we 
define
fidelity 
as the  minimum achievable
Kullback-Leibler (KL)
divergence~\cite{information,mckay} between the synthesized distribution 
of glycans and the target 
glycan distribution.
This KL divergence is a function of the cellular parameters governing
glycan synthesis, such as the number and specificity of enzymes, 
inter-cisternal transfer rates, and number of cisternae
(Sect.\,\ref{sec:optimization}).  
We analyze
the tradeoffs between the number of cisternae and the number and
specificity of enzymes, 
in order to achieve a prescribed target glycan distribution with high
fidelity (Sect.\,\ref{sec:results}). 
Our analysis leads to a number of interesting results, of which we list a
few here:
(i)  In order to construct an accurate representation of a complex
target distribution, such as those observed in real
cells, one needs to have  multiple cisternae and precise enzyme partitioning. 
Low complexity target distributions can be achieved with fewer cisternae.
(ii) This definition of fidelity of the glycan code, allows us to provide a quantitative  argument for
the evolutionary requirement of multiple-compartments. 
(iii) For fixed number of enzymes and cisternae, there is an optimal level of specificity of enzymes that achieves
the complex 
target distribution with high fidelity.  
(iv)  Keeping the number of enzymes fixed, having low specificity or
sloppy enzymes and larger cisternal number could give rise to a diverse
repertoire of functional glycans, a strategy used in  organisms such as plants and algae.

Stated another way, our results imply that the pressure to achieve the target glycan code for a given cell type,
places strong 
constraints on the cisternal number and enzyme specificity~\cite{Linstedt2011}.
This would suggest that a description of the nonequilibrium assembly of a fixed number of Golgi
cisternae must combine the dynamics of chemical processing with membrane dynamics
involving fission, fusion and transport~\cite{himani,sens2013}, opening up a new direction for future research.

\section{Complexity of glycan code in real cells}
\label{sec:hypotest}
\begin{figure}
\begin{minipage}{.6\textwidth}
	\subfloat[]{\includegraphics[scale=.25]{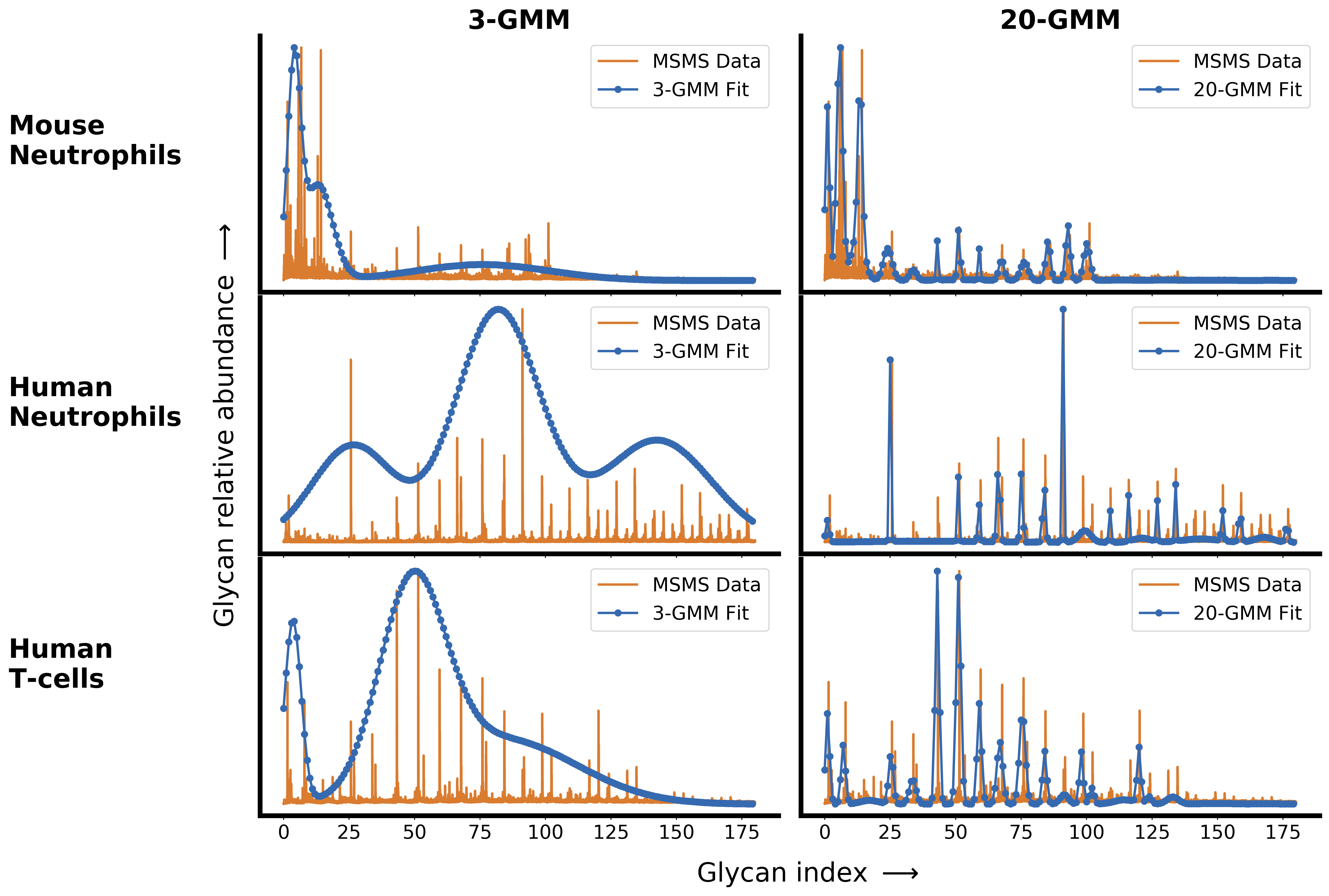}}
\end{minipage}
\begin{minipage}{.3\textwidth}
	\subfloat[]{\includegraphics[scale=.25]{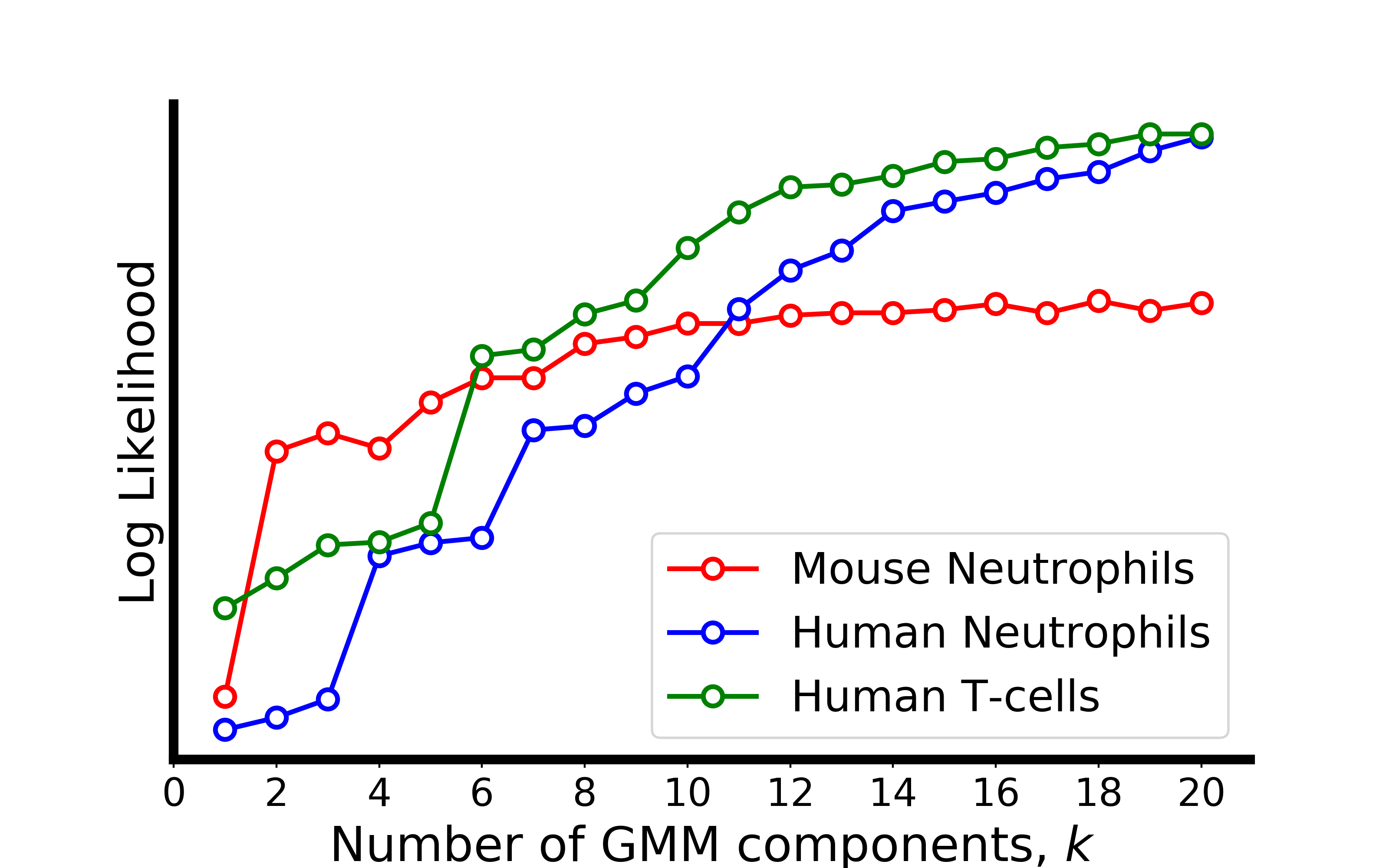}}
\end{minipage}
\caption{Real cells display a complex glycan distribution. (a) Here we take the MSMS data from {\it mouse} neutrophils, {\it human} neutrophils and {\it human}  T-cells
and approximate these using Gaussian Mixture Models (GMM) of less complexity $3$-GMM (left) and more complexity $20$-GMM (right).  (b) The change in log likelihood with increase in the number of GMM components  for  {\it mouse} and {\it human} neutrophils and  {\it human} T-cells, shows a saturation at large enough values of $k$, indicating that these glycan distribution are complex. Details
appear in Appendix\,\ref{sec:empirical_data}. }
\label{mscomplexity} 
\end{figure}

Since each
cell type (in a niche) is identified with a distinct glycan
profile~\cite{gabius2018,varki2017,parashuraman2019}, 
and this glycan profile is noisy because of the stochastic 
noise associated with the synthesis and transport~\cite{parashuraman2019,Bard&Chia2016,Russo2013},
a large number of 
different cell types can be differentiated only if the cells are able to
produce a large set of glycan profiles that are distinguishable in the presence of this noise.
A more complex or richer class of glycan profiles 
is able to support a larger number of well separated profiles, and
therefore, a larger number cell types, or equivalently, a more
complex organism\footnote{A rigorous definition of complexity can be given in terms of the 
 Kullback-Leibler metric~\cite{information,mckay} between two glycan profiles.
We declare that two
profiles are distinguishable only if the Kullback-Leibler distance between
the profiles is more than a given tolerance. This tolerance is an
increasing function of the noise. 
We define the \emph{complexity} of a set of possible glycan profiles as
the size of the largest subset such that the Kullback-Leibler distance of
any pair of profiles is larger than the tolerance.}

In order to implement 
a 
quantitative measure of complexity, we  first
need a
consistent way of smoothening or coarse-graining the discrete glycan
distribution to remove measurement and synthesis noise. In this
paper, we approximate the glycan profile as  
mixture of Gaussian densities with specified number of components that are
supported  
on a finite set of indices~\cite{bacharoglou2010approximation}.
Since the complexity of  $k$-component Gaussian is
an increasing function of $k$, we use the
number of component $k$ and 
complexity interchangeably.

Using this definition we demonstrate that the glycan profiles of typical mammalian cells are very complex. 
We obtain target profiles for a given cell type from Mass Spectrometry
  coupled with determination of molecular structure (MSMS)
  measurements~\cite{msdata}. Fig.\,\ref{mscomplexity} shows the  
the MSMS data from {\it human} T-cells and {\it human} and {\it mouse} neutrophils~\cite{msdata}, and their
coarse-grained representations using Gaussian mixture models (GMM) of differing
complexity - a low complexity $k=3$\,GMM and high complexity $k=20$\,GMM. 
It is clear from Fig.\,\ref{mscomplexity}, that the more complex $k=20$\,GMM
is a better representation 
of the MSMS data as compared to  the less complex $k=3$\,GMM. Indeed the $k=20$ Gaussian mixture model is the 
best compromise between faithfulness of the representation and cost of an
additional component, as seen from the saturation of the 
likelihood function~\cite{mckay}. Details of this systematic
coarse-graining procedure appear in Sect.\,\ref{sec:OptC} and  
Appendix\,\ref{sec:empirical_data}.

Having demonstrated the complexity of the typical glycan distributions associated with a given cell type, we will now describe a general model of 
the cellular machinery that is capable of synthesizing glycans of the desired complexity. We expect that 
cells need a more elaborate mechanism to produce profiles from a more
complex set.

\section{Synthesis of glycans in the Golgi cisternae}
\label{sec:seqchem}

\begin{figure*}
\includegraphics[scale=.45]{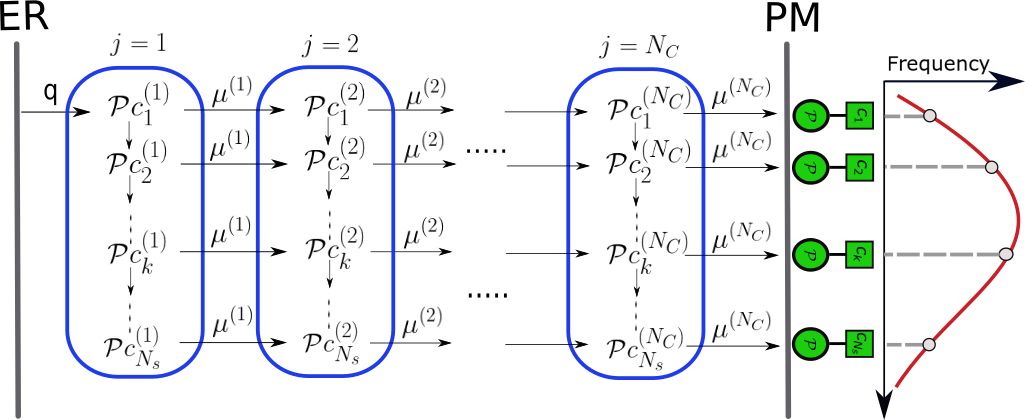} 
\caption{Enzymatic reaction and transport network in the secretory pathway. Represented here is the array of Golgi cisternae (blue) indexed by $j=1, \ldots, N_C$ situated between the ER and PM. 
Glycan-binding proteins ${\cal{P}}c_{1}^{(1)}$ are injected from the ER to cisterna-1 at rate $q$. Superimposed is
transition network of chemical reactions (column) - intercisternal
  transfer (rows), the latter with rates
  $\mu^{(j)}$. ${\cal{P}}c^{(j)}_{k}$ denotes the acceptor substrate in
  compartment $j$ and 
the glycosyl donor   $c_0$ is chemostated in each cisterna. This results
in a frequency distribution of glycans displayed  
at the PM (red curve), that is representative of the cell type.
} 
\label{fig:cartoon}
\end{figure*}

The glycan display at the cell surface is a result of proteins that flux through and undergo sequential chemical modification  in 
the secretory pathway, comprising
an array of 
Golgi cisternae situated between the ER and the PM, as depicted in Fig.\,\ref{fig:cartoon}.
Glycan-binding proteins (GBPs)   
are delivered from the ER to the first cisterna,
whereupon they are processed by the resident enzymes
in a sequence of steps that constitute the N-glycosylation process~\cite{varki}. 
A generic enzymatic reaction in the cisterna 
involves the catalysis of a group transfer reaction in which the monosaccharide moiety 
of a simple sugar donor substrate, e.g. UDP-Gal, is transferred
to the acceptor substrate, by a Michaelis-Menten (MM) type reaction~\cite{varki}

  \be
    \begin{aligned}
      \MoveEqLeft \small{\mbox{Acceptor} +  \mbox{glycosyl donor}  +
        \mbox{Enzyme} }  \\
      & \small{\xrightleftharpoons[\omega_{b}]{\omega_{f}} } \quad \small{\left[\mbox{Acceptor}\cdot \mbox{glycosyl donor} \cdot \mbox{Enzyme}\right] } 
      \quad \small{\xrightarrow[{}]{\omega_{c}}} \quad
      \small{\mbox{glycosylated acceptor} + \mbox{nucleotide} + \mbox{Enzyme}}
    \end{aligned}
    \label{eq:MM}
  \ee

From the first cisterna, the proteins with attached sugars are delivered
to the second cisterna at
a given inter-cisternal transfer rate,
 where further chemical processing catalysed by the enzymes
resident in the second cisterna occurs. This chemical processing and inter-cisternal
transfer continues until the last cisterna, thereupon the fully
processed glycans are displayed at  
the PM~\cite{varki}. The network of chemical processing and inter-cisternal
transfer forms the basis the physical model that we will describe next.

Any physical model of such a network of enzymatic reactions and cisternal
transfer needs to be augmented by reaction and transfer rates and chemical
abundances. 
To obtain the range of allowable values for the reaction rates and
chemical abundances, we 
use 
the elaborate enzymatic reaction 
models, such as the KB2005 model~\cite{UB1997,krambeck:2009,KB2005} (with
a network of $22,871$ chemical reactions and 
$7565$ oligosaccharide structures)
that predict the
N-glycan distribution   
based on the activities and levels of processing enzymes distributed in
the Golgi-cisternae of mammalian cells, and compare these predictions with  
N-glycan mass spectrum data.  For the allowable rates of cisternal
transfer, we 
rely on the
recent study by Ungar and
coworkers~\cite{ungar2019,ungar2016}, whose 
study shows how the overall Golgi transit time and cisternal number, can
be tuned to engineer a homogeneous glycan distribution.

\section{Model Definition}
\subsection{Chemical reaction and transport network in cisternae}
\label{sec:model}

With this background, we now define our quantitative model for chemical processing and transport in the secretory pathway.
We consider an array of $N_C$ Golgi cisternae, labelled by
$j = 1, \ldots, N_C$, between the ER and the PM (Fig.\,\ref{fig:cartoon}).
Glycan-binding proteins (GBPs), denoted as ${\cal{P}}c_{1}^{(1)}$, are
delivered from the ER to cisterna-$1$ at an injection rate  $q$.
It is well established that the concentration of the glycosyl donor in
the $j$-th cisterna is chemostated~\cite{varki,transporter1,transporter2,transporter3}, thus in our model we hold its concentration
 $c^{(j)}_0$
 constant in time for each $j$. The acceptor
${\cal{P}}c_{1}^{(1)}$ reacts with $c_0^{(1)}$ to form the glycosylated
acceptor ${\cal{P}}c_{2}^{(1)}$, following 
 an MM-reaction~\eqref{eq:MM} catalysed by the appropriate enzyme. The
 acceptor ${\cal{P}}c_{2}^{(1)}$ has the potential of being transformed
 into ${\cal{P}}c_{3}^{(1)}$, and  
so on, provided the requisite enzymes are present in that cisterna.  This leads to the 
sequence of enzymatic reactions ${\cal{P}}c_{1}^{(1)} \to {\cal{P}}c_{2}^{(1)} \to \ldots {\cal{P}}c_{k}^{(1)} \to \ldots$, where $k$  enumerates
the sequence of glycosylated acceptors, using a consistent scheme (such as
in~\cite{UB1997}). The glycosylated GBPs are transported from
cisterna-$1$ to cisterna-$2$ at an inter-cisternal transfer rate
$\mu^{(1)}$, whereupon similar enzymatic reactions 
proceed. The processes of intra-cisternal chemical reactions and
inter-cisternal transfer continue to the other cisternae and form 
a network as depicted in Fig.\,\ref{fig:cartoon}. Although, in this paper, we focus 
on a sequence of reactions that form a line-graph, the methodology we
propose extends to tree-like reaction sequences, and more generally to
reaction sequences that form a directed acyclic graphs~\cite{trinajstic2018chemical}.

%
%

Let $N_s$ denote the maximum number of possible glycosylation reactions in
each cisterna $j$, catalysed by  
enzymes labelled as $E^{(j)}_{\alpha}$, with $\alpha = 1, \ldots, N_E$,
where $N_E$ is the total number of enzyme species in each cisterna.  
Since many
substrates can compete for the substrate binding site on each enzyme, one
expects in general that 
$N_s \gg N_E$. The configuration space of the network
Fig.\,\ref{fig:cartoon} is $N_s \times N_C$. For the N-glycosylation
pathway in a typical mammalian cell, 
$N_s=2 \times 10^4$, $N_E=10-20$ and
$N_C=4-8$~\cite{UB1997,KB2005,krambeck:2009,ungar2016}. We account for the 
fact that the enzymes have specific cisternal localisation, by setting
their concentrations to zero in those cisternae where they are not
present.  

Now the action of enzyme $\Eja$ on the substrate ${\cal{P}}\cjk$ in cisterna $j$
is given by
\begin{equation}
  \label{eq:reactionscheme}
  \begin{split}
    {\cal{P}}c_{k}^{(j)} + E_{\a}^{(j)}
    \xrightleftharpoons[\omega_b(j,k,\a)]{\omega_f(j,k,\a)c_0^{(j)}} 
    \left[E_{\a}^{(j)}-{\cal{P}}c_{k}^{(j)}-c_0^{(j)}\right] 
    & \xrightarrow{\omega_c(j,k,\a)}   {\cal{P}}c_{k+1}^{(j)} + E_{\a}^{(j)}
  \end{split}
\end{equation}
In general, the forward, backward and catalytic rates $\omega_{f}$, $\omega_{b}$ and $\omega_{c}$, respectively, 
depend on the cisternal label $j$, the reaction label
$k$, and the enzyme label $\a$, that parametrise the MM-reactions~\cite{enzymology}. 
For instance, structural studies on glycosyltransferase-mediated synthesis of glycans~\cite{Haltiwanger2019},  would suggest that the forward rate $\omega_f$ to depend on
the binding energy of the enzyme $E_{\a}^{(j)}$ to acceptor substrate ${\cal{P}}c_k^{(j)}$ and a
{\it physical variable that characterises cisterna}-$j$.


A potential candidate for such a cisternal variable is
pH~\cite{kellokumpu}, whose value is maintained homeostatically in each
cisterna~\cite{Casey}; 
changes in pH can affect the shape of an enzyme (substrate) or their
charge properties, 
and in general the reaction efficiency of an enzyme has a pH
optimum~\cite{enzymology}. Another possible candidate for a cisternal
variable is 
membrane bilayer thickness~\cite{sergePNAS} -  
indeed both pH~\cite{tsien} and membrane thickness are known to have a gradient across the Golgi
  cisternae.
We take $\omega_f(j,k,\a) \propto \,P^{(j)}(k,\a)$, where $P^{(j)}(k,\a) \in (0,1)$, is the binding probability of enzyme $\Eja$
with substrate ${\cal{P}}c_k^{(j)}$, and define the binding probability $P^{(j)}(k,\a)$ using a biophysical model, similar in spirit to 
the Monod-Wyman-Changeux model of enzyme kinetics~\cite{MWC,Changeux}, of
enzyme-substrate induced fit. 

Let $\bs{l}^{(j)}_{\a}$ and $\bs{l}_k$ denote, respectively, the optimal
``shape''
for enzyme $\Eja$ and the substrate ${\cal{P}}c_k^{(j)}$.  
We assume that the mismatch (or distortion) energy between the substrate
$k$ and enzyme $\Eja$ is 
$\norm{\bs{l}_k - \bs{l}_\a^{(j)}}$,
with a
binding probability given by,
\be
P^{(j)}(k,\a) = \exp\l-\sigma_\a^{(j)} \norm{\bs{l}_k - \bs{l}_\a^{(j)} } \r
\label{eq:binding_prob1}
\ee
where $\norm{.}$ is a distance metric defined on the space of
$\bs{l}_\a^{(j)}$ (e.g., the square of the $\ell_2$-norm would be related to 
an elastic distortion model~\cite{tlusty}) and
the vector $\bs{\sigma} \equiv [\sigma_\a^{(j)}]$ parametrises {\it enzyme specificity}. A large
value of $\sigma_\a^{(j)}$ indicates a highly specific enzyme, a small value
of $\sigma_\a^{(j)}$ indicates a promiscuous or sloppy enzyme.
 It is recognised that the degree of enzyme specificity or sloppiness is an
  important determinant of glycan distribution~\cite{varki,Roseman,hossler,Yang2018}.

As in~\cite{UB1997,krambeck:2009,KB2005}, our synthesis model is mean-field, in that we
ignore stochasticity in glycan synthesis that may arise from low copy 
numbers of substrates and enzymes, multiple 
substrates competing for the same enzymes, and kinetics of
inter-cisternal transfer. Then the 
usual MM-steady state condition on \eqref{eq:reactionscheme}, which
assumes that the concentration of the  
intermediate enzyme-substrate complex does not change with time, implies 
\[ \left[E_{\a}^{(j)}-{\cal{P}}c_{k}^{(j)}-c_0^{(j)}\right]  =
\frac{\omega_f(j,k,\a)\,c_0^{(j)}}{\omega_b(j,k,\a) + \omega_c(j,k,\a)} 
E_{\a}^{(j)} c_{k}^{(j)} .\textbf{\textbf{}}
\]
where $c_k^{(j)}$ is the {\it concentration} of the acceptor substrate
${\cal{P}}c^{(j)}_{k}$ in compartment $j$. 

Together with the constancy of the total enzyme concentration,
$ \L E_{\a}^{(j)} \R_{tot}= 
  E_{\a}^{(j)} + \sum_{k=1}^{N_s} \left[E_{\a}^{(j)}-{\cal{P}}c_{k}^{(j)}-c_0^{(j)}\right]
$,
this immediately fixes the kinetics of product formation (not including inter-cisternal transport),
\be
\label{eq:enz_kinetics}
  \frac{d c_{k+1}^{(j)}}{dt}  =
                    \sum_{\a=1}^{N_E} \frac{V(j, k, \a) P^{(j)}(k,\a) c_{k}^{(j)}}
                     {M(j,k, \a) \l1 + \sum_{k'=1}^{N_s} \frac{P^{(j)}(k',\a)c_{k'}^{(j)}}{M(j,k', \a)}\r} 
\ee
where 
$$M(j,k, \a) = \frac{\omega_b(j,k,\a) + \omega_c(j,k,\a)}{\omega_f(j,k,\a) c_0^{(j)} }P^{(j)}(k,\a)$$
and 
$$V(j, k, \a)=  \omega_c(j,k,\a) \L E_{\a}^{(j)}\R_{tot}\,.$$
From the above, the experimentally measurable parameters $V_{max}$ and
MM-constant $K_M$, for each $(j, k, \a)$ can be easily read out.  
As is the usual case, the maximum velocity $V_{max}$ is not an intrinsic
property of the 
enzyme, because it is dependent on the enzyme concentration $\L
E_{\a}^{(j)}\R_{tot}$; while 
$K_M(j, k, \a) = M(j,k,\a)c_0^{(j)}/P^{(j)}(k,\a)$
is an intrinsic parameter of the enzyme and the enzyme-substrate
interaction. 
The enzyme catalytic efficiency,  the so-called $\text{``$k_{cat}/K_M$"} \propto
P^{(j)}(k,\a)$ and is high for {\it perfect} enzymes~\cite{milo} with
minimum mismatch.

We now add  to this chemical reaction kinetics, the rates of injection
($q$) and inter-cisternal transport $\mu^{(j)}$ from the cisterna $j$ to
$j+1$; in  
Appendix~\ref{sec:seqchemapp} we display the complete set of equations
that describe the changes in the substrate concentrations 
$c_{k}^{(j)}$ with time. These kinetic equations automatically obey the
conservation law for the protein concentration ($p$). 
Rescaling the kinetic parameters in terms of the injection rate $q$,
i.e. $V(j, k, \a) =   {V(j, k, \a)}/{q}$  and 
  $\mu^{(j)}  =  {\mu^{(j)}}/{q}$, we  
see that the steady state concentrations of the glycans in each cisterna
satisfy the following recursion relations (see,
Appendix~\ref{sec:seqchemapp}). In the first cisterna,
\bea 
c_1^{(1)} & = & \f{1}{ \mu^{(1)} +
   \sum_{\a=1}^{N_E} \frac{V(1, 1, \a) P^{(1)}(1,\a) c_{1}^{(1)} }
                     {M(1,1, \a) \l1 + \sum_{k'=1}^{N_s} \frac{P^{(1)}(k',\a)c_{k'}^{(1)}}{M(1,k', \a)}\r } }
      \nn \\ 
c_k^{(1)} & = & \f{
   \sum_{\a=1}^{N_E} \frac{V(1, k-1, \a) P^{(1)}(k-1,\a) c_{k-1}^{(1)} }
                     {M(1,k-1, \a) \l1 + \sum_{k'=1}^{N_s} \frac{P^{(1)}(k',\a)c_{k'}^{(1)}}{M(1,k', \a)}\r } }
                     { \mu^{(1)} +
   \sum_{\a=1}^{N_E} \frac{V(1, k, \a) P^{(1)}(k,\a) c_{k}^{(1)} }
                     {M(1,k, \a) \l1 + \sum_{k'=1}^{N_s} \frac{P^{(1)}(k',\a)c_{k'}^{(1)}}{M(1,k', \a)}\r } }
    \label{eq:concentrations1}\\
c_{N_s}^{(1)} & = & 
\f{\sum_{\a=1}^{N_E} \frac{V(1, N_s-1, \a) P^{(1)}(N_s-1,\a) c_{N_s-1}^{(1)} }
                     {M(1,N_s-1, \a) \l1 + \sum_{k'=1}^{N_s} \frac{P^{(1)}(k',\a)c_{k'}^{(1)}}{M(1,k', \a)}\r } }
                     { \mu^{(1)}} \nn
\eea
and in cisternae $j \geq 2$,
\bea
c_1^{(j)} & = & \f{\mu^{(j-1)} c_1^{(j-1)}}
{ \mu^{(j)} +
   \sum_{\a=1}^{N_E} \frac{V(j, 1, \a) P^{(j)}(1,\a) c_{1}^{(j)} }
                     {M(j,1, \a) \l1 + \sum_{k'=1}^{N_s} \frac{P^{(j)}(k',\a)c_{k'}^{(j)}}{M(j,k', \a)}\r } }
      \label{eq:concentrationsj} \\ 
      c_k^{(j)} & = & \f{ \mu^{(j-1)} c_k^{(j-1)} +
   \sum_{\a=1}^{N_E} \frac{V(j, k-1, \a) P^{(j)}(k-1,\a) c_{k-1}^{(j)} }
                     {M(j,k-1, \a) \l1 + \sum_{k'=1}^{N_s} \frac{P^{(j)}(k',\a)c_{k'}^{(j)}}{M(j,k', \a)}\r } }
                     { \mu^{(j)} +
   \sum_{\a=1}^{N_E} \frac{V(j, k, \a) P^{(j)}(k,\a) c_{k}^{(j)} }
                     {M(j,k, \a) \l1 + \sum_{k'=1}^{N_s} \frac{P^{(j)}(k',\a)c_{k'}^{(j)}}{M(j,k', \a)}\r } } \nn \\ 
c_{N_s}^{(j)} & = &\f{ \mu^{(j-1)} c_{N_s}^{(j-1)} +
   \sum_{\a=1}^{N_E} \frac{V(j, N_s-1, \a) P^{(j)}(N_s-1,\a) c_{N_s-1}^{(j)} }
                     {M(j,N_s-1, \a) \l1 + \sum_{k'=1}^{N_s} \frac{P^{(j)}(k',\a)c_{k'}^{(j)}}{M(j,k', \a)}\r } }
                     {\mu^{(j)}} \nn
\eea
Equations \eqref{eq:concentrations1}-\eqref{eq:concentrationsj} automatically imply that the 
total  steady state glycan
concentration in each cisterna $j = 1, \ldots, N_c$ is given by  
\be
\sum_{k=1}^{N_s} c_k^{(j)} = \f{1}{ \mu^{(j)}} \nn.
\ee
These
nonlinear recursion equations 
 \eqref{eq:concentrations1}-\eqref{eq:concentrationsj} have to  be solved numerically to obtain the
steady state glycan concentrations, $\mathbf{c}\equiv c_k^{(j)}$, as a
function of 
  the independent vectors $\mathbf{M} \equiv [M(j,k,\a)]$,
$\mathbf{V} \equiv [V(j,k,\a)]$, and $\mathbf{L}\equiv [P^{(j)}(k,\a)]$,
 the transport rates $\bs{\mu} \equiv [\mu^{(j)}]$ and specificity,
 $\bs{\sigma} \equiv [\sigma_\a^{(j)}]$.

\section{Optimization Problem}
\label{sec:optimization}
Now, with both the protocol for determining the target glycan distribution and the sequential chemical processing model in hand, we can precisely define the optimization problem referred
to in the Introduction.
Let $\mbf{c}^\ast$ denote the ``target'' concentration
distribution\footnote{We normalize the distribution so that
  $\sum_{k=1}^{N_s} c^\ast_k = 1$.} for a particular cell
type, i.e. the goal of the sequential synthesis mechanism described in
Sect.~\ref{sec:model} is to approximate $\mathbf{c}^\ast$. Let $\bar{\mathbf{c}}$
denote the steady state glycan concentration distribution displayed on the
PM - \eqref{eq:concentrationsj} implies that $\bar{c}_k = \mu^{(N_C)}
c^{(N_C)}_k$, $k = 1, \ldots, N_s$. 
We measure the fidelity between
 the $\bf{c}^\ast$ and $\bar{\bf{c}}$ by the
Kullback-Leibler metric~\cite{information,mckay},

\be
  \label{eq:KL-metric}
  \hspace*{-0.05in}
  D_{KL}(\mbf{c}^\ast \| \bar{\mathbf{c}})  = \sum_{k=1}^{N_s} c^\ast_k \ln
  \Big(\frac{c^\ast_k}{\bar{c}_k}\Big) = \sum_{k=1}^{N_s} c_k^* \ln \Big(
  \frac{c_k^*}{c_k^{(N_C)}\mu^{(N_C)}}\Big)
\ee

Thus, the problem of designing a sequential synthesis mechanism that
approximates $\mathbf{c}^\ast$ for a given enzyme specificity $\bs {\sigma}$,
transport rate $\bs{\mu}$,
number of enzymes $N_E$, and number of cisternae $N_C$ is given by \\
\be
\text{\emph{Optimization A}}: \quad 
  \label{eq:problem-A}
 \min_{\mbf{M} ,\ \mbf{V},\ \mbf{L} \ \geq \mbf{0}}
D_{KL}(\mbf{c}^\ast\| \bar{\mbf{c}})
\ee
There is separation of time scales implicit in Optimization A --
the chemical kinetics of the production of
glycans and their display on the PM 
happens over cellular time scales, while the issues of tradeoffs and
changes of parameters are driven over evolutionary timescales. 

Optimization~A, though well-defined, is a hard
problem, since the steady state concentrations \eqref{eq:concentrationsj}
are not {\it explicitly} known in terms of 
the parameters $(\mbf{M}, \mbf{V}, \mbf{L})$.
In Appendix~\ref{sec:optB_def}, we formulate an alternative problem
\emph{Optimization B} in which the 
steady state concentrations are
defined explicitly in terms of a new parameters $\mbf{R}$ and $\mbf{L}$, and
in Appendix~\ref{sec:equivalence_opt_problem} we prove that Optimization~A
and Optimization~B are exactly equivalent.
This is
a crucial insight  
that allows us to obtain all the results that follow.

In Appendix~\ref{sec:numericalscheme}, we describe the  variant of the
Sequential Quadratic Programming (SQP)~\cite{Boyd}, that we use to 
numerically solve the optimization problem.

\section{Results of optimization}
\label{sec:results}

To start with, 
the dimension of the optimization search
space is extremely large $\approx O(N_s \times N_E \times N_C)$.  
To make the optimization search more manageable, we ignore the
$k$-dependence of the vectors $(\mbf{M}, \mbf{V})$,  (or, alternatively of
$\mbf{R}$, see Appendix~\ref{sec:optB_def} for details). 
The dependence on the reaction rates on the glycosyl substrate is still present
 in the forward reactions via the enzyme-substrate binding probability $P^{(j)}(k,\a)$.
We further assume that shape function is a number, $\bs{l}_\a^{(j)} =
l_\a^{(j)}$ and that $\bs{l}_k = k$. 
Finally we will drop the dependence of the specificity on $\a$ and $j$,
and take it to be a scalar $\sigma$. To fix our model, we will take the
distortion energy that appears in \eqref{eq:binding_prob1}  to be the
linear form $\abs{l_k - l_\a^{(j)}}$. Other metrics, such as $\abs{l_k -
  l_\a^{(j)}}^2$, corresponding to the elastic distortion
model~\cite{tlusty},  do not pose any computational difficulties, and we
see that the  
results of our optimization remain qualitatively unchanged. 

These restrictions significantly reduce the dimension of the optimization search, so much so that 
in certain limits, we can solve the problem 
analytically\footnote{In Appendix~\ref{sec:analytic_calc} we show that \eqref{eq:problem-B} can be solved 
analytically in the limit
$N_s \gg 1$, since the glycan index $k$ can be approximated by a continuous
variable, and  the recursion relations for the steady state glycan
concentrations \eqref{eq:concentrations1}-\eqref{eq:concentrationsj}  
can be cast as a matrix differential equation. This allows us to obtain an
\emph{explicit} expression for the steady state concentration in terms of 
the parameters $(\mbf{R}, \mbf{L})$. }. This helps us obtain some useful
heuristics (Appendix~\ref{sec:model capability}) on how  to tune the
parameters, e.g.\,$N_E$, $N_C$, $\sigma$, and others, in order to generate glycan
distributions $\mbf{c}$ of a given complexity. These heuristics inform our
more detailed 
optimization using ``realistic'' target distributions.

The 
calculations in Appendix~\ref{sec:model capability} imply,
as one might expect, that the synthesis model needs to be more elaborate, 
i.e., needs a larger number of cisternae $N_C$ or a larger number of
enzymes $N_E$,
in order to produce a more complex glycan 
distribution. For a real cell type in a niche, the specific elaboration of
the synthesis machinery, would depend on a variety of  control costs  
associated with  increasing $N_E$ and $N_C$. While an increase in the
number of enzymes would involve genetic and transcriptional costs, the 
costs involved in increasing the number of cisternae could be rather subtle. 

 Notwithstanding the relative control costs of  increasing $N_E$ and
 $N_C$, it is clear from the special case, that increasing the number of
 cisternae  
 achieves the goal of obtaining an accurate representation of the target
 distribution. Let us assume that the target distribution is $c^*_{k} =
 \delta(k-M)$ for a fixed $M \gg 1$, i.e. $c^\ast_k = 1$ when $k = M$, and
 $0$ otherwise,
 and that the $N_E$ enzymes that catalyse the reactions are highly
 specific. In this limit, Optimization A reduces to a simple enumeration
 exercise~\cite{thattai2018}: clearly one needs $N_E=M$, with one enzyme
 species for each of the $k=1, \ldots, M$ reactions, in order to generate
 ${\cal{P}}c_{M}$. For a single Golgi cisterna with a finite cisternal
 residence time (finite $\mu$), the chemical synthesis network will
 generate a significant steady state concentration of lower index glycans
 ${\cal{P}}c_{k}$ with $k<M$, contributing to a low fidelity. To obtain
 high fidelity, one needs  
 multiple Golgi cisternae with a specific enzyme partitioning 
 $(E_1, E_2, \ldots, E_M)$ 
 with $E_j$  enzymes 
 in cisterna~$j = 1, \ldots, N_c$.
 This argument can be   
 generalised to the case where the target distribution is a finite sum of
 delta-functions. The more general case, where the enzymes are allowed to
 have variable specificity, 
 needs a more detailed study, to which we turn to below.

\subsection{Target distribution from coarse-grained MSMS}
\label{sec:target-dist}

As discussed in Sect.\,\ref{sec:hypotest}, we obtain  the target glycan distribution  
from   
glycan profiles for real cells obtained using Mass Spectrometry
coupled with determination of molecular structure (MSMS) measurements~\cite{msdata}.
The raw MSMS data, however, is not suitable as a target distribution. This is
because it is very noisy, with 
chemical noise in the sample and  Poisson noise associated with
detecting discrete events being the most relevant~\cite{MS_noise}. 
This means that many of the small peaks in the raw data are 
not part of the signal, and one has to ``smoothen'' the distribution to remove
the impact of noise.

We use MSMS data from {\it human}  T-cells~\cite{msdata} for our analysis.  As discussed in Sect.\,\ref{sec:hypotest}, the Gaussian mixture models
(GMM) are often used to approximate distributions with a mixed number of modes
or peaks~\cite{mckay}, or in our setting, a given fixed complexity. 
Here, we use a variation of the Gaussian mixture models (see
Appendix~\ref{sec:empirical_data} for details) to create a hierarchy of
increasingly complex distributions to approximate the MSMS raw
data. Thus, the $3$-GMM and $20$-GMM approximations represent the low and high complexity benchmarks, respectively.
In Appendix~\ref{sec:empirical_data}, we show that the likelihood for the glycan
distribution of the {\it human} T-cell saturates at $20$ peaks. Thus,
statistically speaking, the {\it human} T-cell glycan distribution is accurately
approximated by $20$ peaks.

This hierarchy allows us to study the trade-off between the
complexity of the target distribution and the complexity of the synthesis
model needed to generate the distribution as follows.  
Let  $\mbf{T}^{(i)}$ denote the $i^{th}$-GMM approximation for the 
{\it
human} T-cell MSMS data. We sample this target distribution at indices
$k = 1, \ldots, N_s$, that represent the glycan indices, and then
renormalize to obtain the discrete distribution
$\{T^{(i)}_k,k = 1, \ldots, N_s\}$.
 Let $H(\mbf{T}^{(i)}):= -\sum_{k=1}^{N_s} T^{(i)}_k \log T^{(i)}_k $ denote the entropy~\cite{information} of the $i^{th}$-GMM approximation. 
 $H(\mbf{T}^{(i)})$ quantifies statistical information in the target distribution $\mbf{T}^{(i)}$.
 We evaluate the fidelity of the distribution generated by the synthesis model to this target distribution by the
 ratio of the Kullback-Leibler distance to the entropy of the target distribution:
 \be
 {\bar D}(\sigma, N_E, N_C, \mbf{T}^{(i)}) := \f{D(\sigma, N_E, N_C,
 \mbf{T}^{(i)})}{H(\mbf{T}^{(i)})}
 \label{eq:normKL}
 \ee
 
  This normalization
allows us evaluate the fidelity of the synthesis model to the target
distribution $\mbf{T}^{(i)}$ as a fraction of  the total  statistical information in
the target distribution $\mbf{T}^{(i)}$.
 
\subsection
{Tradeoffs between number of enzymes, number of cisternae and enzyme specificity to achieve given complexity}
\label{sec:OptC}

\begin{figure*}
\centering
\subfloat[Less complex target, 3-GMM approximation{\label{fig:DklVsSigma_Nc3GMM}}]{\includegraphics[scale=.3]{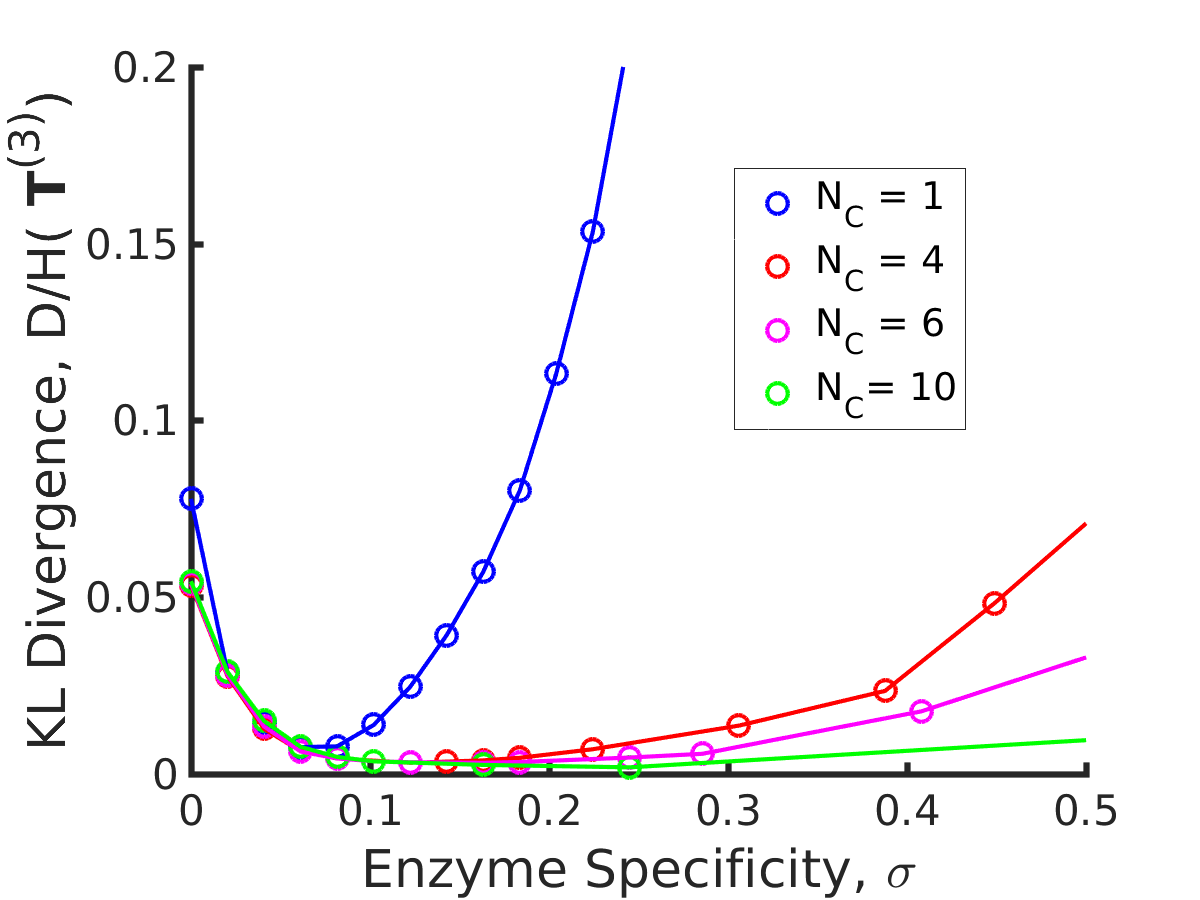}}
\:
\subfloat[More complex target, 20-GMM approximation{\label{fig:DklVsSigma_Nc20GMM}}]{\includegraphics[scale=0.3]{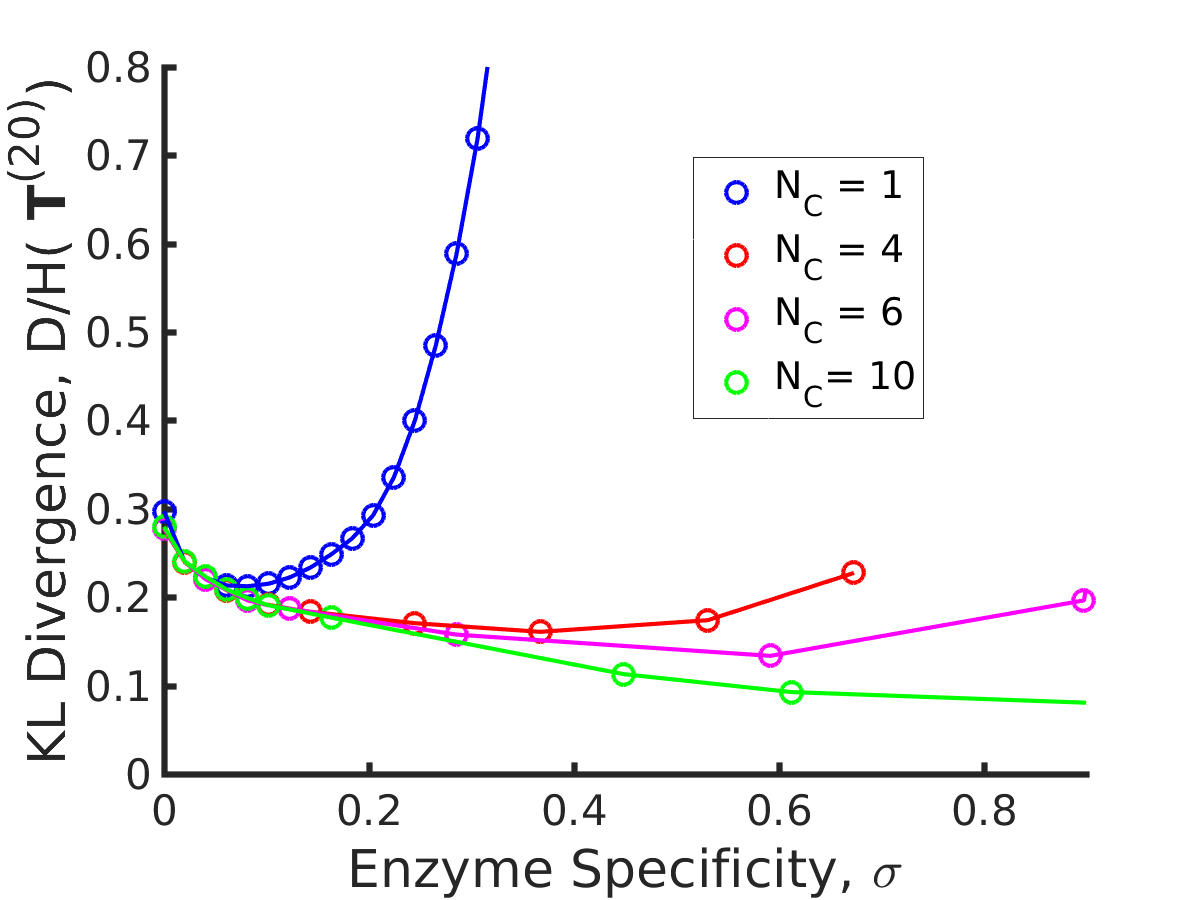}}
\:
\subfloat[Less complex target, 3-GMM approximation{\label{fig:DklVsSigma_Ne3GMM}}]{\includegraphics[scale=.3]{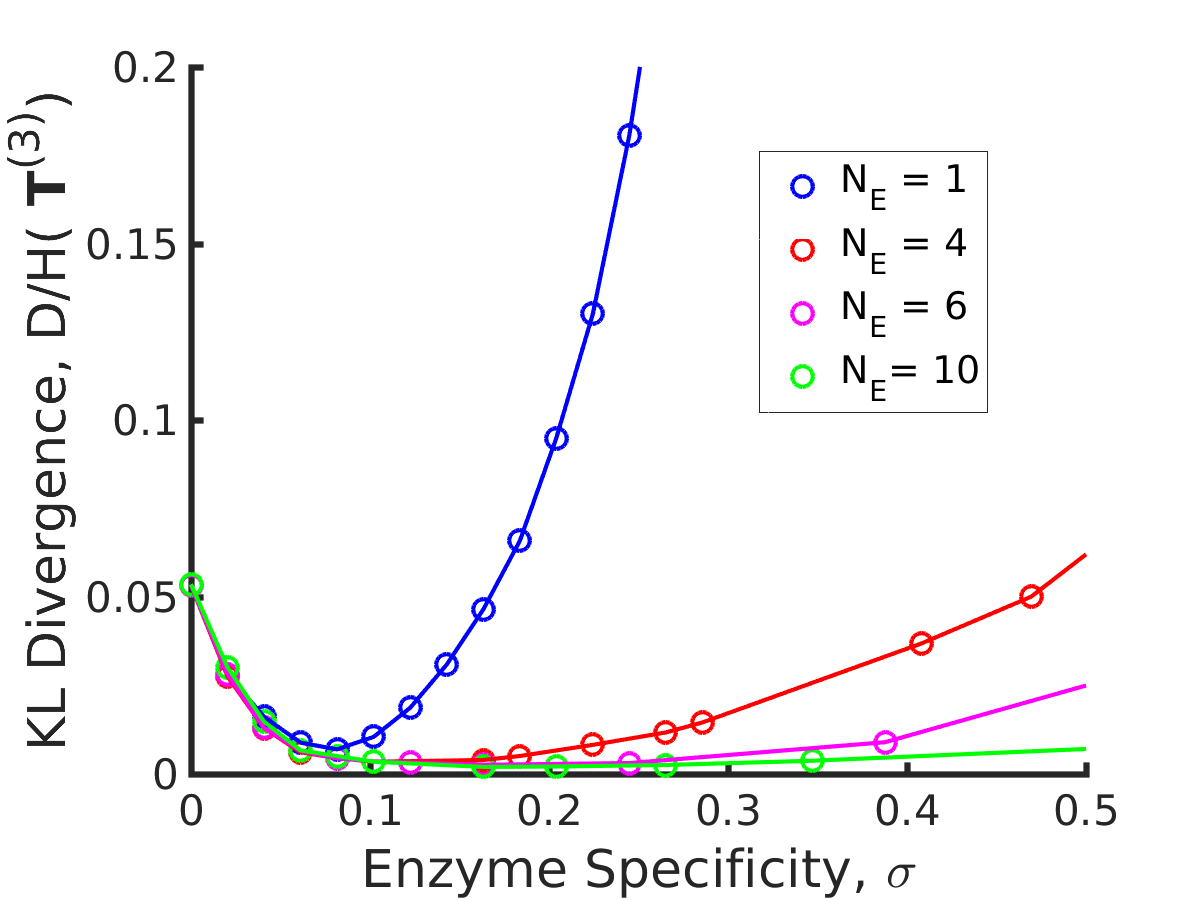}}
\:
\subfloat[More complex target, 20-GMM approximation{\label{fig:DklVsSigma_Ne20GMM}}]{\includegraphics[scale=0.3]{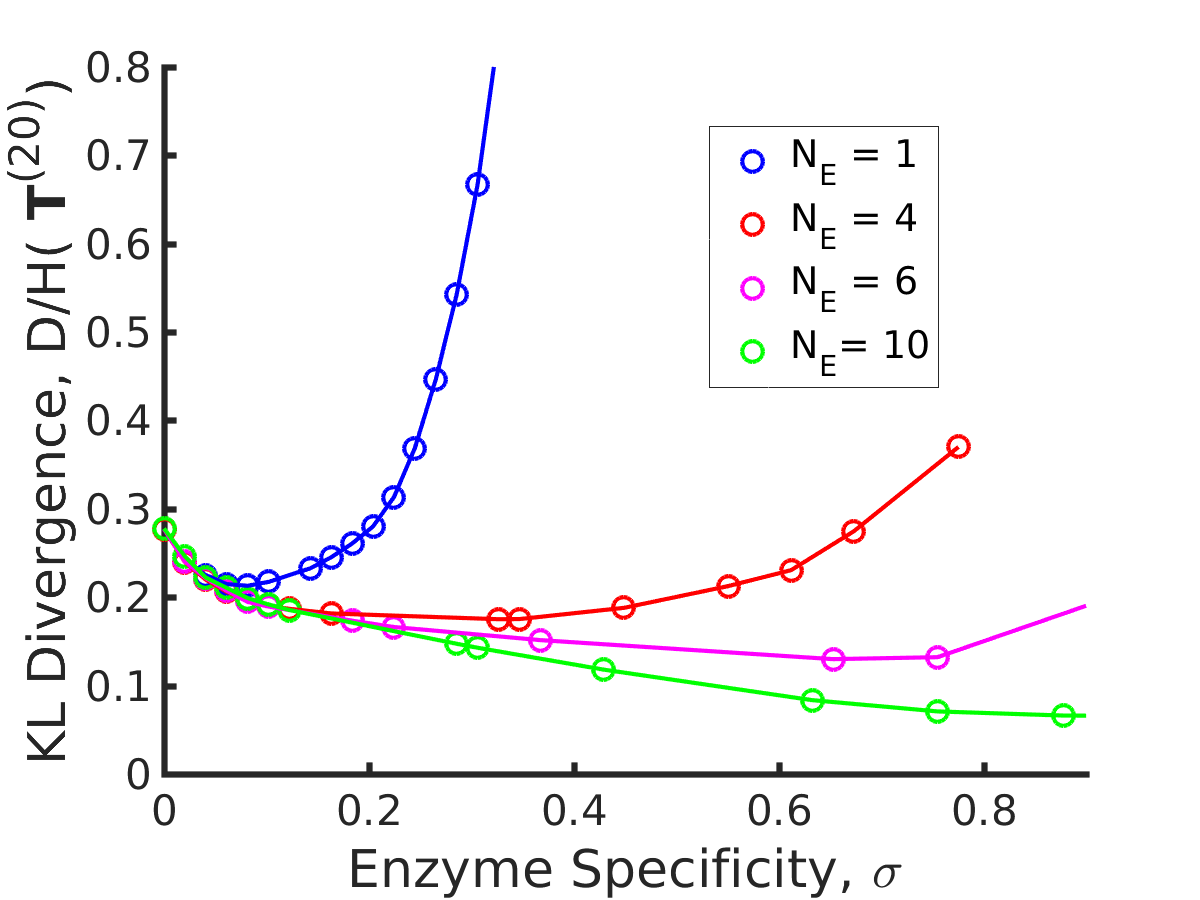}}
\:
\caption{Tradeoffs amongst the glycan synthesis parameters, enzyme specificity $\sigma$, cisternal number  $N_C$ and enzyme number $N_E$,
 to achieve a complex target distribution $\mbf{c}^\ast)$.  (a)-(b) 
Normalised Kullback-Leibler distance $\bar{D}(\sigma,N_E, N_C, \mbf{c}^\ast)$ as function
  of $\sigma$ and $N_C$ (for fixed $N_E=3$), (c)-(d) $\bar{D}(\sigma, N_E,N_C, \mbf{c}^\ast)$ as function
  of $\sigma$ and $N_E$ (for fixed $N_C=3$), with the target distribution $\mbf{c}^\ast$ set to
  the $3$-GMM (less complex) and $20$-GMM (more complex) approximations for the {\it human} T-cell MSMS
  data. $\bar{D}(\sigma,N_E,N_C,\mbf{c}^\ast)$ is a convex function of
  $\sigma$ for each $(N_E,N_C,\mbf{c}^\ast)$,  decreasing in $N_C,N_E$
  for each $(\sigma, \mbf{c}^\ast)$,  increasing in the complexity of
  $\mbf{c}^\ast$ for fixed $(\sigma,N_E, N_C)$. The specificity 
  $\sigma_{\min}(\mbf{c}^\ast,N_E,N_C) = \text{argmin}_{\sigma}
  \{\bar{D}(\sigma,N_E,N_C,\mbf{c}^\ast)\}$  that minimises the error for given
  $(N_E,N_C,\mbf{c}^\ast)$ is an
  increasing function of $N_C,N_E$ and the complexity of the target
  distribution $\mbf{c}^\ast$.  Furthermore, the curvature of
  $\bar{D}(\sigma,N_E,N_C,\mbf{c}^\ast)$ at $\sigma_{\min}(N_E,N_C,\mbf{c}^\ast)$, related to {\it sensitivity},
  is a decreasing function of $N_C,N_E$. 
}
\label{fig:DklVsSigma}
\end{figure*}

\begin{figure*}
\subfloat[Fidelity for less complex target, $\mbf{c^*} =$ 3-GMM \newline approximation {\label{fig:DklvsNeNc3GMM}}]{\includegraphics[scale=.35]{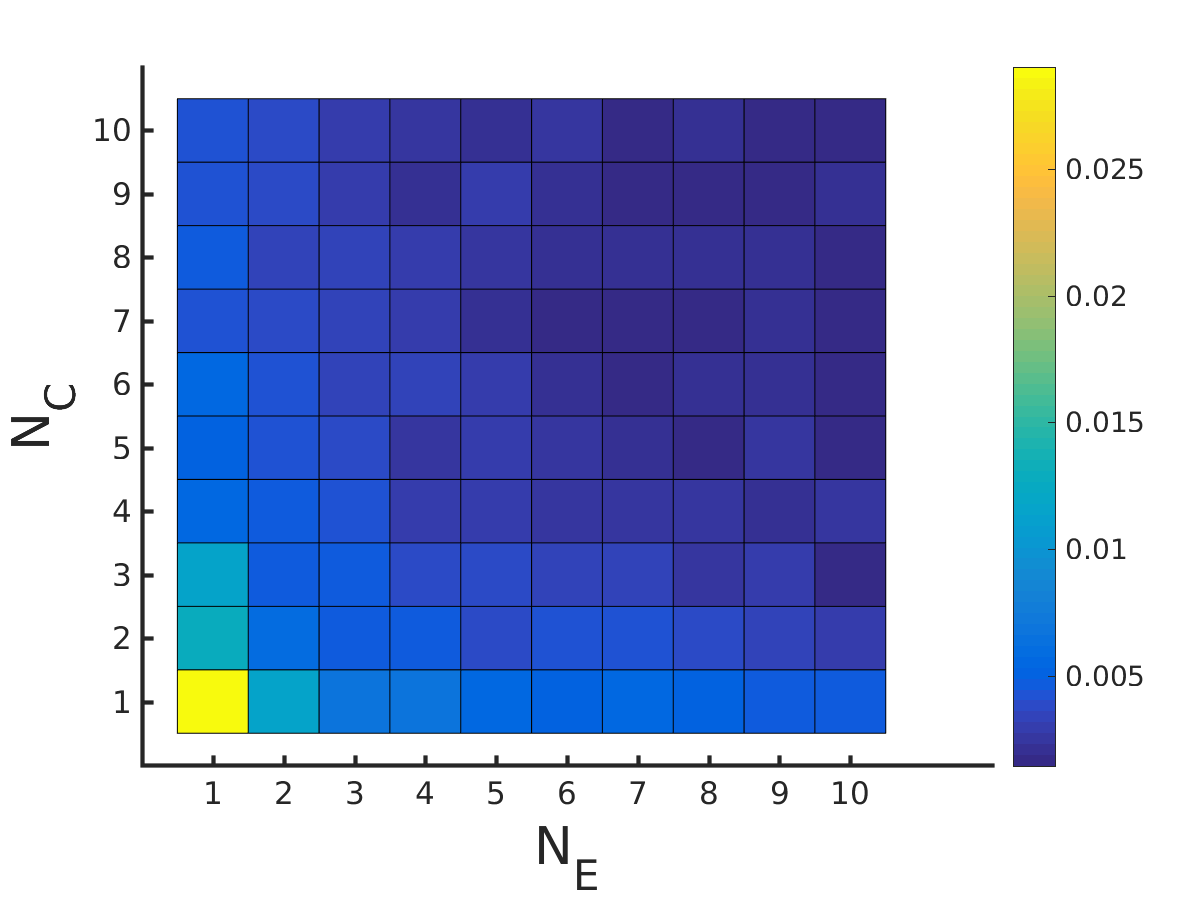}}
\:
\subfloat[Fidelity for more complex target $\mbf{c^*} =$ 20-GMM approximation{\label{fig:DklvsNeNc20GMM}}]{\includegraphics[scale=0.35]{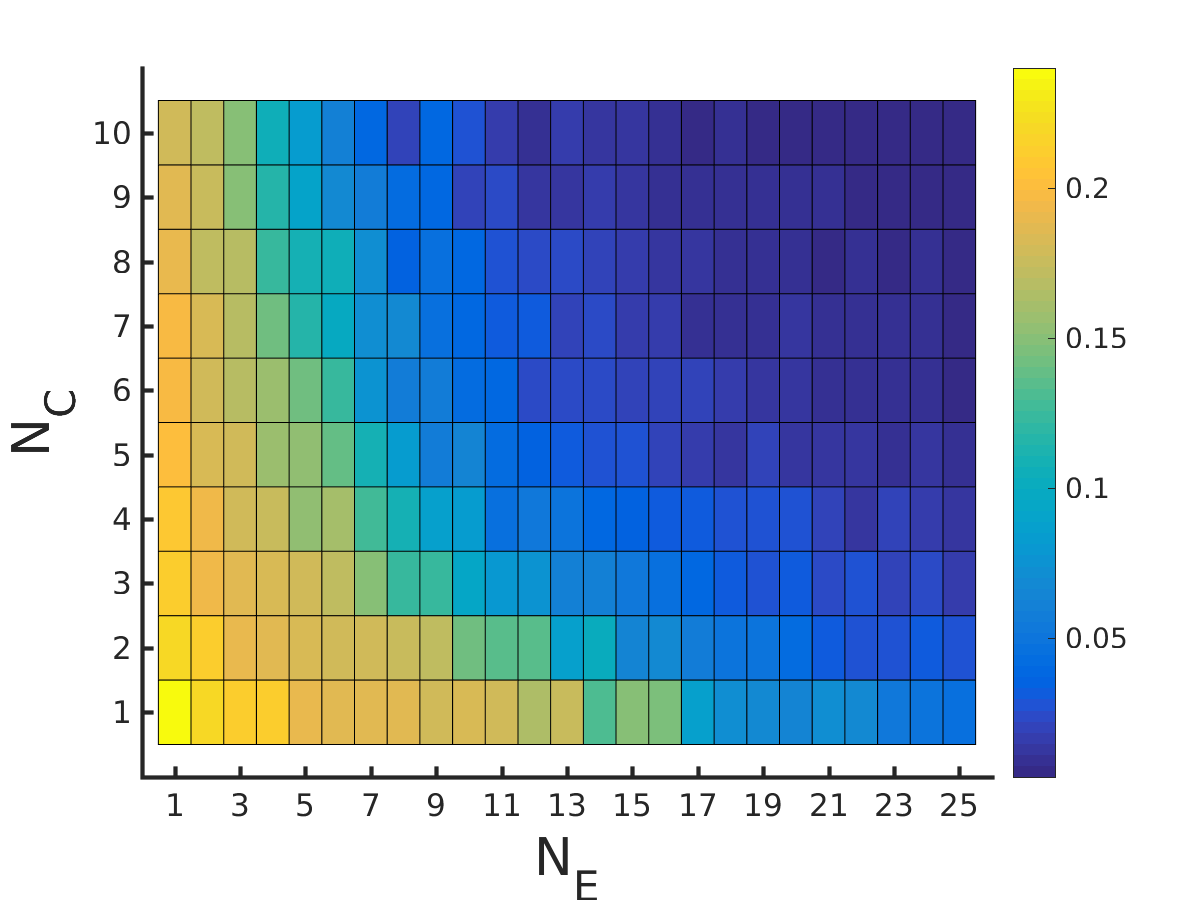}}
\:
\subfloat[Optimal enzyme specificity for less complex target, $\mbf{c^*} =$ 3-GMM approximation{\label{fig:sigmamin3GMM}}]{\includegraphics[scale=.35]{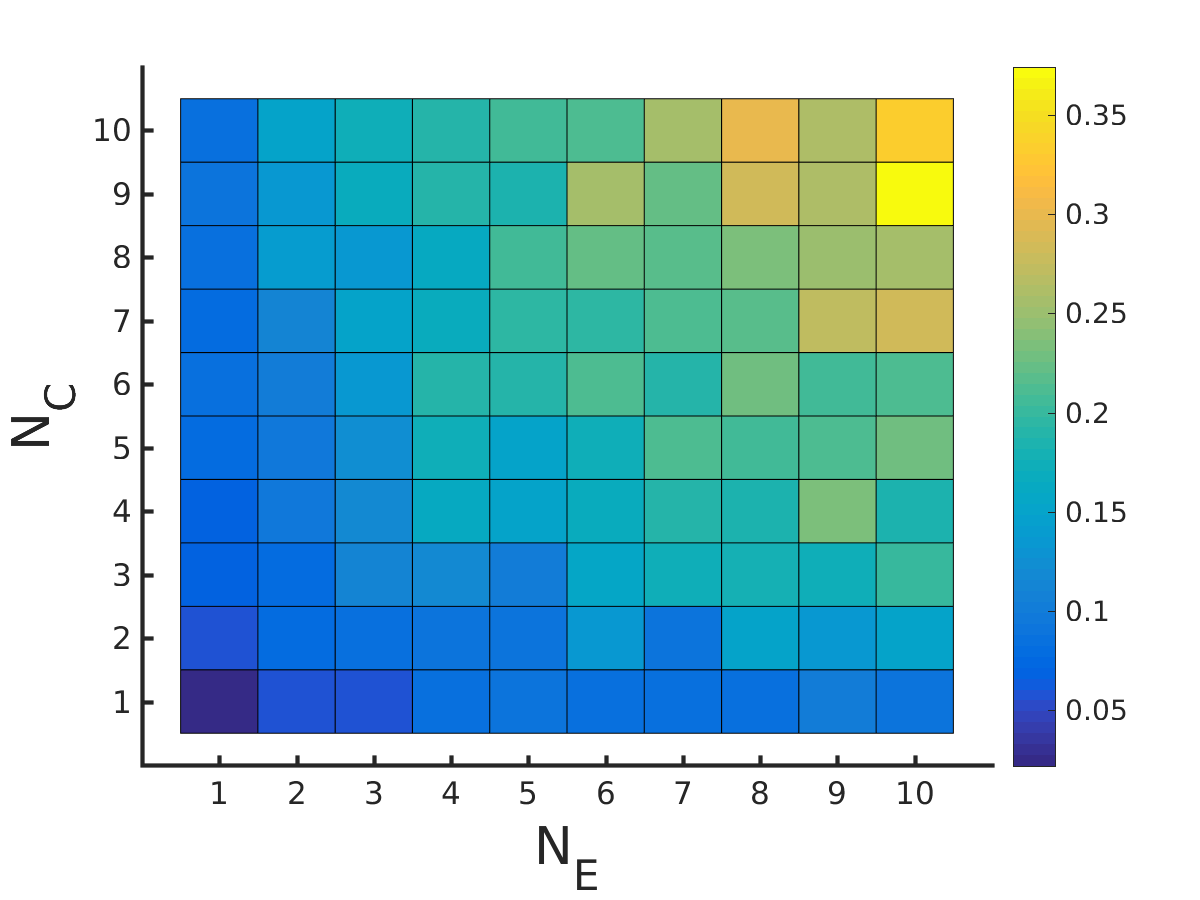}}
\:
\subfloat[Optimal enzyme specificity for more complex target $\mbf{c^*} =$ 20-GMM approximation{\label{fig:sigmamin20GMM}}]{\includegraphics[scale=0.35]{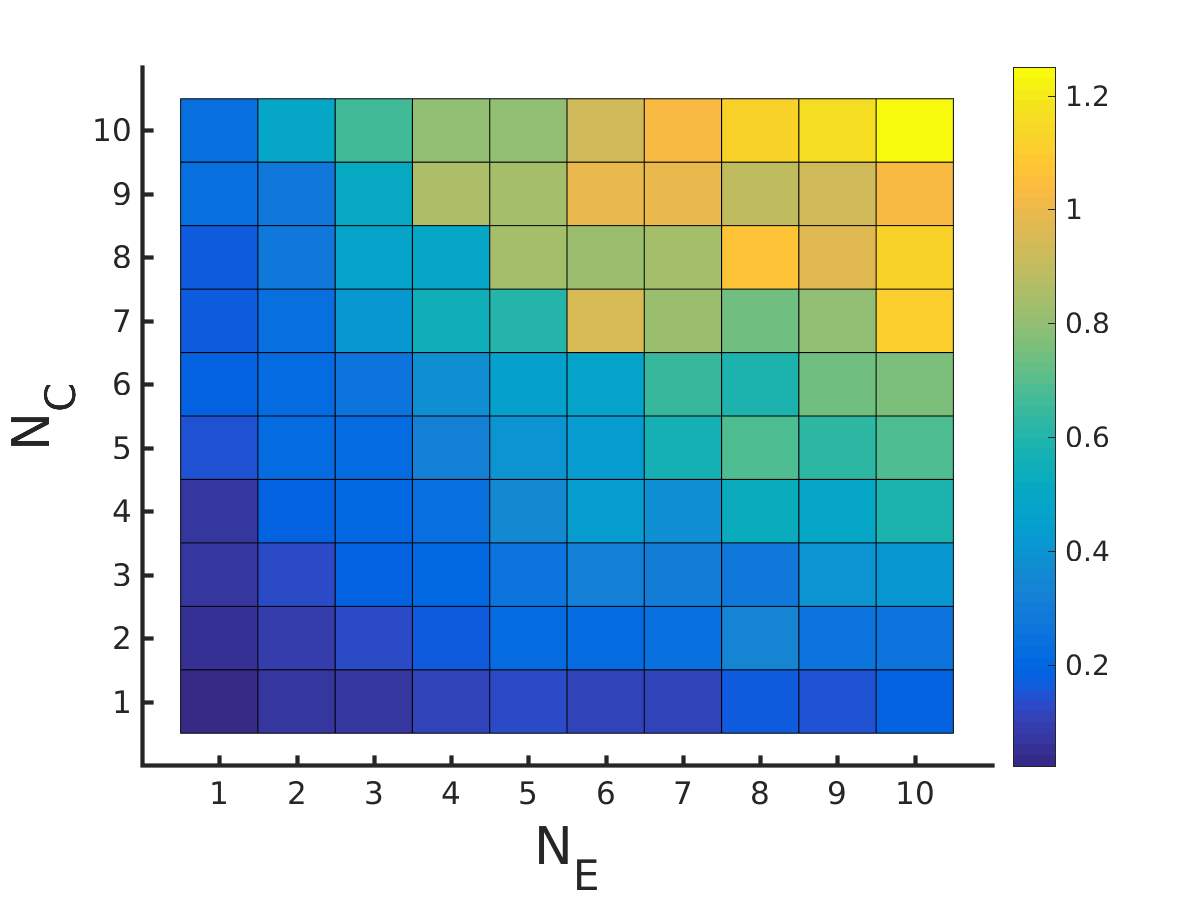}}
\:
\subfloat[Sensitivity for less complex target, $\mbf{c^*} =$ 3-GMM approximation{\label{fig:sensitivity3GMM}}]{\includegraphics[scale=.35]{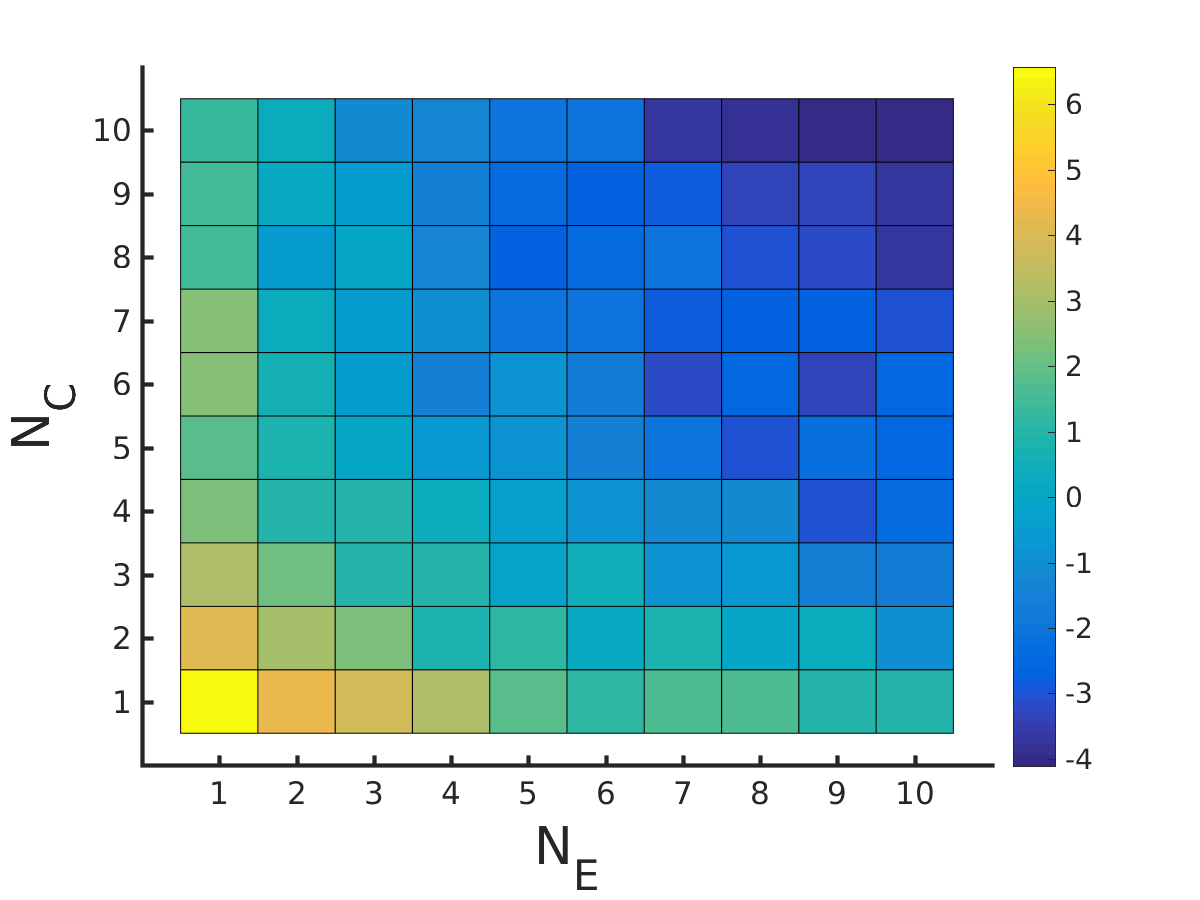}}
\:
\subfloat[Sensitivity for more complex target $\mbf{c^*} =$ 20-GMM approximation{\label{fig:sensitivity20GMM}}]{\includegraphics[scale=0.35]{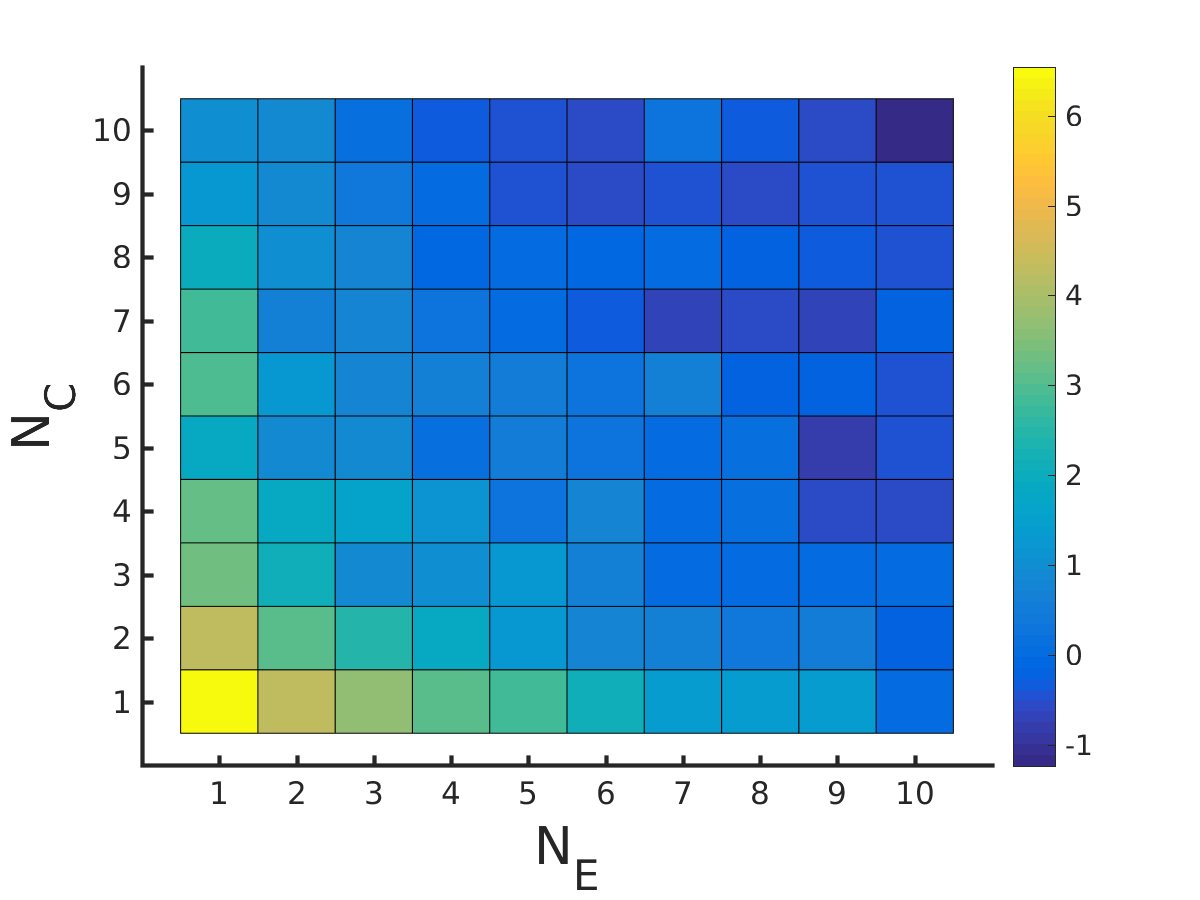}}
\:
\caption{Fidelity of glycan distribution and optimal enzyme properties to achieve a complex target distribution. 
The target $\mbf{c}^\ast$ is taken from $3$-GMM (less complex) and $20$-GMM (more complex)
  approximations of the {\it human} T-cell MSMS data.
   (a)-(b) Minimum normalised KL divergence 
  $\min_{\sigma} \{\bar{D}(\sigma,N_C,N_E,\mbf{c}^\ast)\}$ as a function of $(N_E,N_C)$. 
  More complex distributions require either a larger value $N_E$ or
  $N_C$. The marginal impact of increasing $N_E$ and $N_C$ on
  $\bar{D}$ is approximately equal. 
  (c)-(d) Optimum enzyme specificity  $\sigma_{\min}$ obtained from $\min_{\sigma} \{\bar{D}(\sigma,N_C,N_E,\mbf{c}^\ast)\}$ as a function of
  $(N_E,N_C)$. $\sigma_{\min}$ increases with increasing 
  $N_E$ or $N_C$. To synthesize the more complex $20$ GMM approximation with high fidelity requires enzymes with higher specificity $\sigma_{\min}$ compared to those needed to synthesize the broader, less
  complex $3$-GMM approximation.   
  (e) -(f) Sensitivity $\ln \f{d^2}{d \sigma^2} \bar{D}
  \Big\vert_{\sigma_{\min}}$ of the normalised Kullback-Leibler distance  $\bar{D}(\sigma,N_C,N_E,\mbf{c}^\ast)$  as a function of 
$(N_E, N_C)$. Increasing 
  $N_E$ or $N_C$ decreases this sensitivity implying the specificity $\sigma$
  does not need to be tuned very carefully if $N_E,\; N_C$ are high.
}
\label{fig:NeNcphasespace}
\end{figure*}

We are now in a position to catalogue the main results that follow from an optimization of the parameters of the glycan synthesis machinery to a given target distribution, Figs.\,\ref{fig:DklVsSigma}-\ref{fig:NeNcphasespace}

\begin{enumerate}
\item 
The normalized KL-distance
 $\bar{D}(\sigma,N_E,N_C,\mbf{c}^\ast)$ is a convex function of
 $\sigma$ for fixed values for other parameters
 (Fig.\,\ref{fig:DklVsSigma}
 ), i.e. it first
 decreases with $\sigma$ and then increases beyond a critical value of
 $\sigma_{\min}$. $\bar{D}(\sigma,N_E,N_C,\mbf{c}^\ast)$  is decreasing in $N_C$ and $N_E$ 
 for fixed values of the other parameters, and increasing in the complexity of
 $\mbf{c}^\ast$ for fixed $(\sigma, N_C)$. The marginal contribution of
 $N_C$ and $N_E$ in reducing the normalised distance $\bar{D}$ is
 approximately equal (Figs.\,\ref{fig:DklvsNeNc3GMM},\,\ref{fig:DklvsNeNc20GMM}). The lower complexity
 distributions can be synthesized with high fidelity 
 with small
 $(N_E,N_C)$, whereas
 higher complexity distributions require significantly larger  $(N_E,
 N_C)$, Figs.\,\ref{fig:DklvsNeNc3GMM},\,\ref{fig:DklvsNeNc20GMM}.  
 For a  typical mammalian cell, the number of enzymes in the
 N-glycosylation pathway are in the range
 $N_E=10-20$~\cite{UB1997,KB2005,krambeck:2009,ungar2016}, 
 Fig.\,\ref{fig:DklvsNeNc20GMM} would then suggest that the optimal cisternal
 number would range from $N_C=3-8$~\cite{Linstedt2011}. 

\item 
 The optimal enzyme specificity
 $\sigma_{\min}(\mbf{c}^\ast,N_C) = \text{argmin}_{\sigma}
 \{\bar{D}(\sigma,\bar{N}_E,N_C,\mbf{c}^\ast)\}$,  
 that  minimises the error as function of
 $(N_C,\mbf{c}^\ast)$ with $N_E$ fixed at $\bar{N}_{E}$, is an
 increasing function of $N_C$ and the complexity of the target
 distribution $\mbf{c}^\ast$
 (Figs.\,\ref{fig:DklVsSigma_Nc3GMM},\,\ref{fig:DklVsSigma_Nc20GMM},\,\ref{fig:sigmamin3GMM},\,\ref{fig:sigmamin20GMM}). This is consistent
 with the results in 
Appendix~\ref{sec:model capability} where we established that the width of
 the synthesized distribution is inversely dependent on the specificity
 $\sigma$: since a GMM approximation with fewer peaks has wider peaks,
 $\sigma_{\min}$ is low, and vice versa.  Similar results hold when $N_C$
 is fixed at $ \bar{N}_C$, and $N_E$ is varied (Figs.\,\ref{fig:DklVsSigma_Ne3GMM},\,\ref{fig:DklVsSigma_Ne20GMM},\,\ref{fig:sigmamin3GMM},\,\ref{fig:sigmamin20GMM}).

\item
  Let $\sigma_{\min}(N_C,N_E,\mbf{c}^\ast)$ denote the value of $\sigma$
  that minimizes $\bar{D}(\sigma,N_C,N_E,\mbf{c}^\ast) $. Then the second-derivative
 $\nabla^2_{\sigma_{\min}} \bar{D}(N_C,N_E,\mbf{c}^\ast) = \f{d^2}{d \sigma^2}
 \bar{D}(\sigma,N_C,N_E,\mbf{c}^\ast) \mid_{\sigma = \sigma_{\min}}$  denotes the 
 curvature at $\sigma_{\min}$, and is measure of the sensitivity of
 $\bar{D}(\sigma,N_E,N_C,\mbf{c}^\ast)$ to $\sigma$ for values close to
 $\sigma_{\min}(N_E,N_C,\mbf{c}^\ast)$. 
 $\nabla^2_{\sigma_{\min}} \bar{D}(N_C,N_E,\mbf{c}^\ast)$ is a decreasing function of $N_C$
 (resp. $N_E$) for fixed values of $(N_E,\mbf{c}^\ast)$
 (resp. $(N_C,\mbf{c}^\ast)$), see
 Figs.\,\ref{fig:DklVsSigma},\,\ref{fig:sensitivity3GMM},\,\ref{fig:sensitivity20GMM}. Thus,
 for any target distribution 
 $\mbf{c^\ast}$ there is a minimal value of $(N_E,N_C)$ such that the
 target can be synthesized with high
 fidelity provided the sensitivity $\sigma$ is tightly controlled at $\sigma_{\min}(N_C,N_E,\mbf{c}^\ast)$, and
 there is larger value of 
 $(N_E,N_C)$ such that the target can be synthesized even if the control
 on $\sigma$ is less tight.
\end{enumerate}

Ungar et al.~\cite{ungar2019}  
optimize incoming glycan ratio, transport rate and effective reaction
rates in order to synthesize a narrow  target distribution 
centred around a desired glycan. The ability to produce
specific glycans without much heterogeneity is an important goal in pharma
industry. They define heterogeneity as the total  number of glycans
synthesized, and show that increasing the number of compartments $N_C$
decreases heterogeneity, and increases the concentration of the specific
glycan. They also show that changing transport rate does not affect 
the heterogeneity. Our results are entirely consistent with theirs - we
have shown that $\bar{D}$  decreases as	we increase $N_C$. Thus, if the target
distribution has a single sharp peak, increasing $N_C$ will reduce the
heterogeneity in the distribution.

\subsection{Optimal partitioning of enzymes in cisternae}
\label{sec:EnzymePartition}

\begin{figure*}
\subfloat[]{\includegraphics[scale=0.25]{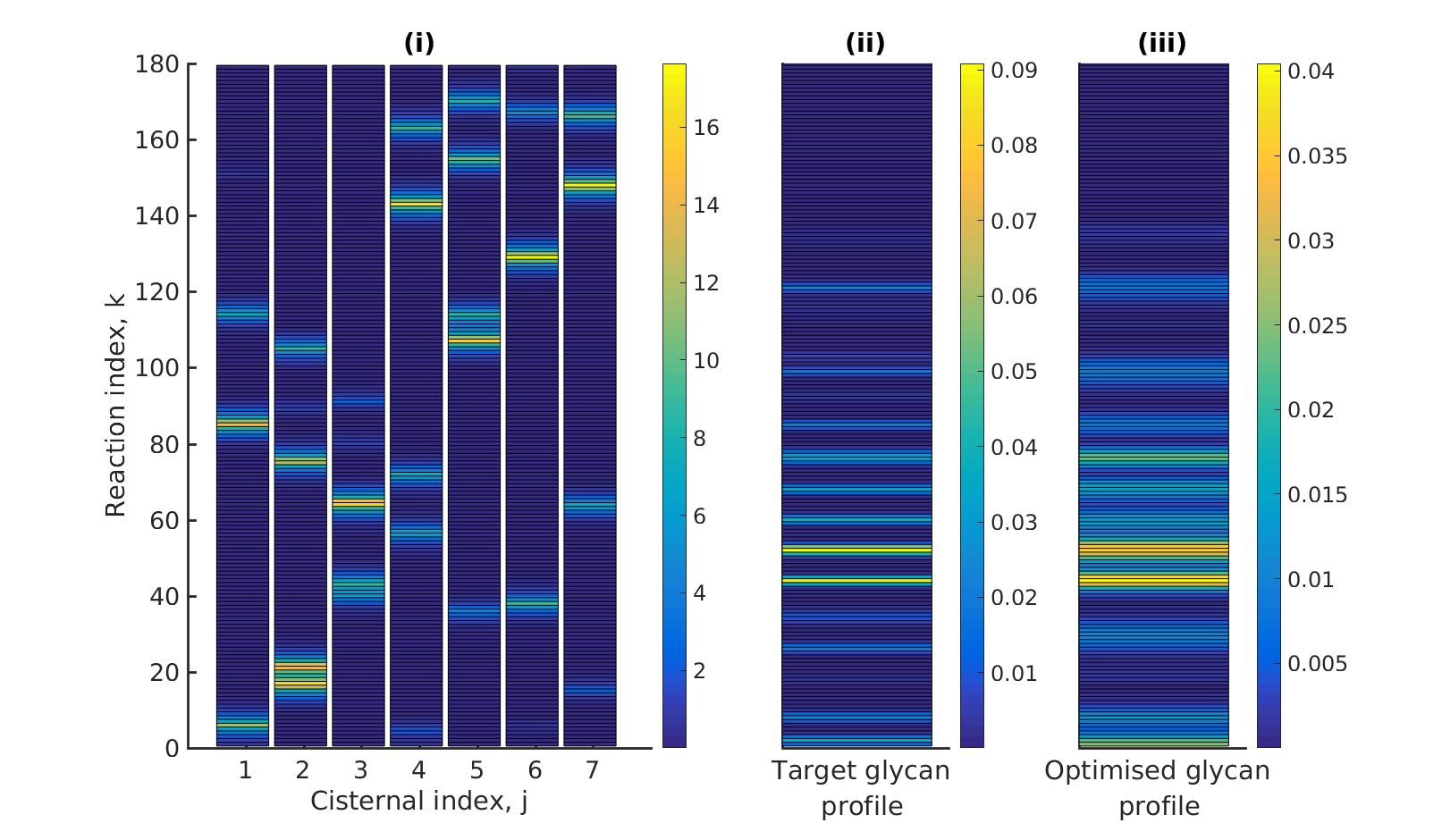}}
\:
\subfloat[]{\includegraphics[scale=0.25]{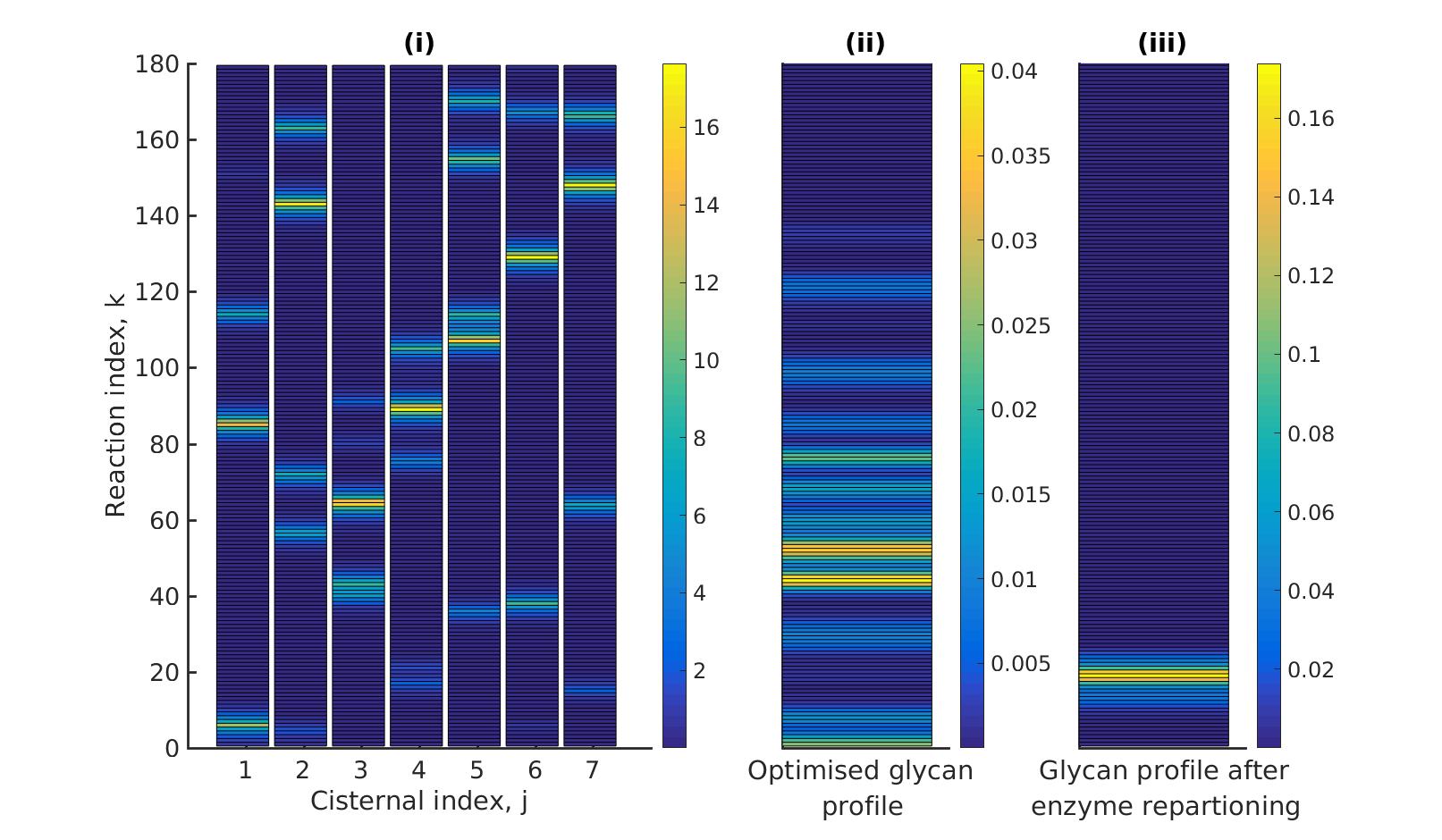}}
\:
\caption{Optimal enzyme partitioning in cisternae. (a) Heat map of the
effective reaction rates in each cisterna (representing the optimal enzyme partitioning) and the steady state concentration in the last compartment ($\mbf{c}^{(N_C)}$) for the 20-GMM target distribution. Here $N_E = 5$, $N_C = 7$, normalised $D_{KL}({\bf T}^{(20)} \| {\bf c}^{(N_C)})/H({\bf T}^{(20)})=0.11$. (b) Effective Reaction rates after swapping the optimal  enzymes of the fourth and second cisternae. The displayed glycan profile is considerably altered from the original profile.}
\label{fig:EnzymePartioning}
\end{figure*}

Having studied the optimum $N_E, N_C, \sigma$ to attain a given target
distribution with high fidelity, we 
ask what is the optimal
partitioning of the $N_E$ enzymes
in these $N_C$ cisternae? Answering this within our chemical reaction model (Sect.\,\ref{sec:model}) requires some care, since it incorporates the following enzymatic features:
(a)~enzymes 
with a
finite specificity $\sigma$ can catalyse several
reactions, although with an efficiency that 
varies with
both the substrate index $k$ and cisternal index $j$, and (b)~every enzyme
appears in each cisternae; however  
their reaction efficiencies depend on the enzyme levels, the enzymatic
reaction rates and the enzyme matching function $\bf{L}$, all of which
depend on the cisternal index $j$. 

Thus, rather than determining the cisternal partitioning of enzymes, we
instead 
identify
chemical reactions that occur with high propensity in
each cisternae. 
For this we define an effective reaction rate $\bar{R}(j,k)$ for
${\cal{P}}c_{k} \to {\cal{P}}c_{k+1}$ in the $j$-th cisterna as
\be
{\bar R}(j,k) = \sum_{\alpha = 1}^{N_E} R^{(j)}_\a P^{(j)}(k,\a).
\ee 
According to our model presented in Sect.\,\ref{sec:model}, the list of
reactions with high effective reaction rates in each cisterna, corresponds
to a  
cisternal partitioning of the perfect enzymes. In a future study, we will
consider a Boolean version of a more complex chemical model,  
 to address more clearly, the optimal enzyme partitioning amongst
 cisternae.

Figure\,\ref{fig:EnzymePartioning}~(a)~(i) shows the heat map of the
effective reaction rates in each cisterna for the optimal $N_E, N_C,
\sigma$ that minimises the normalised KL-distance to the 20~GMM target
distribution $\mbf{T}^{(20)}$ 
(see Fig.\,\ref{fig:EnzymePartioning}~(a)~(ii)). 
The optimized glycan profile displayed in 
Fig.\,\ref{fig:EnzymePartioning}~(a)~(iii) 
is very close to the target. An interesting observation from Fig.\,\ref{fig:EnzymePartioning}a(i)
is that the same reaction can occur in multiple cisternae.

Keeping everything else fixed at the optimal value, we ask whether simply
repartitioning the optimal enzymes amongst the cisternae, alters the
displayed  
glycan distribution. In Fig.\,\ref{fig:EnzymePartioning}~(b)~(i), we have
exchanged the enzymes of the fourth and second cisterna. The glycan
profile after enzyme partitioning (see
Fig.\,\ref{fig:EnzymePartioning}~(b)~((iii)) 
is now 
completely altered (compare Fig.\,\ref{fig:EnzymePartioning}~(b)~(ii) with
Fig.\,\ref{fig:EnzymePartioning}~(b)~(iii)). Thus one may achieve a
different glycan distribution by repartitioning enzymes amongst the 
same number of cisternae~\cite{thattai2018}.

\section{Strategies to achieve high glycan diversity}
\label{sec:Diversity}

\begin{figure*}
\subfloat[]
{\includegraphics[scale=.35]{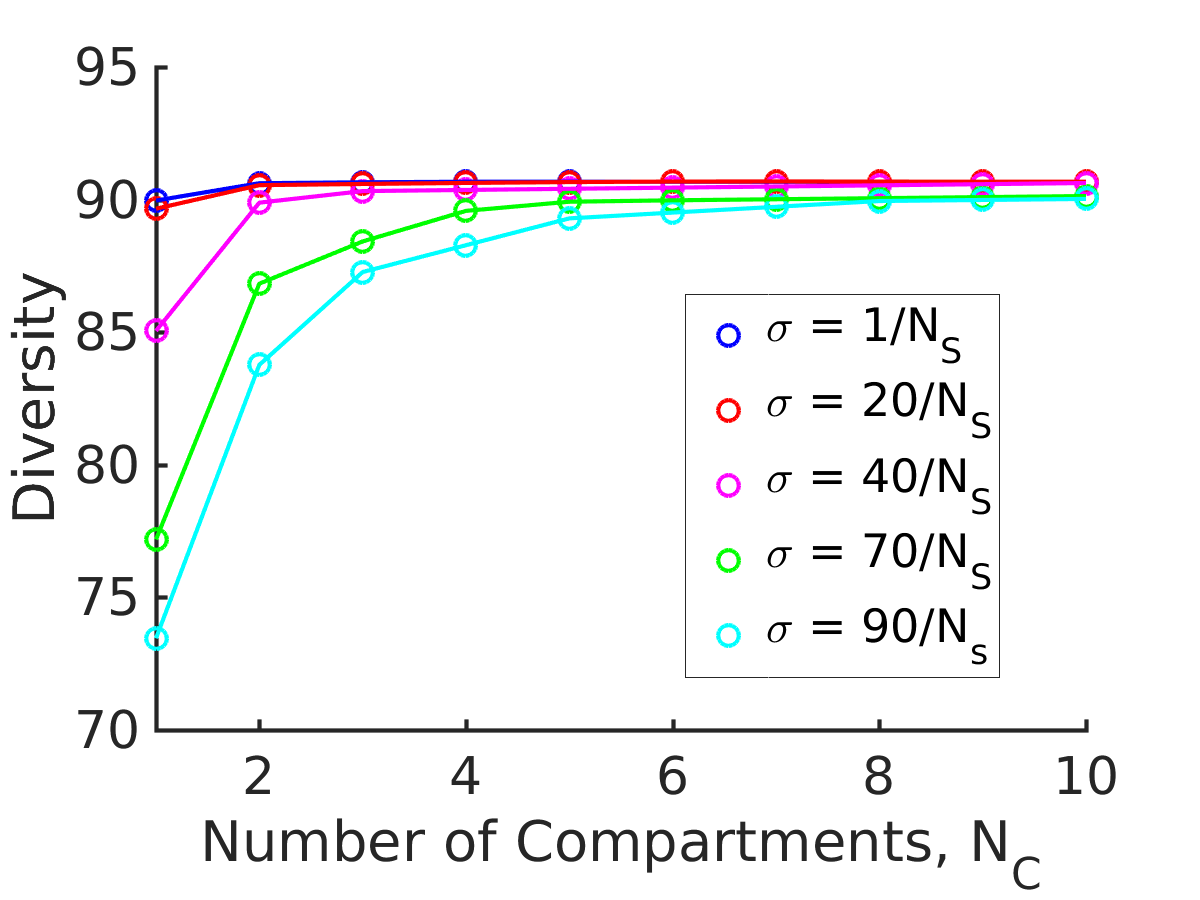}}
\:
\subfloat[]
{\includegraphics[scale=.35]{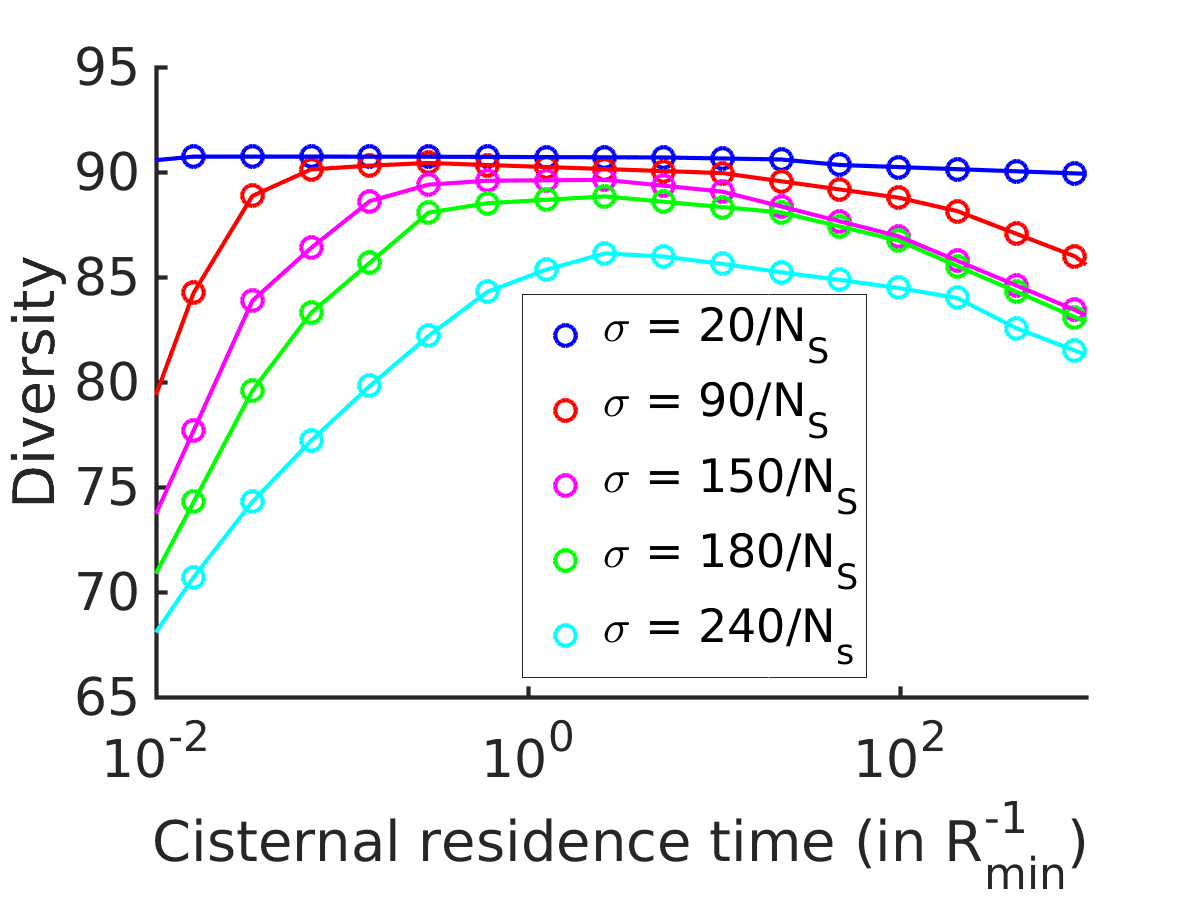}}
\caption{Strategies for achieving high glycan diversity. Diversity versus $N_C$ and transport rate $\mu$ at various values
  of specificity $\sigma$ for fixed $N_E=3$.  (a) Diversity vs. $N_C$ at
  optimal transport rate $\mu$. Diversity initially increases with $N_C$,
  but eventually levels off. 
The levelling off starts at a higher $N_C$ when $\sigma$ is 
increased. These curves are bounded by the $\sigma=0$ curve.  
 (b) Diversity vs. cisternal residence time ($\mu^{-1}$) in units of the reaction time ($R_{\min}^{-1}$) at various value of $\sigma$, for
 fixed $N_C =4$ and $N_E=10$. This has implications for  glycoengineering (see text) where the task is to produce a particular glycan profile with low heterogeneity~\cite{ungar2019,thattai2018}. 
 } 
\label{fig:GMM_CellularCost}
\end{figure*}

So far we have studied how the complexity of the target glycan
distribution places constraints on the evolution of Golgi cisternal number
and enzyme specificity.  We  
now take up another issue, namely, how the physical properties of
the Golgi cisternae, namely cisternal number and inter-cisternal transport
rate, may drive diversification of glycans  
\cite{Varki2011,Dennis2009}. There is substantial correlative evidence to
support the idea that cell types that carry out extensive glycan
processing employ larger numbers of 
Golgi cisternae. For example, the salivary Brunner’s gland cells secrete
mucous that contains heavily O-glycosylated mucin as its major
component~\cite{VanHalbeek1983}.  
The Golgi complex in these specialized cells contain $9-11$ cisternae per stack.
Additionally, several organisms such as
plants and algae secrete a rather diverse repertoire of large, complex glycosylated proteins,
 for a variety of functions~\cite{plants1,plants2,plants3,plants4,plants5,plants6,plants7,plants8,
plants9,plants10}. These organisms possess enlarged Golgi complexes with
multiple cisternae per stack~\cite{BeckerMelkonian1996,golgi_cisternae1,
  golgi_cisternae2,golgi_cisternae3,golgi_cisternae4}.

 In this section, we study 
 how changing the physical parameters 
in our chemical synthesis model can lead to changes in the diversity
 of glycan 
distributions.

We define {\it diversity} as the total number of glycan species produced
above a specified threshold abundance $c_{th}$. This last condition is
necessary 
because very small peaks will not be distinguishable in the presence of
noise.
In
computing the diversity from our chemical synthesis model, we have chosen
the threshold to be $c_{th} = 1/N_s$, 
where $N_s$ is the total number of glycan species. We have checked that the qualitative results do not depend on this choice, Fig.\,\ref{fig:threshold_independence}.


%

Using the sigmoid function $(1+e^{-x/\tau})^{-1}$ as a continuous
approximation to the Heaviside function $\Theta(x)$,
 we define the
following optimization to achieve the maximal diversity for a  
given set of parameter values, $N_E, N_C, \sigma$,

\bea
\lefteqn{\text{Diversity}(\sigma,N_C, N_E) :=} 
 \hspace*{1.5in}
\begin{array}[t]{rl}
  \max_{\bs{\mu}, \mbf{R}, \mbf{L}}  & \sum_{i=1}^{N_s} \big(1+ e^{-N_s(c_i - c_{th})}\big)^{-1} \\
  \text{s.t.} &   R_{\min} \leq R^{(j)}_\alpha \leq R_{\max}, \nn \\
                                     &   \mu_{\min} \leq \mu^{(j)} \leq  \mu_{\max},
\end{array}
\label{eq:mu_R_problem}
\eea

where, as before,  $(\mu_{\max}, \mu_{\min})  = (1, 0.01)$/min, and
$(R_{\max},R_{\min}) =  (20, 0 .018)$/min, and $c_{th} = 1/N_s$ is the threshold. See
Appendix~\ref{sec:param_estimate} for details on the parameter estimation.

The results displayed in Fig.\,\ref{fig:GMM_CellularCost}(a), show that for a
fixed specificity $\sigma$, the diversity at first increases with the
number of cisternae $N_C$, and then saturates at a value that depends on
$\sigma$.  For very high specificity enzymes, one can achieve very high
diversity by appropriately increasing $N_C$. This establishes the link between glycan diversity and 
cisternal number. However, this  link is correlative at best, since there are many ways to achieve high glycan diversity - 
notably by increasing the number of enzymes.

On the other hand, one of the goals of glycoengineering
%
is to produce a particular glycan profile with low heterogeneity~\cite{ungar2019,thattai2018}. For low specificity enzymes, the diversity remains unchanged upon increasing the cisternal residence time.
 For enzymes with high specificity, the diversity typically shows a non-monotonic variation with the cisternal residence time. 
  At small cisternal residence time, the diversity
 decreases from the peak because of early exit of incomplete oligomers. At large cisternal residence time
the diversity again decreases as more reactions are taken to completion. Note that the peak is generally very flat, this is consistent with the results of \cite{ungar2019}.
To get a sharper peak, as advocated for instance by~\cite{thattai2018}, one might need to increase the number of high specificity enzymes $N_E$ further.

\section{Discussion}


The precision of the stereochemistry and enzymatic kinetics of these N-glycosylation reactions~\cite{varki}, has inspired 
 a number of mathematical models~\cite{UB1997,krambeck:2009,KB2005} that predict the
N-glycan distribution   
based on the activities and levels of processing enzymes distributed in
the Golgi-cisternae of mammalian cells, and compare these predictions with  
N-glycan mass spectrum data. Models such as the KB2005 model~\cite{UB1997,KB2005,krambeck:2009} are extremely elaborate (with a network of $22,871$ chemical reactions and
$7565$ oligosaccharide structures) and require many chemical input parameters.
These models have an important practical role to play, that of being
able to predict the impact of the various {\it chemical parameters} on the
glycan distribution, and to evaluate appropriate metabolic strategies to recover the
original glycoprofile.
 Additionally, a recent study by Ungar and coworkers~\cite{ungar2019,ungar2016} shows how
{\it physical parameters}, such as overall Golgi transit time and cisternal number, can be tuned to engineer a homogeneous glycan distribution.
Overall, such models may help predict glycosylation patterns and direct glycoengineering projects
to optimize glycoform distributions.

In this paper, we have been interested in the role of glycans as a marker or molecular code of cell identity~\cite{gabius2018,varki2017,parashuraman2019}. 
In particular, we have studied one aspect of molecular coding, namely  the {\it fidelity} of this glycan code generated by
enzymatic and transport processes located in the secretory apparatus of
the cell. This involves a method of analysis that draws on many different fields, and so it might be useful to 
provide a short summary of the assumptions, methods and results of the paper:
\begin{enumerate}
\item The distribution of glycans at the cell surface is a marker of {\it
    cell-type identity}~\cite{varki,gabius2018,varki2017,parashuraman2019}. This glycan distribution can be very complex; it is believed that
  there is an evolutionary drive for having glycan distributions of
  high {\it complexity}  arising from the following
  considerations, 
  \begin{enumerate}
  \item Reliable cell type identification amongst a large set of different cell types in a complex organism, the
  preservation and diversification of ``self-recognition''~\cite{Drickamer1998}.
  
  \item Consequence of
    pathogen-mediated selection pressures~\cite{varki,varki2017,Gagneux&Varki1999}.
    
    \item Consequence of {\it herd immunity} within a heterogenous population of cells of a community~\cite{WillsGreen1995} or within a single organism~
    \cite{Drickamer1998}.
    
  \end{enumerate}

\item The glycans at the cell surface are the end product of a sequential
  chemical processing via a set of  enzymes resident in the Golgi cisternae, and  transport across cisternae~\cite{varki2017,varki1998,parashuraman2019}. 
  Using a fairly general and tractable model for chemical synthesis and transport, 
  we compute the {\it synthesized}  glycan distribution at the cell surface.  
  Parameters of our synthesis model include the number of enzymes $N_E$,
  specificity of enzymes $\sigma$, number of cisternae $N_C$ and transport
  rates $\mu$. 
  
\item 
We measure the reliability or fidelity of cell identity~\cite{Stanley2011,parashuraman2019,varki1998} in terms of 
  the error between
  synthesized glycan distribution on the cell surface from the its internal ``target''
  distribution using the Kullback- Leibler distance $D_{KL}$~\cite{information,mckay}.
  In our numerical study, we obtain the  {\it target distribution}  for
  the given cell type by suitable coarse-graining of the
  MSMS data for the {\it human} T-cells~\cite{msdata}.  
  We solve a constrained optimization problem for minimising $D_{KL}$, and
  study the tradeoffs between $N_E$,  $N_C$ and $\sigma$. 
\item The results of the optimization to
  achieve a given target complexity
are summarised in
  Figs.~\ref{fig:DklVsSigma}-\ref{fig:NeNcphasespace}. Here, we highlight some its direct consequences:
  \begin{enumerate}
  \item 
    Keeping the number of enzymes fixed, a more elaborate transport mechanism (via control
    of $N_C$ and $\mu$) is essential for synthesising  high complexity
    target distributions~(Figs.~\ref{fig:DklvsNeNc3GMM},\,\ref{fig:DklvsNeNc20GMM}). 
    Fewer cisternae
    cannot be compensated for by  optimising the enzymatic synthesis (via control of
    parameters $\mbf{R, L}$ and $\sigma$). 
  \item Thus, our study suggests that fidelity of the glycan code generation
    provides a  functional control of Golgi
    cisternal number.   It also provides a quantitative  argument for
    the evolutionary requirement of multiple-compartments, by demonstrating
    that the fidelity and sensitivity of the glycan code arising from a
    chemical synthesis that involves multiple cisternae is higher than one
    that involves 
    a single cisterna (keeping everything else fixed)
    (Figs.~\ref{fig:DklvsNeNc3GMM},\,\ref{fig:DklvsNeNc20GMM},\,\ref{fig:sensitivity3GMM},\,\ref{fig:sensitivity20GMM}).
    This feature that with multiple cisternae and precise enzyme partitioning, one may generically achieve a highly accurate representation of the 
target distribution, has been highlighted in an algorithmic model of glycan synthesis~\cite{thattai2018}.


  \item Our study shows that for a fixed $N_C$ and $N_E$, there is an
    optimal enzyme specificity that achieves the lowest distance from a given
    target distribution.    
    As we see in Fig.~\ref{fig:sigmamin20GMM}, this optimal enzyme
    specificity can be very high for highly complex target distributions. 
     \item  
  Organisms such as plants and algae, have a diverse repertoire
    of glycans that are utilised in a variety of functions~\cite{plants1,plants2,plants3,plants4,plants5,plants6,plants7,plants8,
plants9,plants10}. Our study shows that
 it is optimal to
    use low specificity enzymes to synthesize target
    distributions with high diversity (Fig.~\ref{fig:GMM_CellularCost}).
    However, this compromises
    on the complexity of the glycan distribution, revealing a tension
    between complexity and diversity. One way of relieving this tension is
    to have larger $N_E$ and $N_C$.
  \item Consider a situation where the environment, and hence the target
    glycan distribution, fluctuates rapidly. 
    When 
    synthesis parameters cannot change rapidly enough to
    track the environment, high specificity enzymes 
    can lead to a
    \textit{lowering} of the cell's fitness~\cite{specificity_evolution1,specificity_evolution2}. 
    Having slightly sloppy enzymes may  
    give the best selective advantage in a time varying environment. This
    compromise, between robustness in a  
    changing environment and the demand for complexity, is achieved by
    having sloppy enzymes, that allows the system to be more {\it
      evolvable}~\cite{specificity_evolution1,specificity_evolution2}.  
    However, sloppy enzymes are subject to errors from
    synthesising the wrong reaction products. In this case, 
    error correcting mechanisms must be in place to ensure fidelity of the
    glycan code. 
    We leave the role of 
    intra-cellular transport    
    in providing non-equilibrium proof-reading mechanisms to reduce such
    coding errors, and its optimal adaptive strategies and plasticity in a time
   varying environment, as a task for the future.  
  
  \end{enumerate}

\end{enumerate}


Admittedly the chemical network that we have considered here is much simpler than the chemical network associated with all possible
 protein modifications in the secretory pathway. For instance, 
typical N-glycosylation pathways would involve the glycosylation of a variety of GBPs. Further, apart from N-glycosylation, 
there are other glycoprotein, proteoglycan  and glycolipid synthesis pathways~\cite{mboc,varki,parashuraman2019}. 
We believe our analysis is generalisable and that the qualitative results we have arrived at would still hold.

To conclude, our work establishes the link between the cisternal 
machinery (chemical and transport) and optimal coding. We find that the pressure to achieve the target glycan code for a given cell type,
 places strong  constraints on the cisternal number and enzyme specificity~\cite{Linstedt2011}.
An important implication is that a description of the nonequilibrium self-assembly of a fixed number of Golgi
cisternae must combine the dynamics of chemical processing and membrane dynamics
involving fission, fusion and transport~\cite{Linstedt2011,himani,sens2013}. We believe this is a promising direction for future research.

\section{Acknowledgments}
 We thank
M. Thattai, A. Jaiman,  S. Ramaswamy, A. Varki for discussions, and S. Krishna and R. Bhat  
for very useful suggestions on the manuscript. We thank our group members at the Simons Centre for many incisive inputs.
We are very grateful to P. Babu for consultations on the 
MSMS data and literature. 
We acknowledge the computational facilities at NCBS. MR acknowledges a JC Bose Fellowship from DST (Government of India), and thanks Institut Curie
for hosting a visit under the Labex program. This work has received support under the program Investissements d’Avenir launched 
by the French Government and implemented by ANR with the references 
ANR-10-LABX-0038 and ANR-10-IDEX-0001-02 PSL. QV thanks the Simons Centre (NCBS)
for hosting his visit.

\newpage

\appendix

\renewcommand{\thefigure}{A\arabic{figure}}
\setcounter{figure}{0}


\begin{center}
	{\Large \bf  Appendix}
\end{center}

\section{Kinetics of sequential chemical reactions and transport}
\label{sec:seqchemapp}
On including the rates of injection ($q$) and inter-cisternal transport $\mu^{(j)}$ from the cisterna $j$ to $j+1$, into the chemical reaction kinetics, the substrate concentrations
$c_{k}^{(j)}$ change with time as,
\begin{equation}
\label{eq:massaction1}
\begin{split} 
  \frac{dc^{(1)}_1}{dt} = \; &  q  -   \sum_{\a=1}^{N_E} \frac{V(1, 1, \a) P^{(1)}(1,\a) c_{1}^{(1)}}
                     {M(1,1, \a) \l1 + \sum_{k'=1}^{N_s} \frac{P^{(1)}(k',\a)c_{k'}^{(1)}}{M(1,k', \a)}\r} 
   - \mu^{(1)} c_1^{(1)}\\
  \frac{dc_k^{(1)}}{dt} = &
   \sum_{\a=1}^{N_E} \frac{V(1, k-1, \a) P^{(1)}(k-1,\a) c_{k-1}^{(1)}}
                     {M(1,k-1, \a) \l1 + \sum_{k'=1}^{N_s} \frac{P^{(1)}(k',\a)c_{k'}^{(1)}}{M(1,k', \a)}\r} \\
  & \mbox{}  -    \sum_{\a=1}^{N_E} \frac{V(1, k, \a) P^{(1)}(k,\a) c_{k}^{(1)}}
                     {M(1,k, \a) \l1 + \sum_{k'=1}^{N_s} \frac{P^{(1)}(k',\a)c_{k'}^{(1)}}{M(1,k', \a)}\r} 
   \mbox{} - \mu^{(1)} c_k^{(1)}\\   
  \frac{dc^{(1)}_{N_s}}{dt} = &
    \sum_{\a=1}^{N_E} \frac{V(1, N_s-1, \a) P^{(1)}(N_s-1,\a) c_{N_s-1}^{(1)}}
                     {M(1,N_s-1, \a) \l1 + \sum_{k'=1}^{N_s} \frac{P^{(1)}(k',\a)c_{k'}^{(1)}}{M(1,k', \a)}\r}  - \mu^{(1)} c_{N_s}^{(1)}\\ 
\end{split} 
\end{equation}
for cisterna-$1$, and
\begin{equation}
  \label{eq:massaction2}
  \begin{split} 
    \frac{dc^{(j)}_1}{dt} = \; & \mu^{(j-1)} c_1^{(j-1)} -
   \sum_{\a=1}^{N_E} \frac{V(j, 1, \a) P^{(j)}(1,\a) c_{1}^{(j)}}
                     {M(j,1, \a) \l1 + \sum_{k'=1}^{N_s} \frac{P^{(j)}(k',\a)c_{k'}^{(j)}}{M(j,k', \a)}\r} - \mu^{(j)}c^{(j)}_1\\
  \frac{dc_k^{(j)}}{dt} =\; & \mu^{(j-1)} c_k^{(j-1)} + 
 \mbox{  }  \sum_{\a=1}^{N_E} \frac{V(j, k-1, \a) P^{(j)}(k-1,\a) c_{k-1}^{(j)}}
                     {M(j,k-1, \a) \l1 + \sum_{k'=1}^{N_s} \frac{P^{(j)}(k',\a)c_{k'}^{(j)}}{M(j,k', \a)}\r} \\
  &\mbox{} -    \sum_{\a=1}^{N_E} \frac{V(j, k, \a) P^{(j)}(k,\a) c_{k}^{(j)}}
                     {M(j,k, \a) \l1 + \sum_{k'=1}^{N_s} \frac{P^{(j)}(k',\a)c_{k'}^{(j)}}{M(j,k', \a)}\r}  
  - \mu^{(j)} c_k^{(j)} \\
  \frac{dc_{N_s}^{(j)}}{dt} =\; & \mu^{(j-1)} c_{N_s}^{(j-1)} + 
   \sum_{\a=1}^{N_E} \frac{V(j, N_s-1, \a) P^{(j)}(N_s-1,\a) c_{N_s-1}^{(j)}}
                     {M(j,N_s-1, \a) \l1 + \sum_{k'=1}^{N_s} \frac{P^{(j)}(k',\a)c_{k'}^{(j)}}{M(j,k', \a)}\r}   
 - \mu^{(j)} c_{N_s}^{(j)} 
\end{split}
\end{equation}
for cisternae $j=2, 3, \ldots, N_C$. 
These set of dynamical equations \eqref{eq:massaction1}-\eqref{eq:massaction2}, with initial conditions, can be solved to obtain
the concentration $\cjk(t)$ for $t \geq 0$.


Equations \eqref{eq:massaction1}-\eqref{eq:massaction2} automatically obey the conservation law for the protein concentration ($p$), i.e., denoting the protein
concentration in the $j$-th cisterna as $p^{(j)} = \sum_{k'=1}^{N_s} c_{k'}^{(j)}$, we automatically obtain,
\bea
\frac{dp^{(1)}}{dt} & = & q - \mu^{(1)} 
p^{(1)} \nn \\
\frac{dp^{(j)}}{dt} & = &  \mu^{(j-1)} p^{(j-1)} -
\mu^{(j)} p^{(j)} \nn
\eea
for $j = 2, 3, \ldots N_C$.

At steady state, the left hand side of the above equations is set to zero, which after rescaling, gives the nonlinear recursion relations displayed in
(\ref{eq:concentrations1}) and (\ref{eq:concentrationsj}) of the main text.

\section{A computationally simpler optimization equivalent to Optimization A}
\label{sec:optB_def}
Define a new set of parameters,

\be
  \label{eq:R-def}
  R(j,k,\a) = 
         \sum_{\a=1}^{N_E} \frac{V(j, k, \a)}
                     {M(j,k, \a) \l1 + \sum_{k'=1}^{N_s} \frac{P^{(j)}(k',\a)c_{k'}^{(j)}}{M(j,k', \a)}\r} 
\ee

where $\mathbf{c}$ denotes the steady state glycan concentration,
corresponding to a specific   
$(\mbf{M}, \mbf{V}, \mbf{L})$.  Define $\mathbf{v}$ by the following set of linear
equations:

\be 
\label{eq:Bconcentrations1}
\begin{split}
v_1^{(1)} = &\frac{1}{\mu^{(1)} + \sum_{\a
    =1}^{N_E} R(1,1,\a) P^{(1)}(1,\a)} \\ 
 v_k^{(1)} = & \frac{v_{k-1}^{(1)}\sum_{\a
     =1}^{N_E} R(1,k-1,\a) P^{(1)}(k-1,\a)}{\mu^{(1)}
   + \sum_{\a =1}^{N_E} R(1,k,\a)P^{(1)}(k,\a)}
 \\ 
 v_{N_s}^{(1)} = &\frac{v_{N_s-1}^{(1)}  \sum_{\a
     =1}^{N_E}R(1,N_s-1,\a)P^{(1)}(N_s-1,\a)}{\mu^{(1)}}
\end{split}
\ee

for $j = 1$, and

\be
  \label{eq:Bconcentrationsj}
  \begin{split}
v_1^{(j)} = \; & \frac{v_{1}^{(j-1)}\mu^{(j-1)}}{\mu^{(j)} + \sum_{\a
    =1}^{N_E}R(j,1,\a)P^{(j)}(1,\a)}
    \\  
 v_k^{(j)} = \; &\frac{v_{k}^{(j-1)}\mu^{(j-1)}}{\mu^{(j)} + \sum_{\a
     =1}^{N_E}R(j,k,\a)P^{(j)}(k,\a)}\\
 & \mbox{} +  \frac{ v_{k-1}^{(j)}\sum_{\a
     =1}^{N_E}R(j,k-1,\a)P^{(j)}(k-1,\a)}{\mu^{(j)}
   + \sum_{\a =1}^{N_E}R(j,k,\a)P^{(j)}(k,\a)}
  \\ 
 v_{N_s}^{(j)} = \; &\frac{v_{N_s}^{(j-1)}\sum_{\a
     =1}^{N_E}R(j,N_s-1,\a)P^{(j)}(N_s-1,\a)}{\mu^{(j)}}   
  + \frac{v_{N_s}^{(j-1)}\mu^{(j-1)}}{\mu^{(j)}}   
\end{split}
\ee

for $j = 2, \ldots, N_C$.
Then, by the definition of $\mathbf{R}$ in \eqref{eq:R-def}, it
  trivially follows that the steady state concentration $\mbf{c}$ corresponding to
  $(\mathbf{M},\mathbf{V}, \mathbf{L})$ is a solution for
  \eqref{eq:Bconcentrations1}-\eqref{eq:Bconcentrationsj}.

\begin{figure}
\centering
{\includegraphics[scale=0.4]{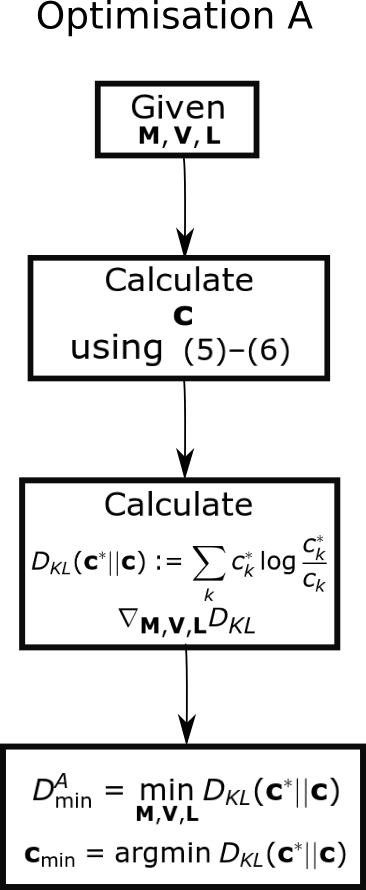}}
\quad
{\includegraphics[scale=0.4]{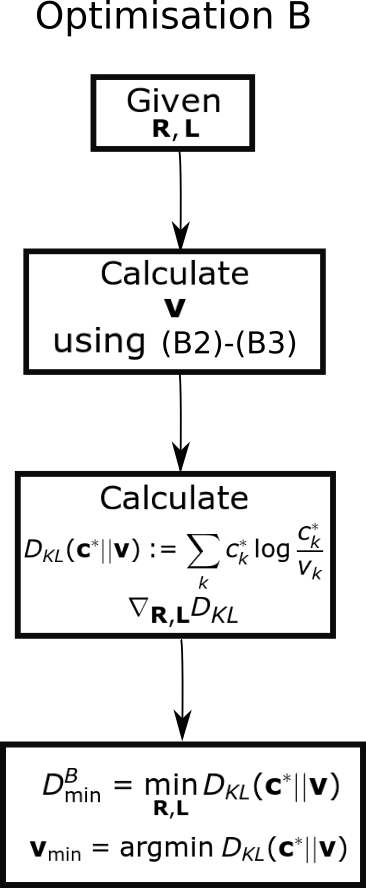} }
\caption{Flow chart showing the optimization schemes for Optimization A and B. We prove that $D_{\min}^A = D_{\min}^B$ by showing the set of all $\bf{c}$ is equal to the set of all $\bf{v}$.
We additionally establish that the optimum $\mbf{v}_{\min}=\mbf{c}_{\min}$.}
\label{fig:Schematics}
\end{figure}

In Appendix~\ref{sec:equivalence_opt_problem} we show that for $\mbf{v}$ obtained from \eqref{eq:Bconcentrations1}-\eqref{eq:Bconcentrationsj} for any parameter $(\mbf{R}, \mbf{L})$, there exists parameter $(\mbf{M,V,L})$ such that \eqref{eq:concentrations1}-\eqref{eq:concentrationsj} are automatically satisfied 
when we set $\mbf{c}=\mbf{v}$, i.e. $\mbf{v}$ is the steady state concentration for $(\mbf{M}, \mbf{V}, \mbf{L})$. 
Thus, the set of all concentration profiles defined by
\eqref{eq:Bconcentrations1}-\eqref{eq:Bconcentrationsj} 
as a function of all possible values of the parameters $(\mbf{R}, \mbf{L})$
is identical to the set defined by
\eqref{eq:concentrations1}-\eqref{eq:concentrationsj} as function of
$(\mbf{M}, \mbf{V}, \mbf{L})$. This is a crucial insight, since it
allows us to search the entire parameter space using
\eqref{eq:Bconcentrations1}-\eqref{eq:Bconcentrationsj}, where the
concentration is known explicitly in terms of $(\mbf{R}, \mbf{L})$.
See Figure~\ref{fig:Schematics} for a flow chart of the two optimization schemes.

To pose this new optimization problem, it is convenient to define $\bar{v}_i = \mu^{(N_c)}v^{(N_c)}_i$. Then, it follows that

{\it Optimization B}: 

\be
  \label{eq:problem-B}
  D(\sigma,N_E,N_C,\mbf{c}^\ast)  :=  \min_{\mbf{R} \geq \mbf{0},\ \mbf{L}}
  D_{KL}(\mbf{c}^\ast\| \bar{\mbf{v}})
\ee
 
is equivalent to \eqref{eq:problem-A}. Since $\mbf{v}$ is explicitly
known as a function of $(\mbf{R}, \mbf{L})$, optimization B \eqref{eq:problem-B} is a
more tractable optimization problem than~\eqref{eq:problem-A}. Note that in this setting,
the function $D(\sigma,N_E,N_C,\mbf{c}^\ast)$ \eqref{eq:problem-B} is independent of the
rates $\bs{\mu}$.

While this optimization is easy to implement,  we note that the parameters
(e.g., reaction rates, specificity) are not constrained  
to take only physically relevant values; a legitimate concern is that 
the absence of such
physico-chemical constraints might drive this optimization to physically
unrealistic solutions.

There are two possible ways to impose these parameter constraints. One is
to impose constraints on the ``microscopic'' chemical parameters, such as
the rate of 
individual reactions $R(j,k,\a)$ and the inter-cisternal transport
rate ${\mu}^{(j)}$. These take into consideration constraints arising from
molecular enzymatic processes.  
The other is to impose constraints on ``global'' physical parameters, such as the total transport time across the Golgi cisternae and the
average  enzymatic reaction time.
Here, we impose constraints on the microscopic reaction and transport parameters.

{\it Optimization C} :
\bea
\lefteqn{D(\sigma,N_C, N_E,{\bf{c^*}}) :=} 
\hspace*{1.3in}
\begin{array}[t]{rl}
  \min_{\bs{\mu}, \mbf{R}, \mbf{L}}  & D_{KL}(\mbf{c}^\ast \| \bar{\mbf{v}}) \\
  \text{s.t.} &   R_{\min} \leq R(j,k,\a) \leq R_{\max}, \nn \\
                                     &   \mu_{\min} \leq \mu^{(j)} \leq  \mu_{\max}.
\end{array}
\label{eq:mu_R_problem}
\eea
The upper and lower bounds on the rates $\mbf{R}$ and $\bs{\mu}$ are estimated in Appendix~\ref{sec:param_estimate} : $\mu_{\max} = 1$/min (resp.\ $\mu_{\min} = .01$/min) and $R_{\max} = 20$/min (resp.\ $R_{\min} = .018$/min).

\section{Equivalence of Optimizations $A$ and $B$}
\label{sec:equivalence_opt_problem}

Let 
\[
  \mathcal{A} =  \left\{
    [c_k^{(j)}]_{j,k}  : 
    \begin{array}{l}
      M(j,k,\a) \geq 0,
      V(j,k,\a) \geq 0, 
      l_\a^{(j)} \geq 0,\\
      \text{$[c_{k}^{(j)}]_{jk}$ given by 
      \eqref{eq:concentrations1} and \eqref{eq:concentrationsj}}
    \end{array}
  \right\}
\]

denote the set of concentrations achievable in Optimization $A$. Similarly, let

\[
  \mathcal{B} =  \left\{ [v_k^{(j)}]_{j,k} :
    \begin{array}{l}
      R(j,k,\a) \geq 0, l_\a^{(j)} \geq 0\\
      \text{$[v_k^{(j)}]_{j,k}$ given by \eqref{eq:Bconcentrations1}
      and \eqref{eq:Bconcentrationsj}}
    \end{array}
  \right\}
\]

denote the set of concentrations achievable in the Optimization $B$.

Our task is to show that $\mathcal{A} = \mathcal{B}$.
Suppose $[c_k^{(j)}]_{j,k} \in \mathcal{A}$. Let $[M(j,k,\a)]$,
$[V(j,k,\a)]$ and $[l_\a^{(j)}]$ be the corresponding parameters. Define  
\[
  R(j,k,\a) = 
         \sum_{\a=1}^{N_E} \frac{V(j, k, \a)}
                     {M(j,k, \a) \l1 + \sum_{k'=1}^{N_s} \frac{P^{(j)}(k',\a)c_{k'}^{(j)}}{M(j,k', \a)}\r}  \; \geq 0
\]
Then $[c_k^{(j)}]_{j,k} \in \mathcal{B}$.

Suppose $[v_k^{(j)}]_{j,k} \in \mathcal{B}$. Let $[R(j,k,\a)]$, $[l_\a^{(j)}]$
denote the corresponding parameters. Since $\sum_{k=1}^{N_s}  v_k^{(j)} =
1/\mu^{(j)} < \infty $, it follows that $\sum_{k=1}^{N_s} P^{(j)}(k,\a)
v_k^{(j)} < 1/ \mu^{(j)} < \infty$. 
Thus, there exists parameters $[M(j,k,\a)]$,
$[V(j,k,\a)]$ and $[l_\a^{(j)}]$
such that  
\be
  R(j,k,\a) = 
         \sum_{\a=1}^{N_E} \frac{V(j, k, \a)}
                     {M(j,k, \a) \l1 + \sum_{k'=1}^{N_s} \frac{P^{(j)}(k',\a)v_{k'}^{(j)}}{M(j,k', \a)}\r} 
\ee
Therefore, $[v_k^{(j)}]_{j,k}$ satisfy \eqref{eq:concentrations1} and
\eqref{eq:concentrationsj}, i.e.  $[v_k^{(j)}]_{j,k} \in \mathcal{A}$. 

Moreover, suppose $\mbf{v}$ satisfies
\eqref{eq:Bconcentrations1}-\eqref{eq:Bconcentrationsj} for a given set of
parameters $(\mbf{R}, \mbf{L})$. Then there exist $(\mbf{M}, \mbf{V}, \mbf{L})$
such that $\mbf{v}$ satisfies
\eqref{eq:concentrations1}-\eqref{eq:concentrationsj}, i.e. $\mbf{v}$ is
the steady state concentration for $(\mbf{M}, \mbf{V}, \mbf{L})$. 

\section{Analytical solution when $N_s \gg 1$}
\label{sec:analytic_calc}
It is possible to obtain analytical expressions for the steady state glycan distribution, in the limit
$N_s \gg 1$, when the glycan index $k$ can be approximated by a continuous variable. In this case, \eqref{eq:concentrations1}-\eqref{eq:concentrationsj} 
can be cast as differential equations,
\bea
\lefteqn{\f{dc^{(1)}_k}{dk}  \approx  c^{(1)}_k - c^{(1)}_{k-1}} \nn \\
&& \hspace*{-0.25in} =   \left(\frac{\sum_{\a =1}^{N_E}R(1,k-1,\a)\exp(-\sigma \abs{k-1 -
      l_\a^{(1)}})}{\mu^{(1)} + \sum_{\a =1}^{N_E}R(1,k,\a) \exp(-\sigma
    \abs{k - l_\a^{(1)}})} -1 \right)  
c^{(1)}_{k-1} \nn \\  
&& \hspace*{-0.25in} \approx -\left(\frac{ \mu^{(1)} + \f{d}{dk} \sum_{\a
      =1}^{N_E}R(1,k,\a)\exp(-\sigma \abs{k - l_\a^{(1)}})}{\mu^{(1)} +
    \sum_{\a =1}^{N_E}R(1,k,\a)\exp(-\sigma \abs{k - l_\a^{(1)}})}
\right) c^{(1)}_k, \nn \\
\label{DifferentialEquantion1} 
\eea
and
\bea
\lefteqn{\f{dc^{(j)}_k}{dk}  \approx  c^{(j)}_k - c^{(j)}_{k-1}} \nn \\
&= & \frac{\mu^{(j-1)}}{\mu^{(j)} + \sum_{\a
 =1}^{N_E}R(j,k,\a)\exp(-\sigma \abs{k - l_\a^{(j)}})} c^{(j-1)}_k
\nn \\ 
&& \hspace*{-0.25in}  \mbox{}  -  \left( \frac{ \mu^{(j)} + \f{d}{dk}\sum_{\a
      =1}^{N_E}R(j,k,\a)\exp(-\sigma \abs{k- l_\a^{(j)}})}{\mu^{(j)} +
    \sum_{\a =1}^{N_E}R(j,k,\a) \exp(-\sigma \abs{k - l_\a^{(j)}})}
\right) c^{(j)}_k \nn \\ 
\label{DifferentialEquationNc}  
\eea  
for $j =2, \ldots, N_C$. In \eqref{DifferentialEquantion1} and
\eqref{DifferentialEquationNc}, 
\bea
\lefteqn{\f{d}{dk}\sum_{\a =1}^{N_E}R(j,k,\a)\exp(-\sigma \abs{k-
    l_\a^{(j)}})} \nn \\
&& = \sum_{\alpha = 1}^{N_E} R(j,k,\a) \sigma \exp(-\sigma \abs{k-
  l_\a^{(j)}}) (1- 2\mathbb{I}(k \geq l_\a))   
 +\; R'(j,k,\a) \exp(-\sigma \abs{k-
  l_\a^{(j)}}) 
\eea
where the indicator function $\mathbb{I}(\cdot)$ is equal to $1$ if the
argument is true, and zero otherwise and $R'(j,k,\a)$  is the derivative of $R(j,k,\a)$ with respect to $k$.

Define a vector function $C(k) \in \mathbb{R}^N_c$ of the continuous
variable $k$ by $C(k) = [c^{(1)}_k, c^{(2)}_k,\ldots
c^{(N_C)}_k] $. Then~\eqref{DifferentialEquantion1} and
\eqref{DifferentialEquationNc} can be written as:
\be 
\f{d C(k)}{dk} = M(k) C(k) 
\label{MatrixEqn}
\ee
where the matrix $M(k)$ is given by 
\bea
M(k) & = & 
\begin{bmatrix}
A^{(1)}(k) &  0              & 0              & 0                      & \ldots 0 \\
B^{(2)}(k) & A^{(2)}(k)  & 0              & 0                      & \ldots 0 \\
0              & B^{(3)}(k)  & A^{(3)}(k) & 0                      & \ldots  0 \\
\vdots      & \vdots        & \vdots      & \vdots               & \vdots   \\
0             & \ldots          &0              & B^{({N_C})}(k)  & A^{({N_C})}(k)\\
\end{bmatrix} \nn \\
\eea
with
\bea
A^{(j)}(k) & =  &  -  \frac{\mu^{(j)}+ \f{d}{dk} \sum_{\a
    =1}^{N_E}R(j,k,\a)  \exp(-\sigma \abs{k - l_\a^{(j)}})}{\mu^{(j)} +
  \sum_{\a =1}^{N_E}R(j,k,\a) \exp(-\sigma \abs{k - l_\a^{(j)}})}   \nn
\\ 
B^{(j)}(k) & = & \frac{\mu^{(j-1)}}{\mu^{(j)} + \sum_{\a
    =1}^{N_E}R(j,k,\a) \exp(-\sigma \abs{k - l_\a^{(j)}})} \nn 
\eea
The functions $A^{(j)}(k)$ and $B^{(j)}(k)$ involve absolute value and
indicator functions; therefore, the differential equation has to be
solved in a piecewise manner assuming continuity of solution $C(k)$.

The general solution of \eqref{MatrixEqn} 
\be
C(k) = C_0 \exp \left( \Omega(k)  \right) 
\ee
is written in terms of the Magnus Function $ \Omega(k) = \sum_{n=1}^{\infty}
\Omega(n,k)$, obtained from the
Baker-Campbell-Hausdorff formula~\cite{Magnus},
\beas
\Omega(1,k) & = & \int_0^k M(k_1) dk_1 \nn   \\
\Omega(2,k) & = & \f{1}{2} \int_0^k dk_1\int_0^{k_1} dk_2 \left[ M(k_1),M(k_2) \right]    \nn \\
\Omega(3,k) & = & \f{1}{6} \int_0^k dk_1\int_0^{k_1} dk_2 \int_0^{k_2}
dk_3 \left[ M(k_1), \left[ M(k_2),M(k_3) \right]  \right]   
       \mbox{}  + \left[ M(k_3), \left[ M(k_2),M(k_1)
  \right]  \right]  \nn \\
  \ldots \ldots
\eeas
where  $\left[M(k_1),M(k_2) \right] := M(k_1) M(k_2) - M(k_2) M(k_1)$
is  the commutator, and the higher order terms in $\ldots$ contain higher order nested
commutators. 

Here, we establish conditions under which the series
$\sum_{n=1}^{\infty}\Omega(n,k)$ that defines solution $C(k)$
to the differential
equation \eqref{MatrixEqn} converges. We also solve~\eqref{MatrixEqn} for
some special cases.

The commutator 
\beas
\lefteqn{[M(k_1), M(k_2)] = }
\hspace{1.1in} 
\begin{bmatrix}
0              &  0              & 0              & 0         &  \ldots
&  0 \\ 
a_{21}      & 0              & 0               & 0         &  \ldots    &  0 \\
a_{31}      & a_{32}      & 0               & 0          & \ldots    & 0 \\
0               & a_{42}     & a_{43}      & 0           & \ldots     & 0 \\  
\vdots      & \vdots        & \vdots      & \vdots               & & \vdots   \\
0             & \ldots       &   &a_{n,n-2}              & a_{n,n-1}
& 0\\ 
\end{bmatrix}
\eeas  
where
\beas
a_{i,i-1} &= & A^{(i-1)}(k_2) B^{(i)}(k_1) + A^{(i)}(k_1) B^{(i)}(k_2) 
   - A^{(i-1)}(k_1) B^{(i)}(k_2) + A^{(i)}(k_2) B^{(i)}(k_1)  \nn \\
 a_{i,i-2} & = & B^{(i-1)}(k_2) B^{(i)}(k_1) - B^{(i-1)}(k_1) B^{(i)}(k_2)
\eeas

The general form of  $\Omega(n,k)$ is given by~\cite{Magnus}
\be
\begin{split}
\Omega(n,k) = & \f{z_n}{n!} \int_0^k dk_1\int_0^{k_1}dk_2\ldots
\int_0^{k_{n-2}}dk_{n-1}  
\int_0^{k_{n-1}}dk_n \sum_{l} W_l M(k_{p^l_1}) M(k_{p^l_2})\ldots
M(k_{p^l_n}) 
\end{split} 
\ee
where $(p^{(l)}_1, p^{(l)}_2 \ldots p^{(l)}_n )$ is  a permutation of
$(1,2,3,\ldots n)$, $W_l \in \lbrace-1, 1\rbrace$, and $z_n \in {1
  ,\ldots} n$. 
  

Let $\bar{A}  =  \max_{k,l,m}\abs{M_{l,m}(k)}$. Define
\be
\bar{M} =
\begin{bmatrix}
\bar{A} &  0         & 0       & 0  \ldots  &    0 \\
\bar{A} & \bar{A}    & 0       & 0  \ldots  &    0 \\
0      	& \bar{A}    & \bar{A} & 0  \ldots  &    0 \\
\vdots  & \vdots     & \vdots  &            & \vdots \\
0       & \ldots     & 0       & \bar{A}    & \bar{A}    \\
\end{bmatrix}
\nn
\ee
We can bound all the matrix elements of $\Omega(n,k)$ in the following way
\bea
\Omega_{lm}(n,k) & \leq & z_n \bar{M}^n_{l,m} \int_{0}^k dk_1
\int_{0}^{k_1} dk_2 \ldots \int_{0}^{k_{n-1}} dk_n  \nn\\ 
&  =  & z_n \bar{M}^n\Big|_{lm} \f{k^n}{n!}
\eea
The matrix 
\[
\bar{M}^n = 
\begin{bmatrix}
a_{11} &  0       & 0       & 0        & \ldots    &  0 \\
a_{21} & a_{22}   & 0       &  0       &  \ldots   &  0 \\
a_{31} & a_{32}   & a_{33}  & 0        & \ldots    & 0 \\
a_{41} & a_{42}   & a_{43}  & a_{44}   & \ldots    & 0 \\  
\vdots & \vdots   & \vdots  & \vdots   &           & \vdots \\
a_{n1} & \ldots   &         & a_{n,n-2}& a_{n,n-1} & a_{nn}\\
\end{bmatrix}
\]  
where 
$a_{lm} =  S_{lm}(n) \bar{A}^n$
for appropriately defined polynomials $S_{l,m}(n)$. Thus, it follows that
$ \Omega_{lm} \leq z_n S_{lm}(n) (A^*)^n \f{k^n}{n!}$ and
$\Omega_{l,m}(k)  \leq \sum_{n=1}^{\infty} z_n S_{l,m}(n) (A^*)^n
\f{k^n}{n!}$. Consequently, the series will converge if $\bar{A}k < 1$,
i.e.  $k \leq \frac{1}{\bar{A}}$. 
Assuming $\mu^{(j)} = \mu \;\forall j$, we can bound $\bar{A}$ as   
\bea 
\bar{A}  \leq \max_{j,k} \left( \frac{\mu+ \sigma \sum_{\a
    =1}^{N_E}R(j,k,\a)  \exp(-\sigma \abs{k - l_\a^{(j)}})}{\mu +
  \sum_{\a =1}^{N_E} R(j,k,\a)\exp(-\sigma \abs{k - l_\a^{(j)}})} \right. 
  +  \left. \f{\sum_{\a =1}^{N_E} R'(j,k,\a) \exp(-\sigma \abs{k - l_\a^{(j)}})}{\mu +
  \sum_{\a =1}^{N_E}R(j,k,\a)\exp(-\sigma \abs{k - l_\a^{(j)}})} \right)
\eea
Since the parameters $\mu,\; \sigma,\; R(j,k,\a),\; l^{(j)}_\a$ and $N_E$ are finite and
positive, and $R'(j,k,\a)$ is finite, $\bar{A}$ has a finite upper bound, implying $k$ is always greater than zero, and the series has finite radius of convergence.

While in principle we can obtain the glycan profile for any $N_E$ and $N_C$ with arbitrary accuracy, assuming $R(j,k,\a) = R_{\a}^{(j)}$, we provide explicit formulae for 
a few representative cases : (i) $(N_E = 1, N_C = 1)$ and (ii) $(N_E = 1, N_C= 2)$.

\begin{figure}
\subfloat[
]{\includegraphics[scale=.3]{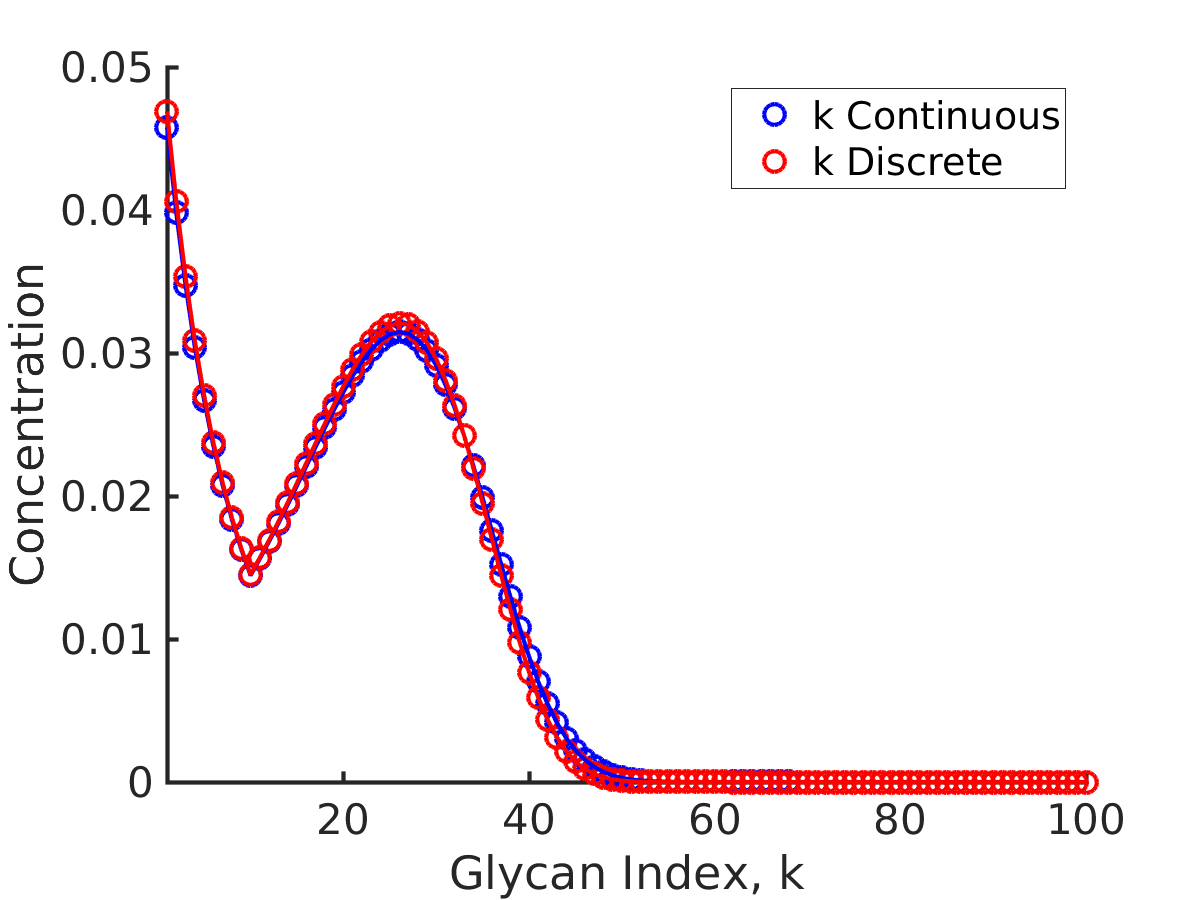}} 
\;
\subfloat[]{\includegraphics[scale=.3]{Figures/Appendix4_Figure1a.png}} 
\:
\caption{Glycan concentration profile calculated from the model using (a) formula \eqref{analyticC_nC1} for $N_E=N_C=1$  and 
(b) formulae \eqref{Analytic_Nc2_kl1l2}-\eqref{Analytic_Nc2_l2l1k} for $N_E=1, N_C=2$.} 
\label{fig:analytics}
\end{figure}

(i) $N_E=1, N_C = 1$: The solution of the differential equation is
given by  
\be
\hspace*{-0.18in}
c(k) =
\left\{
  \begin{array}{ll}
    c_0 e^{-k}   \left(\f{\mu + R \exp(-\sigma(l-k))}{\mu + R \exp(-\sigma
    l)}\right)^{(1/\sigma)-1} & k \leq l\\ 
    c(l) e^{-(k-l)} \left(\f{\mu + R }{\mu + R \exp(-\sigma (k-l))}
    \right)^{(1/\sigma) +1}  &  k>l
  \end{array}
\right.
\label{analyticC_nC1}
\ee
A representative concentration profile is plotted in
Fig.\,\ref{fig:analytics}(a). 
The concentration profile consists of two distinct components: an
initial exponential decay, and then an exponential rise
and fall concentrated around $l$. The relative weight
of these two components is controlled by the sensitivity
$\sigma$ and the rate $R$. Such explicit formulae can be obtained for any $N_E >1$, as long as $N_C=1$.

(ii) $N_E = 1, N_C=2$: The concentration profile $c^{(2)}$ in cisterna~2 can be obtained from the following calculation.
Let $l^{(j)}$ denote the ``length'' of the enzyme
in cisterna~$j = 1, 2$.
  For $k \leq \min\{l^{(1)},l^{(2)}\}$
\be
\begin{split}
c^{(2)}(k)  = \;&  c_0 \mu^{(1)} e^{-k}  \Big(\frac{ \mu^{(2)} + R^{(2)}
    \exp(-\sigma(l^{(2)}-k))}{\mu^{(1)} + R^{(1)} e^{-\sigma l^{(1)}}}
\Big)^{(1/{\sigma})-1}     
  \int_0^k \f{ (\mu^{(1)} + R^{(1)} \exp(-\sigma (l^{(1)} -
  k)))^{(1/\sigma) -1} }{ (\mu^{(2)} + R^{(2)}
  \exp(-\sigma(l^{(2)}-k)))^{1/\sigma} }  dk  \\ 
& \mbox{} + c^{(2)}(0) e^{-k}  \left(\f{\mu^{(2)} + R^{(2)} e^{-\sigma (l^{(2)} -
      k) }}{\mu^{(2)} + R^{(2)} e^{-\sigma l^{(2)} } } \right)^{(1/\sigma) -
  1}
\end{split}
\label{Analytic_Nc2_kl1l2}
\ee
Next, consider the case where $l^{(1)} \leq l^{(2)}$. Then, for $ l^{(1)}< k \leq l^{(2)}$
\bea
c^{(2)}(k) &  = &  c^{(1)}(l^{(1)}) \mu^{(1)}  e^{-(k-l^{(1)})} (\mu^{(1)}
+ R^{(1)})^{(1/\sigma) +1 }  
  (\mu^{(2)} +   R^{(2)} \exp(-\sigma(l^{(2)}-k)) )^{(1/\sigma) - 1}  \nn \\  
& & \int_{l^{(1)}}^k  \f{(\mu^{(2)} + R^{(2)} 
  \exp(-\sigma(l^{(2)}-k)))^{-1/\sigma}}{(\mu^{(1)} + R^{(1)}\exp(-\sigma
  ( k-l^{(1)} )))^{(1/\sigma) +1 }}  dk    \nn \\ 
&+&  c^{(2)}(l^{(1)}) e^{-(k-l^{(1)})  } \left(\f{\mu^{(2)} + R^{(2)}
    e^{-\sigma (l^{(2)} - k) }}{\mu^{(2)} + R^{(2)} e^{-\sigma (l^{(2)} -
      l^{(1)}) } } \right)^{(1/\sigma)-1}  
\label{Analytic_Nc2_l1kl2} 
\eea
and for  $l^{(1)} \leq l^{(2)} < k$,
\bea
c^{(2)}(k) &  = &  c^{(1)}(l^{(1)}) \mu^{(1)} e^{-(k- l^{(1)})}   
   \left (\frac{\mu^{(1)} + R^{(1)} }{ \mu^{(2)} + R^{(2)}
    \exp(-\sigma(k-l^{(2)}))} \right)^{(1/\sigma)+1}  \nn \\   
&  &   \int_{l^{(2)}}^k  \f{(\mu^{(2)} + R^{(2)}
  \exp(-\sigma(k-l^{(2)})))^{1/\sigma}}{(\mu^{(1)} + R^{(1)}\exp(-\sigma (
  k - l^{(1)})))^{(1/\sigma) +1}}  dk   \nn \\ 
&+&  c^{(2)}(l^{(2)}) e^{-(k-l^{(2)})  } \left(\f{\mu^{(2)} + R^{(2)}  }
  {\mu^{(2)} + R^{(2)} e^{-\sigma (k - l^{(2)} ) }}\right)^{(1/\sigma)+1}  
\label{Analytic_Nc2_l1l2k} 
\eea
Next, the case where $l^{(1)} \geq l^{(2)}$. For $l^{(2)}< k \leq l^{(1)}$,
\bea
c^{(2)}(k) &  = &  c_0 \mu^{(1)} e^{-k}  \frac{(\mu^{(1)} + R^{(1)}
  e^{-\sigma l^{(1)}} )^{1 - (1/\sigma)} }{(\mu^{(2)} + R^{(2)}
  \exp(-\sigma(k-l^{(2)})))^{(1/\sigma) + 1} }     
 \int_{l^{(2)}}^k  \f{(\mu^{(1)} + R^{(1)}\exp(-\sigma ( l^{(1)} -
  k)))^{(1/\sigma) -1}}{(\mu^{(2)} + R^{(2)}
  \exp(-\sigma(k-l^{(2)})))^{-1/\sigma}}  dk \nn\\   
&+&  c^{(2)}(l^{(2)}) e^{l^{(2)} - k } \left(\f{\mu^{(2)} + R^{(2)}  }
  {\mu^{(2)} + R^{(2)} e^{-\sigma (k - l^{(2)} ) }}\right)^{(1/\sigma)+1}     
\label{Analytic_Nc2_l2kl1} 
\eea
For $^{(2)} \leq l^{(1)} < k$, 
\bea
c^{(2)}(k) &  = &  c^{(1)}(l^{(1)}) \mu^{(1)} e^{-(k- l^{(1)})}    
   \left (\frac{\mu^{(1)} + R^{(1)} }{ \mu^{(2)} + R^{(2)} \exp(-\sigma(k-l^{(2)}))} \right)^{(1/\sigma)+1}  \nn \\  
&  &   \int_{l^{(2)}}^k  \f{(\mu^{(2)} + R^{(2)} \exp(-\sigma(k-l^{(2)})))^{1/\sigma}}{(\mu^{(1)} + R^{(1)}\exp(-\sigma ( k - l^{(1)})))^{(1/\sigma) +1}}  dk   \nn \\
&+&  c^{(2)}(l^{(1)}) e^{-(k-l^{(1)})  } \left(\f{\mu^{(2)} + R^{(2)} e^{-\sigma (l^{(1)} - l^{(2)} ) }  } {\mu^{(2)} + R^{(2)} e^{-\sigma (k - l^{(2)} ) }}\right)^{(1/\sigma) +1}   
\label{Analytic_Nc2_l2l1k}
\eea
The integrals in \eqref{Analytic_Nc2_kl1l2} to \eqref{Analytic_Nc2_l2l1k}
can evaluated numerically.
The result of the numerical computation is shown in Fig.\,\ref{fig:analytics}.

\section{Capability of the chemical network model to generate complex distributions}
\label{sec:model capability}


Is our glycan synthesis model capable of generating concentration distributions of arbitrary complexity?
 In what way do we need to change the parameters
$N_E, N_C, \sigma, \ldots$, in order to generate glycan distributions $\mbf{c}$ of a given complexity? The purpose of this section, is to obtain some heuristics for this
task.

We show in Appendix~\ref{sec:analytic_calc} 
that \eqref{eq:problem-B} can be solved 
analytically in the limit
$N_s \gg 1$, because in this limit the glycan index $k$ can be approximated by a continuous
variable, and  the recursion relations for the steady state glycan
concentrations \eqref{eq:concentrations1}-\eqref{eq:concentrationsj}  
can be cast as a matrix differential equation. This allows us to obtain an
\emph{explicit} expression for the steady state concentration in terms of 
the parameters $(\mbf{R}, \mbf{L})$. 

\begin{figure*}
\subfloat[]
{\includegraphics[scale=0.3]{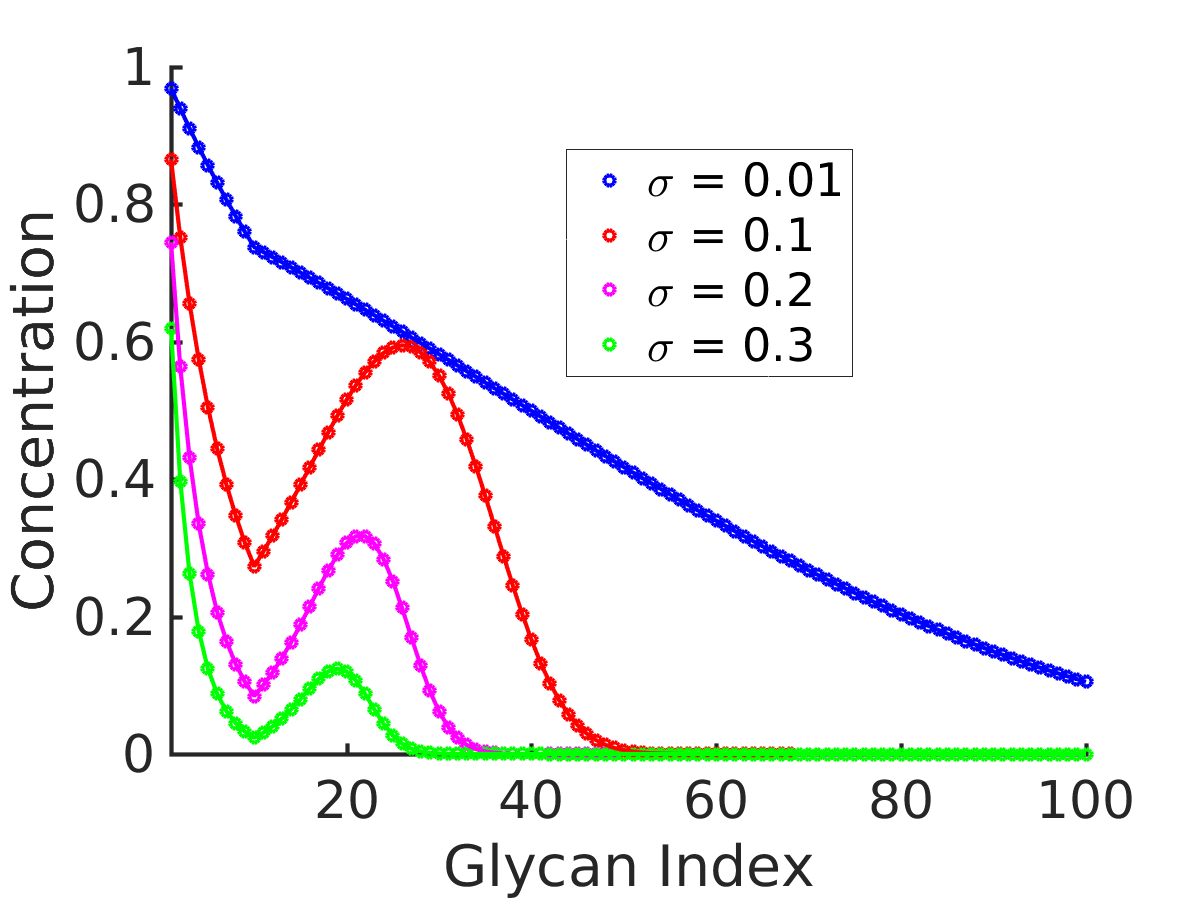}} 
\;
\subfloat[]
{\includegraphics[scale=0.3]{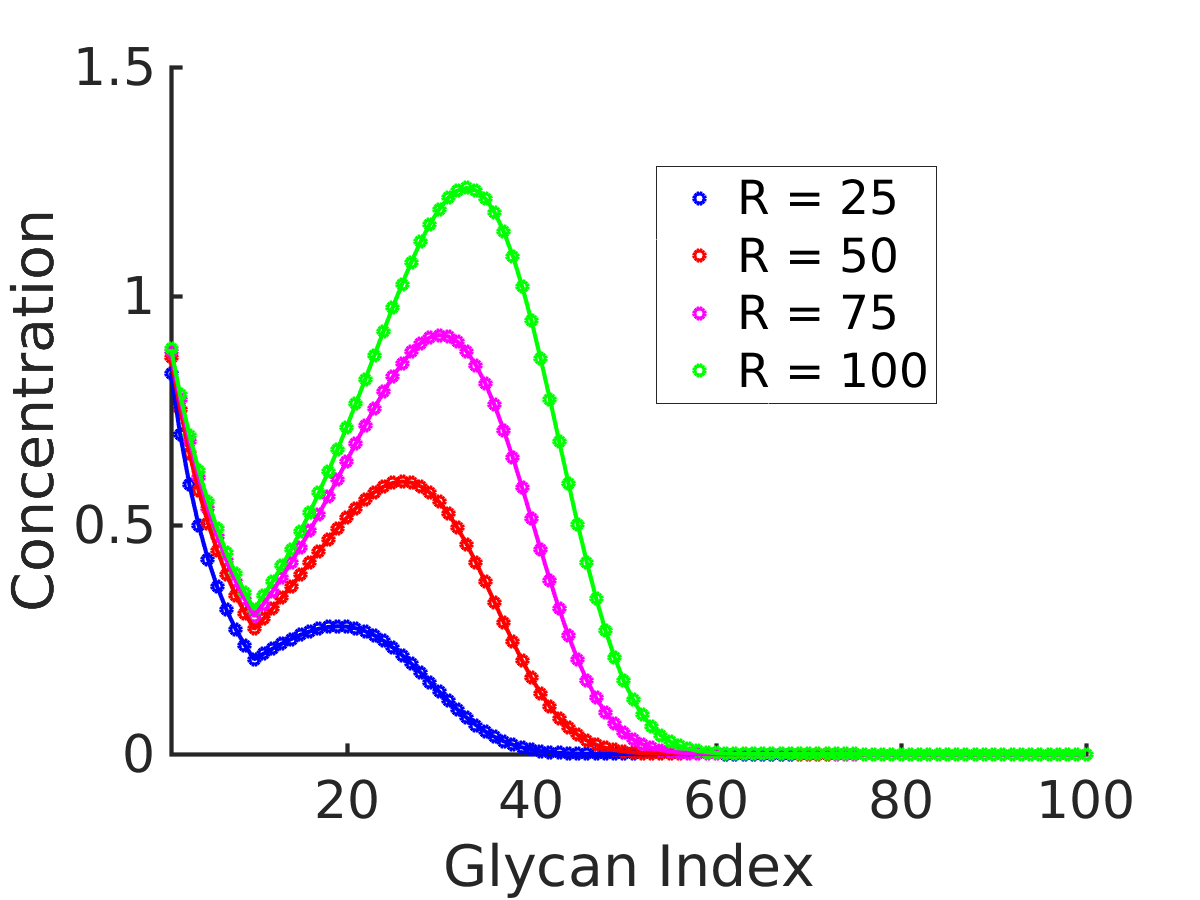}}
\\
\subfloat[]
{\includegraphics[scale=0.3]{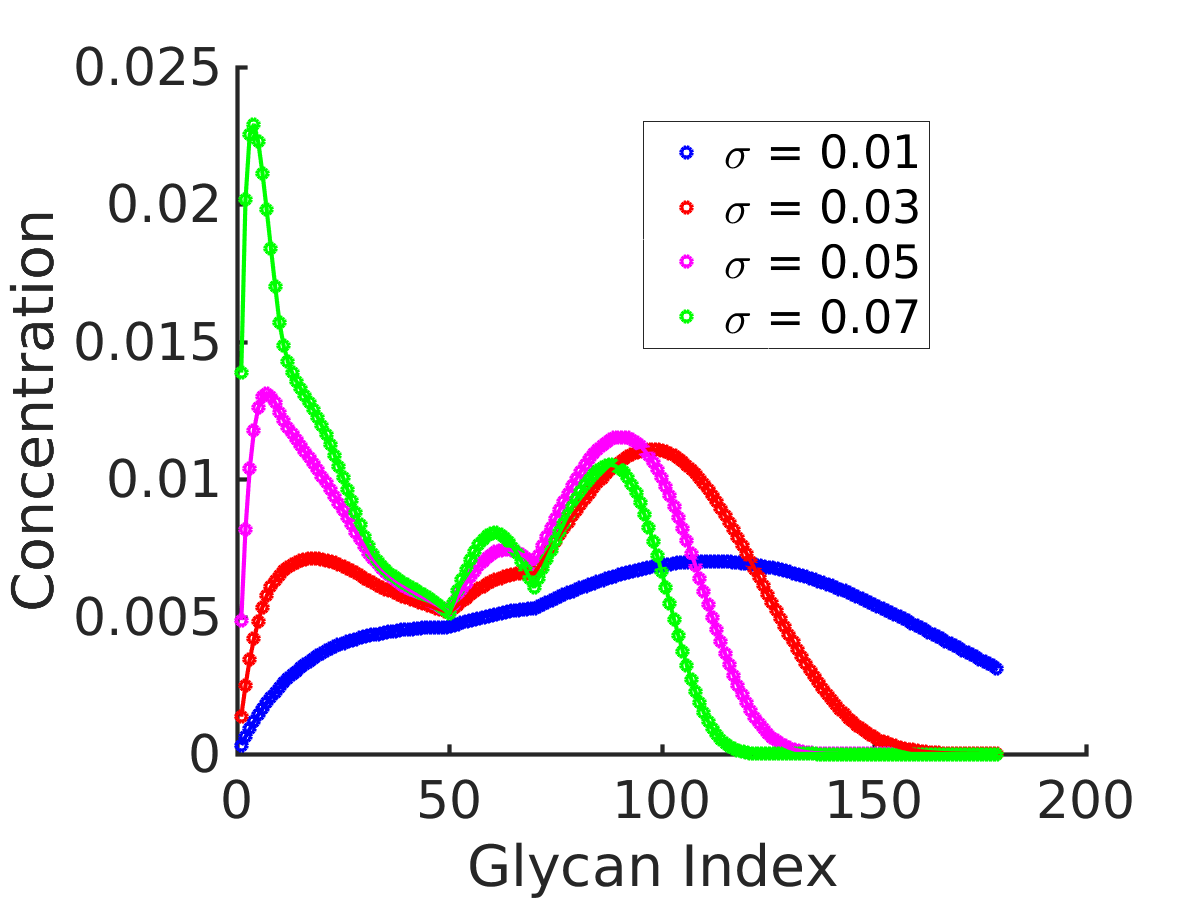}}
\:
\subfloat[]
{\includegraphics[scale=0.3]{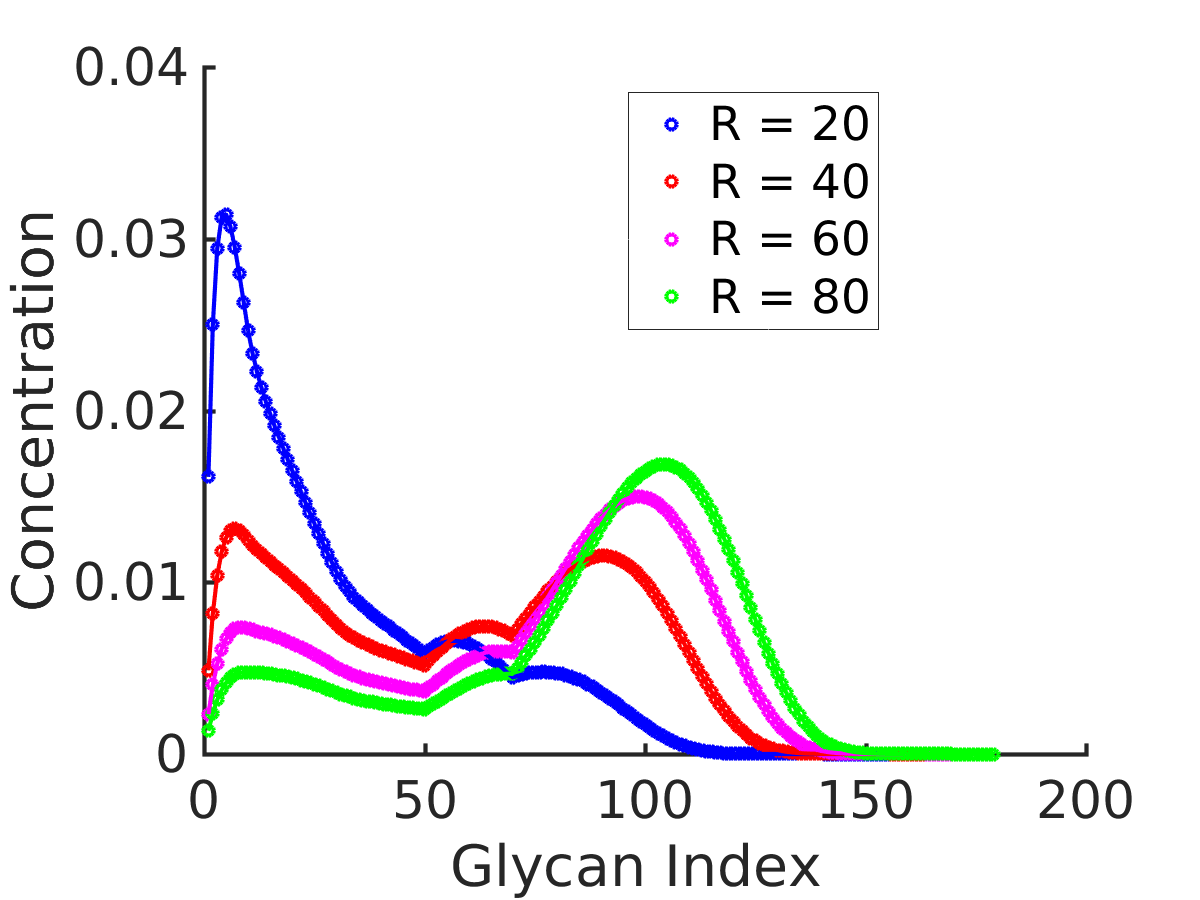}}
\caption{Glycan profile $\{c_k: k = 1, \ldots, N_s\}$  as a function of
    specificity $\sigma$~(Fig.\,(a),\,(c)), and reaction rates
    $R$~(Fig.\,(b),\,(d)). \\
    Fig.\,(a):  $N_E =N_C =1, (R=50, \mu = 1,  l= 10)$. $c_k$ decreases
    exponentially with $k$ for very low and very high $\sigma$; however, the decay rate is
    lower at low $\sigma$. For intermediate values of $\sigma$, the
    distribution has \emph{exactly} two peaks, one of which is at $k=0$,  
    and eventually decays exponentially. The width of the distribution 
    is a decreasing function of 
    $\sigma$. \\
    Fig.\,(b): $N_E =N_C =1, (\sigma =0.1, \mu = 1, l= 10)$. At low $R$,
    $c_k$ is concentrated at low $k$. The proportion of  
    higher index  glycans in an increasing function of $R$.  \\
    Fig.\,(c): $N_E = N_C = 2, (R = 40, \mu = 1, [l^{(1)}_{1}, l^{(1)}_2,
      l^{(2)}_1, l^{(2)}_2] = [10,30, 50,70])$. As $\sigma$ increases, the
    distribution becomes more complex -- from a single peaked
    distribution at low $\sigma$ to a maximum of  
    four-peaked distribution at high $\sigma$. The peaks gets sharper, and more
    well defined as $\sigma$ increases. \\
    Fig.\,(d): $N_E=N_C = 2, (R = 40, \mu = 1, [l^{(1)}_{1}, l^{(1)}_2,
      l^{(2)}_1, l^{(2)}_2] = [10,30, 50,70])$. As in the plots in
      Fig.\,(b), increasing $R$ 
      shifts the peaks towards higher index
    glycans and the proportion of higher index glycan
    increases. 
}
\label{fig:1peak}
\end{figure*}

We derive our heuristics from a semi-analytical treatment in the limit $N_s \gg 1$ (Appendix~\ref{sec:analytic_calc}), which apart from being simple to implement in general, provides an explicit formula for $c_k$ for the case $N_E=N_C=1$ (\ref{analyticC_nC1}).
  Figures\,\ref{fig:1peak}(a)-(d)
  show the glycan profile $c_k\,\,vs.\,k$ as one varies the enzyme specificity $\sigma$, the reaction rates $R$ and transport rates $\mu$, for two different values of $N_E$ and $N_C$.
The results in the plots lead us to the following general observations:
\begin{itemize}
\item Very low specificity enzymes cannot  generate complex glycan
    distributions. Keeping everything else fixed, intermediate or high
    specificity enzymes can 
    generate glycan distributions of higher complexity by increasing $N_E$ or
    $N_C$ (Figs.\,\ref{fig:1peak}(a),(c)).



\item Decreasing the specificity $\sigma$ or increasing the rates $R$
    increases the proportion of higher index glycans.  Keeping
    everything else fixed, changes in the rate $R$ have a stronger impact
    on the relative
    weights of the higher index glycans to lower index glycans. The
    relative weight of the higher index glycans increases with increasing
    $N_E$ and $N_C$ (Figs.\,\ref{fig:1peak}(b)-(d)).
    
\item Keeping everything else fixed, decreasing enzyme specificity increases the spread of the distribution around the peaks (Figs.\,\ref{fig:1peak}(a),(c)).

\end{itemize}

\section{Parameter estimation} 
\label{sec:param_estimate}

The typical transport time of glycoproteins across the
Golgi complex is estimated to be in the range $15$-$20$
mins.~\cite{UB1997}, which corresponds to the transport rate, $\mu = .18$/min. 
We bound the transport rate for our optimization between 0.01/min and 1/min.

Next, we estimate the range of values for the chemical reaction rates. The injection rate 
$q$ is in the range $100 - 1500$ pmol/$10^6$
cell 24 h~\cite{UB1997,krambeck:2009}. For our calculation we
set $q = 387.30$ pmol/$10^6$ cells 24 hr = $0.27$ pmol/$10^6$ cells min, the
geometric mean of $100$ and $1500$. We set the range for the enzymatic
rate $R$ to be
\[
  R_{\min} = \min_{\a} \Big\{\f{V^{(\a)}_{\max}/\nu}{K^{(\a)}_M + \f{1}{\nu}\frac{q}{\mu}}
  \Big\} \leq R \leq
  R_{\max} = \max_{\a} \Big\{ \f{V^{(\a)}_{\max}/ \nu}{K_M^{(\a)}} \Big\}.
\]
where $K^{(\a)}_M$ and $V^{(\a)}_{\max}$ denote the Michaelis constants and $V_{\max}$ of the $\a$-th enzyme.
The conversion from 1 pmoles/$10^6$ cells to concentration can be obtained
by taking cisternal volume ($\nu$) to be 2.5$\mu m^3$
\cite{UB1997,krambeck:2009}.  
This gives

\be
 \text{1 pmoles/$10^6$ cells} = \f{10^{-12} \text{moles}}{10^6 \times 2.5
   \times 10^{-18} \times 10^3 \text{litre} } = 400 \mu M.     
\ee

In Table \ref{table:R} we report the parameters for the $8$ enzymes taken
from Table 3 in \cite{UB1997}. From these parameters it follows that  

\begin{eqnarray*}
  R_{\min} & = & \min_{\a} \Big\{\f{V^{(\a)}_{\max}/ \nu}{K^{(\a)}_M + \f{1}{\nu}\frac{q}{\mu}} \Big\} \\
                 & = & \frac{V^{(7)}_{\max}/ \nu}{K_{M}^{(7)} + \f{1}{\nu}\f{q}{\mu}} = \frac{.16 \times 400 \mu M /\text{min} }{3400 \mu M + 149.4 \mu M } = 0.018 \text{min}^{-1}
                 \\
  R_{\max} & = &  \max_{\a} \Big\{ \f{V^{(\a)}_{\max} / \nu}{K_M^{(\a)}} \Big\} \\
                   & = &\frac{V^{(1)}_{\max} /\nu}{K^{(1)}_M} = 
                 \frac{5 \times 400 \mu M/ \text{min}}{100 \mu M} = 20 \text{min}^{-1} 
\end{eqnarray*}

\begin{table}[ht]
\centering 
\begin{tabular}{|c|c|c|}
\hline                         
$\a$ & $K^{(\a)}_M$ & $V^{(\a)}_{\max}$ \\
      &($\mu$mol)&  (pmol/$10^6$ cell-min) \\ 
[0.5ex]
\hline \hline 
1 & 100  & 5    \\
2 & 260 &  7.5  \\
3 & 200 &  5  \\
4 & 100 &  5 \\  
5 & 190 & 2.33  \\
6 & 130 &  .16   \\
7 & 3400 & .16 \\
8 & 4000 & 9.66 \\[1ex] 
\hline                   
\end{tabular}
\caption{Enzyme parameters taken from Table\,3 in~\cite{UB1997} that we use to calculate the bounds on the reaction rate $R$. Here $K^{(\a)}_M$ and $V^{(\a)}_{\max}$ denote the Michaelis constant and $V_{\max}$ of the $\a$-th enzyme.}
\label{table:R} 
\end{table}

\section{Constructing target distributions for glycans of a given cell type} 
\label{sec:empirical_data}

The target distribution of the glycans on the cell surface
is obtained via mass spectrometry.
The x-axis of mass spectroscopy~(MS) graphs is mass/charge of the
ionised sample molecules and the y-axis is relative intensity
corresponding to each mass/charge value, taking the highest intensity as
$100\%$. 

This relative intensity roughly correlates with the relative abundances of
the molecules in the sample. 

This raw MS data is noisy and cannot be directly used as the target distribution
in our optimization problem. There are three major sources of noise in the MS
data~\cite{MS_noise}: the chemical noise in the sample, the Poisson
noise associated with detecting discrete events, and the Nyquist-Johnson noise
associated with any charge  system. We propose a simple model that
accounts for the chemical noise and the Poisson sampling noise. 
Using this noise model and the available MS data, we generate parametric bootstrap
samples of glycan measurements, and fit a Gaussian Mixture Model (GMM) on
this sample to approximate the 
glycan distribution. This
GMM probability distribution is used as the target distribution in our
numerical experiments.

\begin{figure}
\centering
{\includegraphics[scale = 0.5]{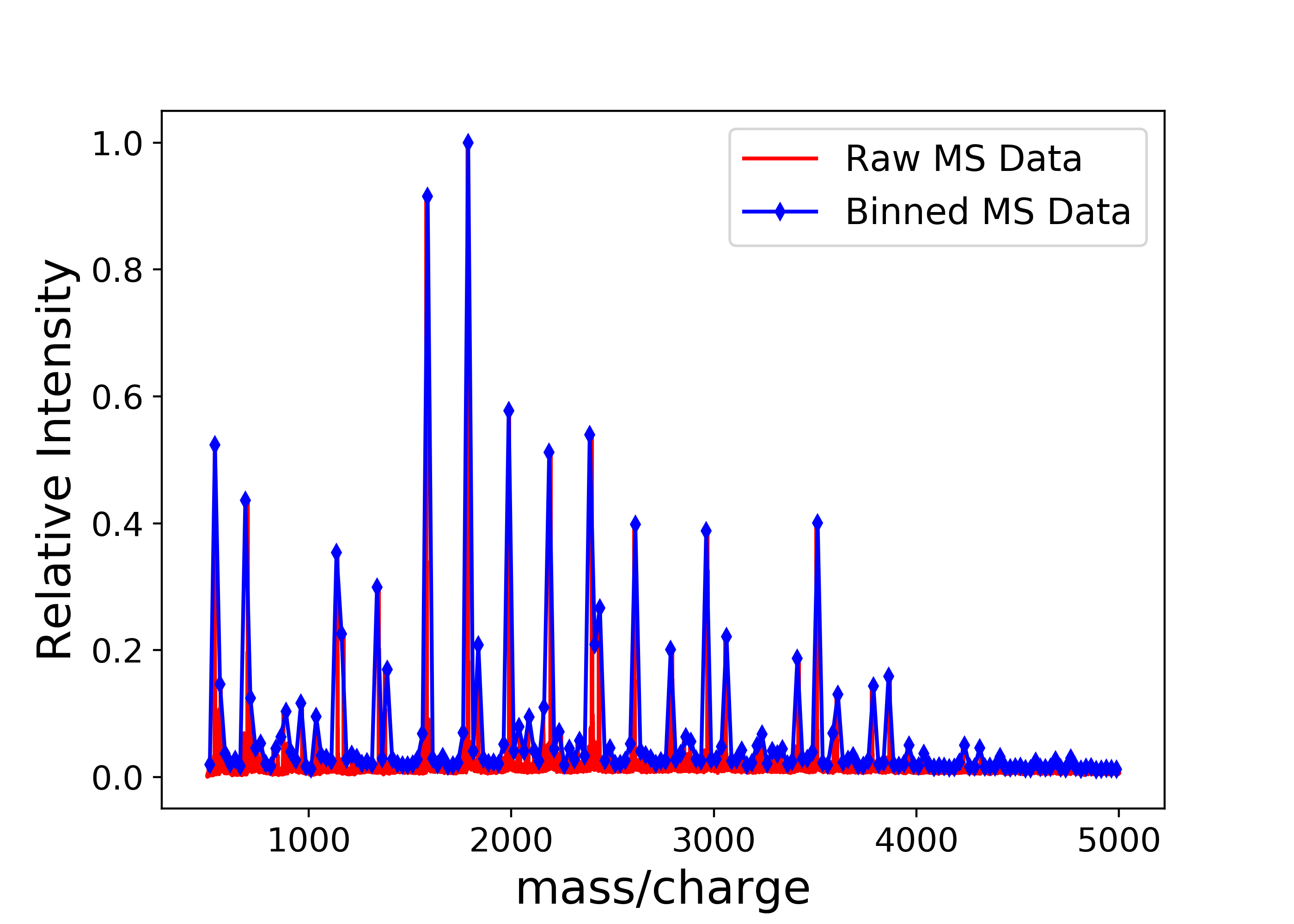}}
\caption{The binned MS data (blue) approximates the raw MS data (red) very well. We use this binned data for GMM approximation of the MS data.}
\label{fig:binning}
\end{figure}

The MS data obtained from \cite{msdata} had mass ranging between 500 to 5000 Daltons with intensity reported at every 0.0153 Daltons. We first bin this MS data into 180 bins and take the maximum value within each bin as the value of intensity for that bin. Fig.\,\ref{fig:binning} shows the raw MS data and the binned distribution.         

\begin{figure*}
\subfloat[]{\includegraphics[scale=0.3]{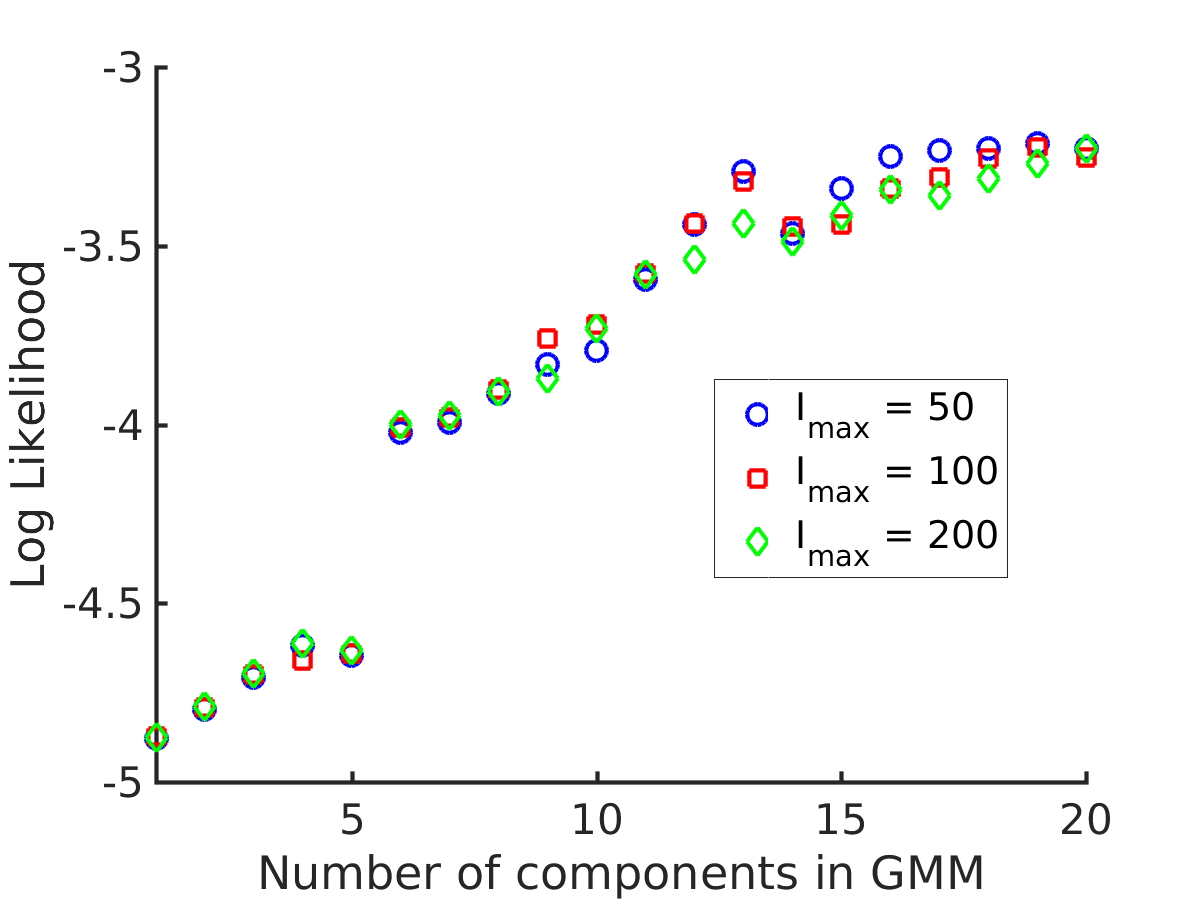}}
\:
\subfloat[]{\includegraphics[scale=0.3]{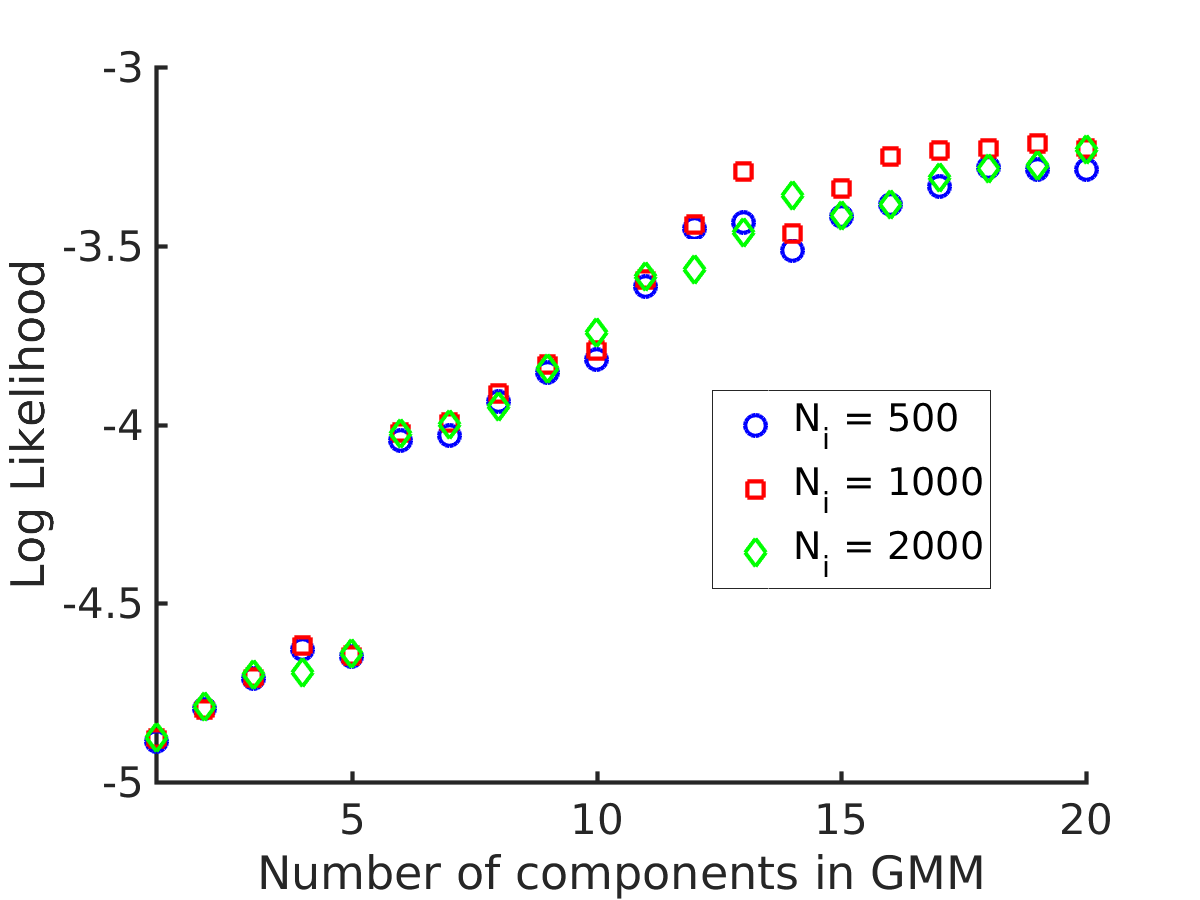}}
\caption{Log likelihood vs. number of components ($N$) in the GMM. We see that the log likelihood saturates at around $N=20$, thus $20$-GMM is a very good representation of the MS-data from {\it human} T-cells. The different symbols are for (a) different values of  the maximum intensity $I_{max}=50, 100, 200$ and (b) different values of the number of i.i.d samples $N_i = 500, 1000, 2000$, showing 
the insensitivity of the log likelihood to the value of $I_{max}$ and $N_i$.  } 
\label{fig:loglikelihood}
\end{figure*}

Let $\bar{I}_k$ represents the relative intensity of the $k$-th bin in the binned MS
graph. We generate a sample population of glycans using the MS data in the
following way:
\begin{enumerate}
\item Poisson sampling noise: The MS data does not have absolute count information. We assume an arbitrary maximum count $I_{\max}$, and define the
intensity $I_k = I_{\max} \bar{I}_k$.
 The plots in  
Fig.\,\ref{fig:loglikelihood}(a) show that the results are not sensitive to
  the specific value of $I_{\max}$. 
\item Chemical noise: The sample used for MS analysis also contains small
  amounts of molecules that are not glycans. These appear as the very small peaks in the
  MS data. We assume that the
  probability $p_k$ that the peak at index $k$ corresponds to a glycan is
  given by
  \[
    p_k = 1 - e^{- \f{{I}_k}{I_{max}}} = 1 - e^{-\bar{I}_k}
  \]
 which adequately suppresses this chemical noise.
\item Bootstrapped glycan data:
  The count $n_k$ at the glycan index $k$ is distributed according to the
  following distribution:
  \[
      n_k = \left\{
        \begin{array}{lll}
          0 & (1-p_k) & n = 0\\
          n & p_k e^{-I_k} \f{(I_k)^n}{n!} & n \geq 1.
        \end{array}
      \right.
    \]  
  We assume that the MS data was generated from $N$ different cells. Thus,
  the total count at glycan index $k$ is given by the sum of $N_i$
  i.i.d. samples distributed according to the distribution above. We
in Fig.\,\ref{fig:loglikelihood}    (b) show that results are insensitive to $N_i$.
\end{enumerate}
Next, we interpret the counts as samples from a ``spatial" distribution
$f$. We approximate this distribution as a Gaussian mixture, i.e. 
$f(x) = \sum_{\ell= 1}^{N} \gamma_{\ell} \,\eta(x\mid
\mu_{\ell},\sigma_{\ell})$, where $\eta(x\mid \mu,\sigma)$ denotes the
density of a normally distributed random variable with mean $\mu$ and standard deviation
$\sigma$ at the location $x$. In this setting, we assume that each count 
is a sample from the distribution $\eta(x \mid \mu_{\ell}, \sigma_{\ell})$
with probability $\gamma_{\ell}$. Thus, each count is classified as coming
from one of the Gaussian components.

\section{Numerical scheme for performing the non-convex optimization} 
\label{sec:numericalscheme}

We solve Optimization C using the numerical
scheme detailed below. The optimization problem consists of minimising a
non-convex objective with linear box constraints. We use the 
MATLAB FMINCON function to solve this optimization. We use Sequential Quadratic Programming (SQP),
a gradient based iterative optimization scheme for solving
optimizations with non-linear differentiable objective and
constraints. Since our problem is non-convex and SQP
only gives local minima, we initialise the algorithm with
many random initial points. We use SOBOLSET function of MATLAB to generate space filling pseudo random
numbers. We have taken $1000$ initialisations for each $N_E, N_C$ and $\sigma$ value. We have taken $50$ equally spaced points
between $0$ and $1$  to explore the $\sigma$-space for Fig.~\ref{fig:DklVsSigma}. Some minor
fluctuations in $D$ due to non-convexity of the objective function in the  final results were smoothed out by
taking the convex hull of the $D$ vs. $\sigma$ graph. The results for $\sigma_{min}(N_E,N_C)$ and $D(\sigma_{min},N_E,N_C)$ (Fig.~\ref{fig:NeNcphasespace}) were obtained by adding $\sigma$ to the optimization vector and then performing the optimization. The sensitivity results (Figs.~\ref{fig:sensitivity3GMM} and~\ref{fig:sensitivity3GMM}) were obtained by approximating the $D$ vs $\sigma$ graph around $\sigma_{min}$ with a parabola, the coefficient of the quadratic term being the curvature of the graph at $\sigma_{min}$. 

A similar numerical scheme was used to optimize diversity. 



\begin{figure*}
\subfloat[$c_{th} = \f{1}{N_s}$]{\includegraphics[scale=.25]{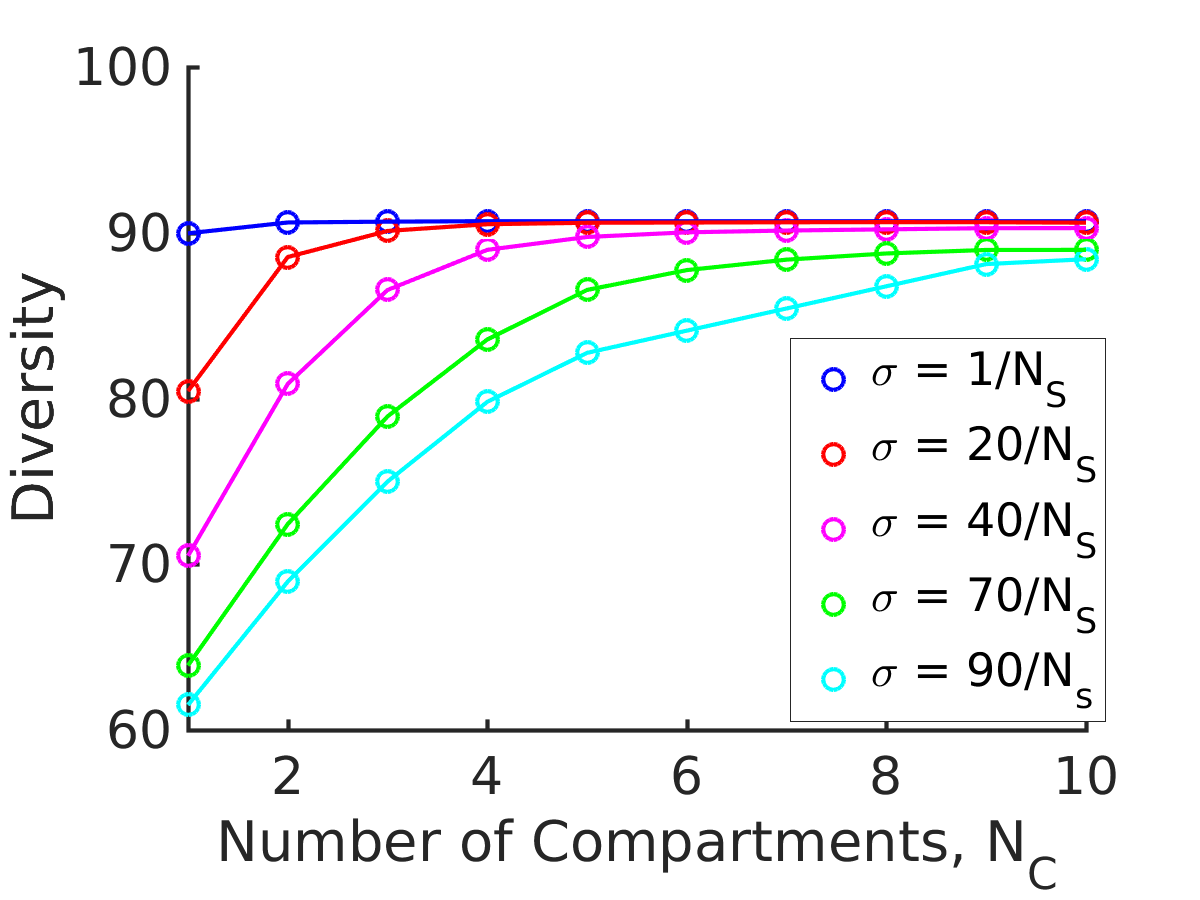}}
\:
\subfloat[$c_{th} = \f{1}{2 N_s}$]{\includegraphics[scale=0.25]{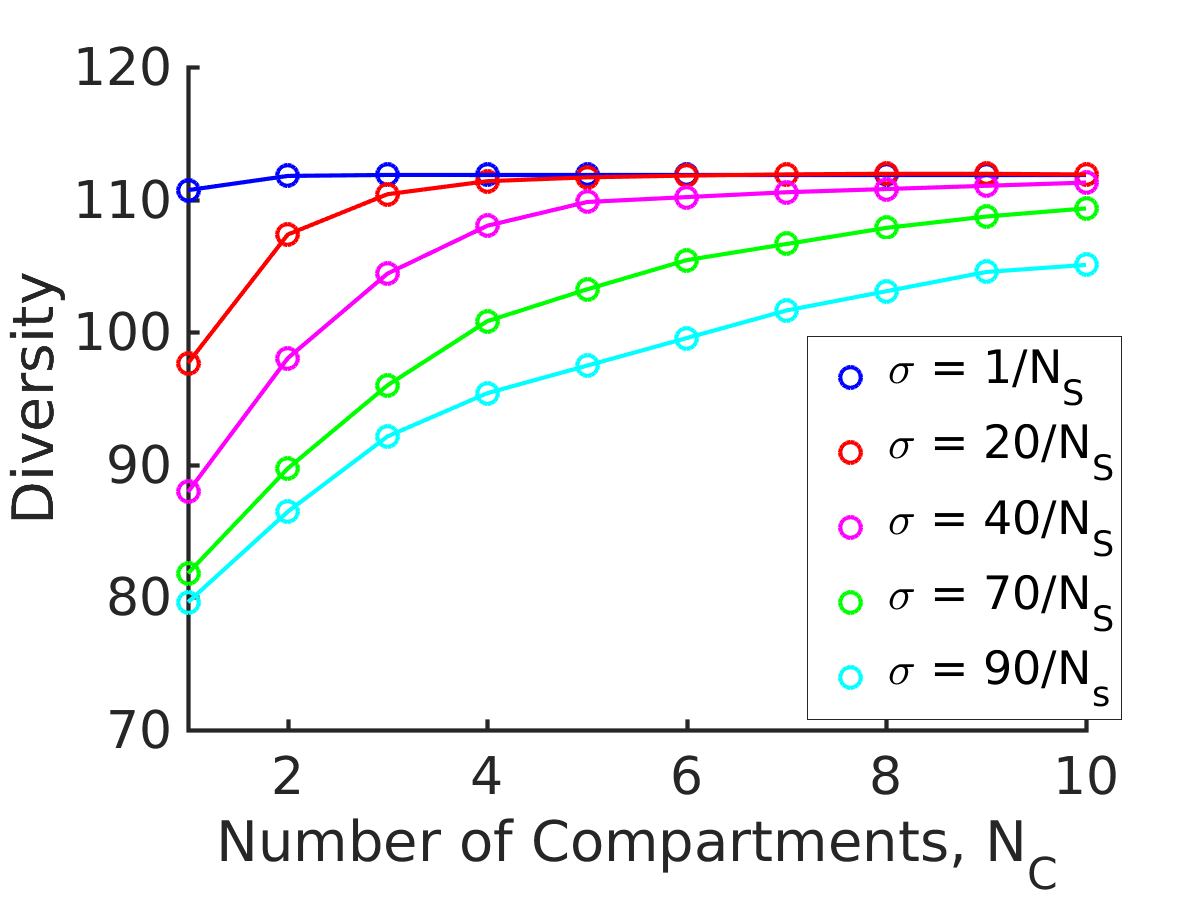}}
\:
\subfloat[$c_{th} = \f{1}{4 N_s}$]{\includegraphics[scale=0.25]{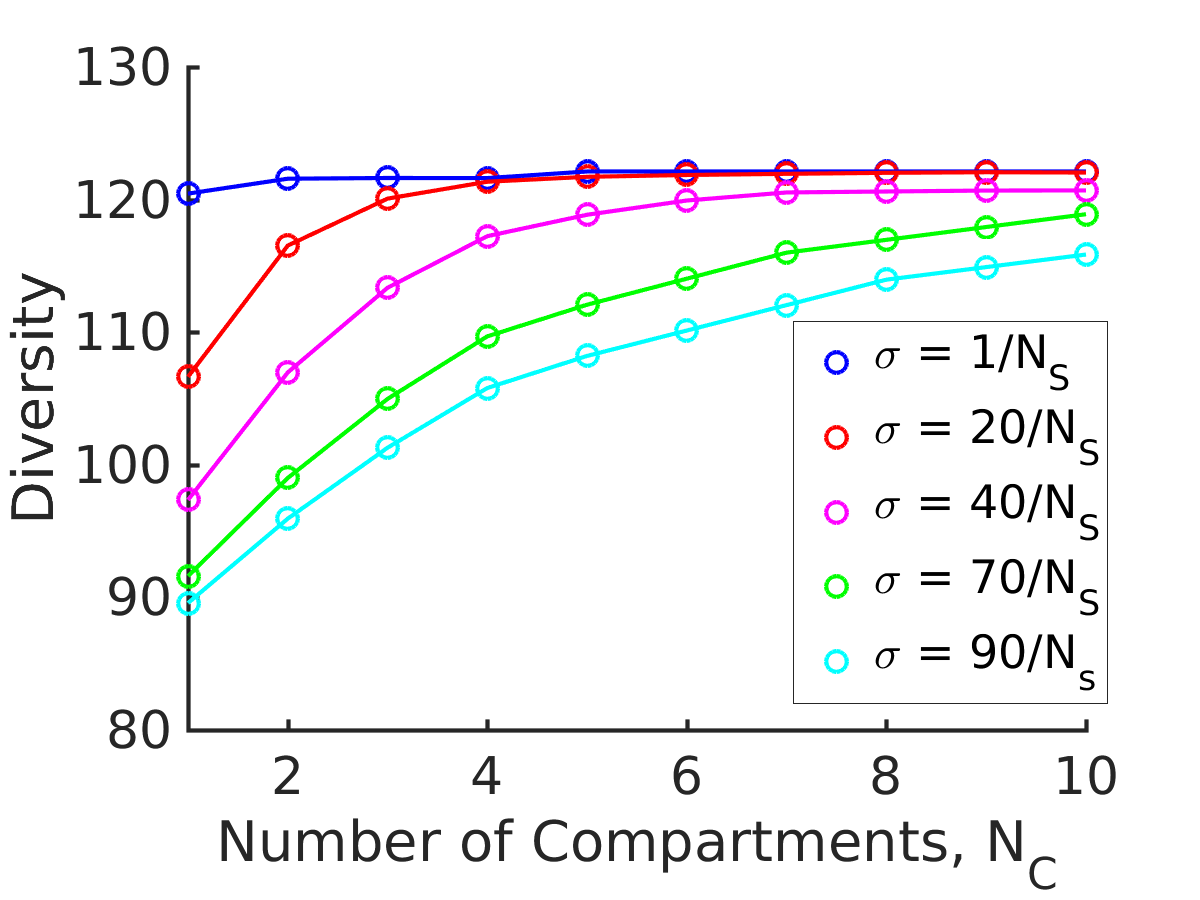}}
\caption{Diversity vs. $N_C$ for different values of $\sigma$ keeping $N_E=1$ fixed, for three different values of the threshold, $c_{th}=\f{1}{N_s},\;\f{1}{2 N_s},\;\f{1}{4 N_s}$. Changing the value of the threshold $c_{th}$, only changes the saturation value of the diversity curve. }
\label{fig:threshold_independence}
\end{figure*}

%

\bibliography{Refrences.bib}

\end{document}